\documentclass[12pt,preprint]{aastex}

\usepackage{graphicx}
\usepackage{amsmath}

\shorttitle{Background False Positives from {\it Kepler} Data}
\shortauthors{S. Bryson et al.}

\newcommand{\includeFigureFile}{}

\begin{document}

\title{Identification of Background False Positives from {\it Kepler} Data}

\author{
Stephen~T.~Bryson\altaffilmark{1},
Jon~M.~Jenkins\altaffilmark{2},
Ronald~L.~Gilliland\altaffilmark{3},
Joseph~D.~Twicken\altaffilmark{2},
Bruce~Clarke\altaffilmark{2},
Jason~Rowe\altaffilmark{2},
Douglas~Caldwell\altaffilmark{2},
Natalie~Batalha\altaffilmark{1,4},
Fergal~Mullally\altaffilmark{2},
Michael~R.~Haas\altaffilmark{1},
Peter~Tenenbaum\altaffilmark{2}
}

\altaffiltext{1}{NASA Ames Research Center, Moffett Field, CA 94035; steve.bryson@nasa.gov}
\altaffiltext{2}{SETI Institute/NASA Ames Research Center, Moffett Field, CA 94035}
\altaffiltext{3}{Center for Exoplanets and Habitable Worlds, The Pennsylvania State University, University Park, PA Ê16802}
\altaffiltext{4}{Department of Physics and Astronomy, San Jose State University, San Jose, CA 95192}
%

\begin{abstract}

      The {\it Kepler Mission} was launched on March 6, 2009 to perform a photometric 
survey of more than 100,000 dwarf stars to search for Earth-size planets 
with the transit technique. The reliability of the resulting planetary candidate list
relies on the ability to identify and remove false positives.  Major sources of astrophysical false positives 
are planetary transits and stellar eclipses on background stars.  We describe several new techniques for the
identification of background transit sources that are separated from their target stars, 
indicating an astrophysical false positive.  These techniques use only Kepler photometric data.
We describe the concepts and construction
of these techniques in detail as well as their performance and relative merits.  

\end{abstract}




\section{Introduction}\label{intro}


The {\it Kepler} mission is designed to determine the frequency of Earth-size planets in and near the habitable zone of 
solar-like stars via the detection of photometric transits \citep{bor10, mission}.  
{\it Kepler} surveys more than 100,000 late-type dwarf stars in the solar 
neighborhood with visual magnitudes between 8 and 16 for $> 4$
years looking for transits of planets around those stars.  There are several astrophysical 
phenomena that can cause a false-positive detection that mimics a planetary transit on a target star. 
Approximately 40\% of the transit-like signals detected by {\it Kepler} that have been deemed
{\it Kepler} Objects of Interest (KOIs) have been determined to be 
due to false positives.

\begin{figure}[htbp]
\begin{center}
\includegraphics[trim = 0 0 0 0.5cm, clip, scale=1]{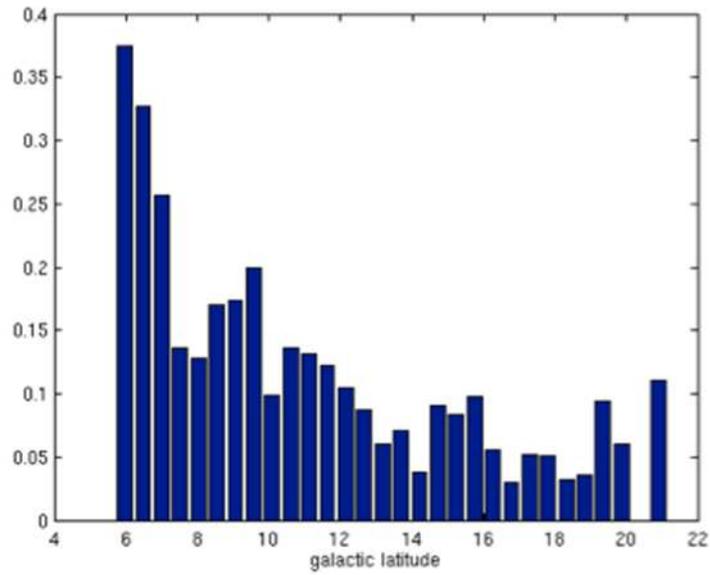}
\caption{The distribution of the fraction of transit signal sources that are offset from the target star, indicating
a background false positive.  For low
Galactic latitude almost 40\% of all {\it Kepler} KOIs are background false positives, while for mid to high Galactic latitudes 
the fraction drops to about 10\%.  This figure is based on data from \citet{bat12}.
}
\label{fig:APOvsGalLat}
\end{center}
\end{figure}

To increase the reliability of the determination of which
KOIs are planetary candidates, it is important to identify as many of these
false-positives as possible.  In many cases, the identification of false-positive KOIs is based on
{\it Kepler} data alone, because these KOIs have transit signals that are 
too small for conventional ground-based followup.
This paper describes several distinct but complimentary methods for using {\it Kepler} data to detect 
cases where the source of a transit-like event 
is offset from the target star's position on the sky.  These background false positives
make up a substantial fraction of all false positives, with most of the other false positives being due to
grazing eclipsing stellar companions associated with the target star.  At low Galactic latitudes, 
{\it background false positives} account for almost 40\% of 
all {\it Kepler} transit-like signals, with the fraction dropping to about 10\% at high Galactic latitudes
(see Figure~\ref{fig:APOvsGalLat}).
Background false positives are detected in {\it Kepler} data by observing that the pixels that
change during transit are offset from those that contain the image of the target star.  Such cases are referred
to as {\it active pixel offsets} (APOs).  The methods described in this paper cannot detect all
background transit sources: for example when the transit source is extremely close to the
target star on the sky.  However they can identify a large percentage of background false positives.
We believe that by identifying false positives that have an observable offset, 
the techniques described in this paper reduce the background false positive rate in the
planetary candidate catalog to below 10\%.

The techniques described in this paper rely on pixel data returned from the Kepler spacecraft.  Without 
this pixel data the identification of background transit sources is much more difficult.  Indeed, for dim target 
stars or for small planets with low SNR transits, ground-based followup typically 
will not suffice to identify background false positives.  In such cases, background false positive
identification would be impossible using stellar light curves alone.  Without the pixels, the star
hosting the transit signal cannot be determined.  Without knowing the star hosting the transit, 
the object causing the transit cannot be characterized.  Therefore the availability of the pixel 
data used to create the stellar light curves is a critical component of the
success of any transit survey.  This insight should drive the design of future transit survey missions.

In the rest of this section we discuss background false positives in general, their identification via pixel analysis and how
that identification is used in the vetting of {\it Kepler} planet candidates.  The bulk of this paper describes
several techniques for performing pixel-level analysis to identify background false positives.
In \S\ref{centroiding} we describe the photometric centroid technique, and in \S\ref{pixel differencing} the use of difference 
images to localize the transit signal source.  Pixel correlations are described in \S\ref{pixel correlation}.  We briefly address the special case of 
saturated targets in \S\ref{saturated targets}.  \S\ref{results} presents several perspectives on how well these techniques 
perform, with special emphasis on comparing the photometric centroid and difference image techniques.  

Throughout this paper we use several example KOIs
\citep{bor11a, bor11b, tcert, bat12, Burke13}.  Some of these KOIs are now valid
candidates, while others have been determined to be false positives.
We give particular attention to two examples to illustrate our techniques: KOI-221, which is a {\it Kepler} target where the transit 
source location is observationally coincident with the target, and KOI-109, which is a {\it Kepler} target for which the transit source 
is clearly offset from the target star.  
The list of KOIs and their properties can be found at the NASA Exoplanet Archive\footnote{http://exoplanetarchive.ipac.caltech.edu}
while the light curves and pixel data for all {\it Kepler} targets can be found at the 
Mikulski Archive for Space Telescopes\footnote{http://archive.stsci.edu/kepler}.


\subsection{Background False Positives}\label{causes}

There are several astrophysical phenomena that can mimic a planetary
transit on a specified target star.  \citet{brown03} 
distinguishes 12 combinations of giant planets and stars in eclipsing 
and transiting systems that can produce light curves mimicking a planet 
transiting a solitary primary star. Six of the combinations do not involve 
planets at all, and four others distort the transit light curve so that the 
size of the planet is indeterminate. 

In this paper we are concerned with  
those phenomena which are due to astrophysical sources that are not
associated with the target star.  These primarily include eclipsing 
binaries or large planet transits on stars that have flux in the pixels
used to create the target star's light curve.  Because of dilution from the 
target star, even deep background eclipsing binaries often cannot be identified from
the target star's light curve alone.  Analysis at the pixel level is required to identify the location
of the transit signal source.  We are particularly interested in cases
where the transit signal's source is sufficiently separated from the target star
that we can measure a statistically significant offset between the target star
and the transit source.

Additional sources of false positives that can be detected by the methods described in this
paper include 

\begin{itemize}
\item Very wide multiple star systems, where the transit source is gravitationally bound
to the target star.  When the separation between the target star and the companion hosting
the transit signal source is large enough the methods described in this paper can detect the offset.

\item Optical ghosts and electronic crosstalk \citep{instrum} from planetary transits or eclipsing binaries
elsewhere on the {\it Kepler} focal plane.  When the image of the ghost or crosstalk falls on the 
target star's pixels but is sufficiently 
separated from the target star these sources can be detected by the methods described in this paper.
In addition, optical ghosts can have very non-stellar morphologies.  Transit signals due to optical ghosts
will exhibit these morphologies in several of the techniques described in this paper.
\end{itemize}

Our basic strategy is to measure the location of the transit source on the sky, compare that to the location of the
target star, and declare the transit signal a false positive if the transit source location is significantly 
offset (more then three standard deviations, written $> 3\sigma$)
from the target star location based on reliable data. 
All the methods of computing these offsets described in this paper use $\chi^2$ minimizing (least-squares) methods.
Assuming Gaussian statistics, these offsets form a two-degree-of-freedom $\chi^2$ distribution, that 
have offsets  $> 3\sigma$ due to random fluctuations about 1.11\% of the time.
As we will show 
in this paper, offset uncertainties follow an approximately Gaussian distribution in a statistical sense, through 
the uncertainty around individual targets may not be Gaussian.

\subsection{Pixel Analysis to Identify the Location of the Transit Source}

As mentioned in Section~\ref{causes}, the background binary causing a transit signal can be very faint, indeed significantly
fainter than the general background or the wings of the target star, and still mimic a planetary transit.  
Consider the case of an aperture that contains only a target star with constant flux $F$ and a background binary
with other negligible sky background.
If the background binary
is $\Delta m$ magnitudes fainter than the target star, then the flux ratio of the background star to the 
target star is $\Delta F = \left( 100 \right) ^ {-\Delta m/5}$.  If the background binary 
has a fractional eclipse depth $d_{\mathrm{back}}$, 
then the total flux out of transit is $F^{\mathrm{out}} = F + F \Delta F$.  In transit the total flux is 
$F^{\mathrm{in}} = F + (1 - d_{\mathrm{back}}) F \Delta F$.
Therefore the fractional observed depth in the aperture is 
\begin{equation*}
d_{\mathrm{obs}} = 1 - \frac{F^{\mathrm{in}}}{F^{\mathrm{out}}} 
	= 1 - \frac{1 + (1 - d_{\mathrm{back}}) \Delta F}{1 + \Delta F}
	= \frac{d_{\mathrm{back}} \Delta F}{1 + \Delta F}.
\end{equation*}
In the case of a 14th magnitude target star and a 22nd magnitude background eclipsing binary with $d_{\mathrm{back}} = 0.5$
we get $d_{\mathrm{obs}} = 315$ ppm.  A transit of this depth is easily detected in {\it Kepler} data and would mimic 
the transit of a small planet, though the 22nd magnitude background star would not be readily apparent in the {\it Kepler} data.

There are several ways to use {\it Kepler} pixel data to measure the distance from the target star to the transit source.  We 
focus on three classes of techniques, each of which have their strengths and weaknesses.  As we describe in detail below, 
none of these techniques work
well in all circumstances due to systematic error sources that vary from technique to technique 
and situation to situation, 
but we find that the combination of these techniques covers the majority of cases where there
is sufficient signal to noise ratio (SNR) to measure the transit source location.  Our focus is on techniques that can be reliably automated due to 
the large number of objects in the {\it Kepler} data.  We would also, when possible, like to associate the transit source 
with a known star.  Therefore we describe techniques that provide an estimate of the transit source location on the sky rather than
simply determining if the transit source is at the target star location.

\begin{figure}[htbp]
\begin{center}
\includegraphics[scale=0.6]{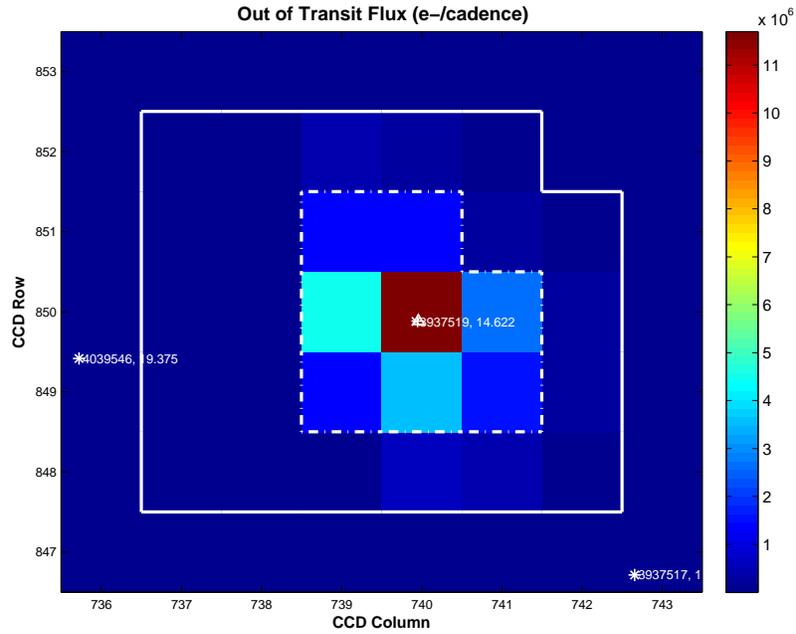} 
\caption{Pixels collected for a {\it Kepler} target.  All collected pixels are outlined by the solid white line.
The photometric optimal aperture is outlined with a white dot-dashed line.  
The pixel values are shown by the pixel color.  Asterisks give the location of known 
stars in the field, including those just outside the collected pixels.  For each star the Kepler Input Catalog number
and Kepler magnitude are given.}
\label{fig:collectedPixels}
\end{center}
\end{figure}

{\it Kepler} collects pixels specific to each target \citep{bryson_TAD}.  A subset of these pixels, called the photometric optimal aperture,
is summed to create the light curve for the target (see Figure~\ref{fig:collectedPixels}).  
The pixel analysis in this paper uses either the optimal aperture plus one halo
of pixels, defined as any pixel adjacent to the optimal aperture (the photometric centroid technique described in ~\S\ref{centroiding}), 
or all pixels collected for a target (the difference image technique described in ~\S\ref{pixel differencing}).  
For most targets, {\it Kepler} pixel data is 
collected once every {\it long cadence} (29.4 minutes), and for a subset of targets 
data is collected once every {\it short cadence} (0.98 minutes).  In this paper we limit our discussion to 
long cadence observations. 

All of the methods described in this paper identify spatially separated false positives by comparing pixel values during in-transit 
cadences to values of the same pixels during out-of-transit cadences.  

Analysis of {\it Kepler} pixels to identify the location of the transit relative to the target star has to solve three problems:

\begin{itemize}
\item {\bf Analyzing the Pixels Within a Cadence} There are various ways that the transit source location can be inferred from 
pixel data.  Some of these methods require the identification of cadences that occur during transit and cadences that do not. 

\item {\bf Combining the Cadences Within a Quarter} The {\it Kepler} spacecraft rotates 90 degrees about the
photometer boresite every $\sim 93$ days \citep{mission}.  Each $\sim 93$ day period is referred to as a Quarter.  While 
the {\it Kepler} focal plane is approximately symmetric under these 90 degree rolls, a star falls on different
CCDs at difference pixel coordinates in different quarters.  How in-transit and out-of-transit cadences within a quarter are
selected and combined varies from technique to technique.  

\item {\bf Combining the Cadences Across Quarters} Some of the techniques we discuss operate within
a single quarter and will deliver different results from quarter to quarter.  
These results for each quarter must be combined to provide an overall measurement.  
\end{itemize}

There are three classes of methods that we use to solve these problems:

\begin{itemize}
\item {\bf Photometric Centroid Shift} Detection of a shift in the photometric centroid
of the flux in the pixels (see \S\ref{centroiding}) that is correlated with the transit signal.  
This centroid shift can be used to estimate the location of the transit source as described in \S\ref{centroiding}.

\item {\bf Difference Imaging} By constructing the difference of the in- and out-of-transit pixel images, a direct image of the transit 
source can be constructed as described in \S\ref{pixel differencing}.  
The centroid of this image provides a direct measurement of the location of the transit source.

\item {\bf Pixel Correlation Images} When the transit signal can be detected in individual pixels via correlation with the photometric transit signal, 
an image can be constructed where the value of each pixel is given that correlation value as described in \S\ref{pixel correlation}.  
This is an alternative method of creating a direct image of the transit source, whose centroid provides the transit source location.
\end{itemize}

These methods assume that the only source of flux variation is the object creating the transit signal.  When this 
assumption is not satisfied these methods will be subject to systematic error.  Such systematic error will, however, 
be different for the different techniques, so when these methods give inconsistent results we have an indication that 
systematic error is present.
  
In contrast to the photometric centroid method, which is based on measured centroid shifts, 
the difference and pixel correlation image methods produce images that directly show the transit 
source.  While the location of the transit source can then easily be determined via photometric centroids of these images,
we use a more robust centroid method based on fitting the {\it Kepler} Pixel Response Function (PRF)
\citep{bryson10}.  The PRF characterizes how light from a single star is spread across several pixels, so 
it is essentially the system point spread function, comprised of the optical point spread function convolved with
pixel structure and pointing behavior over a {\it Kepler} long cadence.  Given a star's location on the pixels (including
sub-pixel position) the PRF provides the contribution of that star's flux to the nearby pixel values.  \S\ref{section:prf_fitting} describes how 
the PRF is fit to pixel images to determine the location of the transit source.

These three methods are in principle very similar, but have different responses to systematics and noise, 
transit SNR, and field crowding.  The use of all three methods provides increased sensitivity and confidence
in the identification of background false positives, particularly when the transit 
SNR is low.

\subsection{The Role of Offset Analysis in Planet Candidate Vetting}\label{section:vetting_role}

The techniques described in this paper are used to decide whether or not a detected transit signal
belongs on the {\it Kepler} planetary candidate list.  
These techniques have been applied to {\it Kepler} planetary candidate vetting
\citep{bor11a, bor11b, tcert, bat12, Burke13} with improved reliability and accuracy over time.
The approach that eventually evolved is to identify those targets that 
show a significant offset between the target star and the transit source
relying primarily on the difference imaging method.  Those targets that have a borderline significant source offset or have other
cause for concern are examined using all the methods described in this paper, including manual examination of the pixels.
Targets that have a confirmed offset from the transit source are identified as false positives.  This disposition has changed
over time for a small number of targets, as the techniques described in this paper have become more refined and as 
more data becomes available, resulting in greater measurement precision.
The details of how these analyses were applied 
are described in papers detailing the release of planetary candidate lists \citep{bor10,bor11a,bor11b,bat12,Burke13}.
We give here a brief history of this evolution.  
\citet{bor10} used photometric centroid time series analyzed via the cloud plots described in \S\ref{section:fw_centroids} and
an early version of difference images.  These difference images were visually examined rather than centroided, so 
offsets from the target star on the order of a pixel (4 arcsec) or larger were identified.  \citet{bor11a} and \citet{bor11b}
used the difference image method including PRF centroiding described in \S\ref{pixel differencing} without the
multi-quarter averaging, so each quarter was examined individually.  Difference imaging with the multi-quarter averaging 
(\S\ref{section:multiq_averaging}), joint-multi-quarter PRF fits (for low SNR targets) (\S\ref{section:joint_multiq_fit}) and pixel correlation images
(\S\ref{pixel correlation}) were used in \citet{bat12}. \citet{Burke13} relied more strongly on multi-quarter averaged difference imaging and 
photometric offsets.  Joint-multi-quarter PRF fits and pixel correlation images were
disabled in \citet{Burke13} because of computational limitations.  These limitations will be overcome in the future by moving the
{\it Kepler} analysis pipeline to supercomputer platforms.

Because of the evolution towards the techniques described in this paper, the quality of background false positive identification has 
changed over time.  Therefore the tables published in the early papers listed above have less accurate background false positive identification
than the later papers.  This is reflected in the tables in the {\it Kepler} archives, so care must be taken when performing statistical
analysis with these tables.  At the time of this writing an effort is underway to re-check all KOIs using the methods described in this 
paper, as well as improved light curve analysis to identify non-background false positives such as grazing binary stars.

\section{Source Location from Photometric Centroid Shifts} \label{centroiding}

\emph{Photometric centroids} compute the ``center of light'' of the pixels associated with a target.
When a transit occurs, the photometric centroid will shift, even when the transit is on the target star
(the ideal case of a transit on a target star exactly in the center of a symmetric aperture with uniform background,
which is required for there to be no centroid shift, is never realized in practice).  As described in this section, 
we use this shift to infer the location of 
the transit source, from which we can compute the transit source offset from the target star.
This method works well when the target star is crowded by many field stars, but suffers 
from high sensitivity to variable flux not associated with the transit such as stellar variability and
photometric noise.  As described in \S\ref{section:fw_error}, 
due to the implementation of the {\it Kepler} processing pipeline this
method tends to over-estimate the distance of the transit source from the target star when the transit source
is at the edge of the target star's pixels.

\subsection{Computing Pixel Centroids}
\label{section:fw_centroids}

The most traditional method for estimating the position of a light source is that of photometric centroids, also known as
flux-weighted centroids.  Photometric centroids measure the ``center of light'' of all flux in the pixels.  While 
photometric centroids do not
exactly measure the location of any particular star, it will be shown below that under idealized circumstances they
can be used to compute the location of a transit source. 

The row and column photometric centroids of the pixels for each target are computed for each cadence as
\begin{eqnarray}
  C_{\mathrm{row}} = \frac{\sum_{j = 1}^N  r_j b_j}{\sum_{j = 1}^N b_j}, \qquad  C_{\mathrm{column}} = \frac{\sum_{j = 1}^N c_j b_j}{\sum_{j = 1}^N b_j}
\end{eqnarray}
where $b_j$ is the flux in pixel $j$ at row and column $\left(r_j, c_j\right)$.  
If we denote the covariance matrix of the pixel values $b_j$ as $\mathcal{C}_{ij}$ 
(so the uncertainties in the pixel values are the square root of the diagonals: 
$\sigma_j = \sqrt{\mathcal{C}_{jj}}$), then the standard propagation of errors gives the uncertainty in the photometric row centroid as
\begin{eqnarray}
  \sigma_{C_\mathrm{row}} =\sqrt{ \frac{\sum_{j = 1}^N \sum_{i = 1}^N  r_i  \mathcal{C}_{ij} r_j }{\left( \sum_{j = 1}^N b_j\right)^2}
  	+ \frac{\left(\sum_{j = 1}^N  r_j b_j \right)^2}{\left(\sum_{j = 1}^N b_j\right)^4} \sum_{i = 1}^N \sum_{j = 1}^N  \mathcal{C}_{ij} }
\end{eqnarray}
with a similar formula for the uncertainty in the column centroid.  We see that the sensitivity of the
centroid value $\sigma_{C_\mathrm{row}}$ is proportional to the square root of the elements of the
covariance matrix $\mathcal{C}_{ij}$, in particular to the uncertainty in the pixel values $\sigma_j$, 
divided by the total flux in the pixels $\sum_{j = 1}^N b_j$.  Therefore, photometric centroids are very sensitive to 
variations in pixel value, in particular to shot noise and stellar variability.

For photometric 
centroids computed in the {\it Kepler} pipeline, $j$ ranges over the optimal aperture 
plus a single ring of pixels (sometimes called a {\it halo}).  The result is a time series containing the row and column centroids, 
called {\it centroid time series}.  The centroid shift is defined as the 
centroid value for cadences out of transit, $C^{\mathrm{out}}$, subtracted from the centroid value for
cadences in transit $C^{\mathrm{in}}$: $\Delta C = C^{\mathrm{in}} - C^{\mathrm{out}}$.  We assume shifts in different cadences are
uncorrelated, so these shifts have an uncertainty given by
$\sigma_{\Delta C}^2 = \sigma_{C^{\mathrm{in}}}^2 + \sigma_{C^{\mathrm{out}}}^2$.

It is very important to distinguish between the {\it centroid shift}, which measures how far the centroid
moves between in- and out-of-transit cadences, and the {\it source offset}, 
which measures the separation of the target star from the transit source.  As we will describe 
below, the centroid shift and source offset are related, but measure very different things.  The
centroid shift measures the change in the photometric centroid due to all changes in flux in the aperture.
The source offset is derived from the centroid shift, but measures the separation between the target star and the transit source 
(which may or may not be a different star).  In particular, because there is always background flux and field stars, the
centroid shift $\Delta C$ will always be non-zero even when the transit signal is on the target star.  In such cases the 
centroid shift can be relatively large while the source offset
may be very close to zero.  

Low-frequency secular trends due to small, slow changes such as
differential velocity aberration, small pointing drifts and thermally induced focal length changes
are common in centroid time series \citep{characteristics_handbook}.  These trends are removed prior to the analysis described in 
this section, for example by local median filtering using a window of 48 cadences.  

To facilitate combining the centroids across quarters, the centroid time series is converted to 
celestial right ascension (RA) and declination (Dec)
using the {\it Kepler} focal plane geometry model in combination with 
motion polynomials that capture local variations in the focal plane geometry model \citep{tenenbaum10}.  
In these coordinates the centroid shift
$\Delta C$ is expressed as seconds of arc.

When the centroid shift $\Delta C$ is large enough, it can be taken to indicate that the transit source is not on the 
target star.  Using $\Delta C$ directly to make this determination must be done with great care, however.  
$\Delta C$ will be smallest when the target star is the source of the transit, 
the target star is isolated, residual background flux is small after background correction, 
and the target star is near the geometric center of the centroided pixels.  This is rarely the case, however, so even when the target star
is the source of the transit there will be a non-trivial centroid shift.  A larger centroid shift that
is correlated with the time of transit is an indicator that the transit source may not be the target star.  Determining
whether a centroid shift indicates that the transit source is not the target star
is difficult, however, and depends on the details of other flux sources in the target's pixel aperture.
In \S\ref{section:fw_source_estimate} we describe how to use the centroid shift to estimate the location of the source of the centroid 
signal, which is a more robust method for determining whether the transit source is the target star than using the centroid shift alone.

\begin{figure}[htbp]
\begin{center}
\includegraphics[scale=0.4]{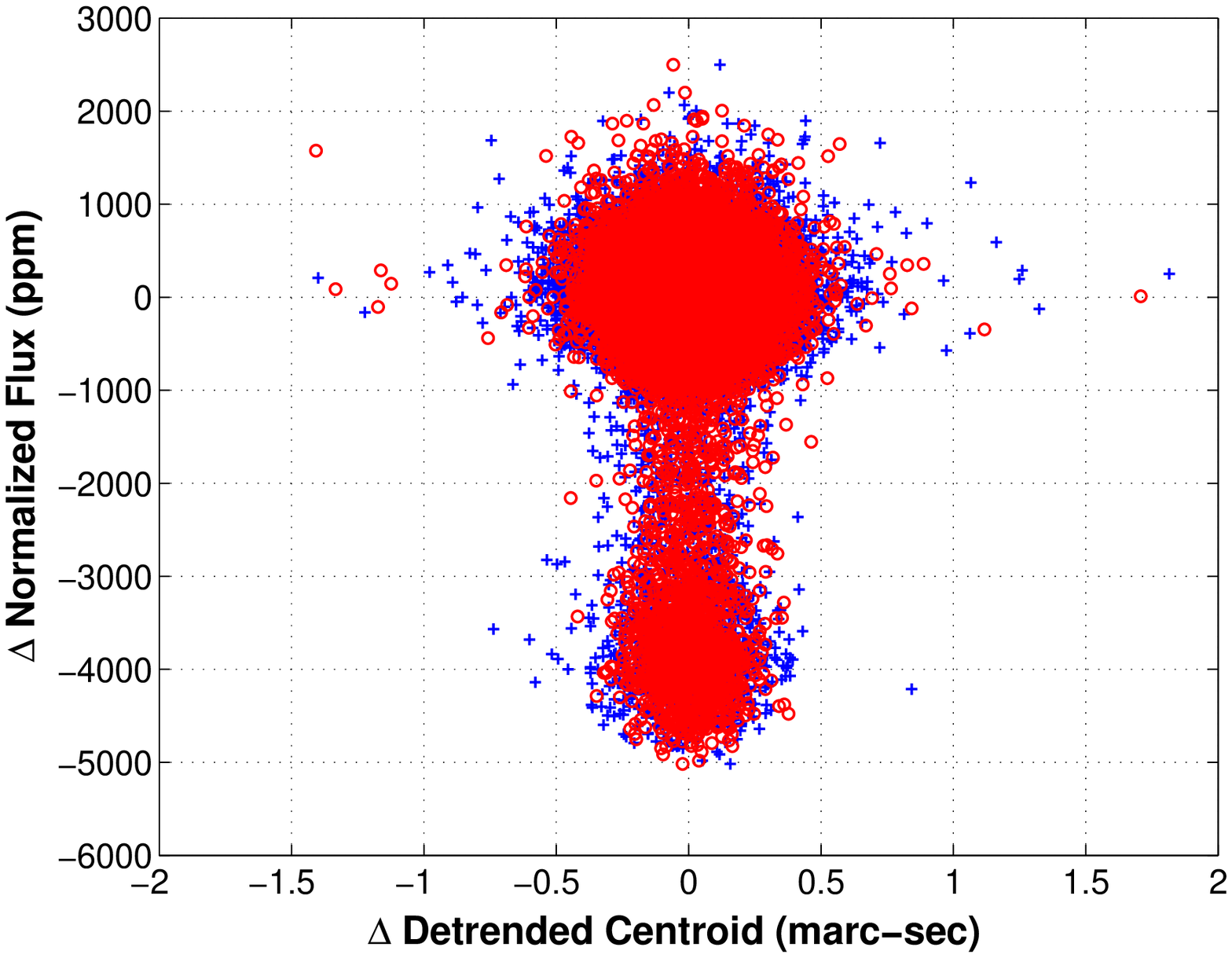}
\includegraphics[scale=0.4]{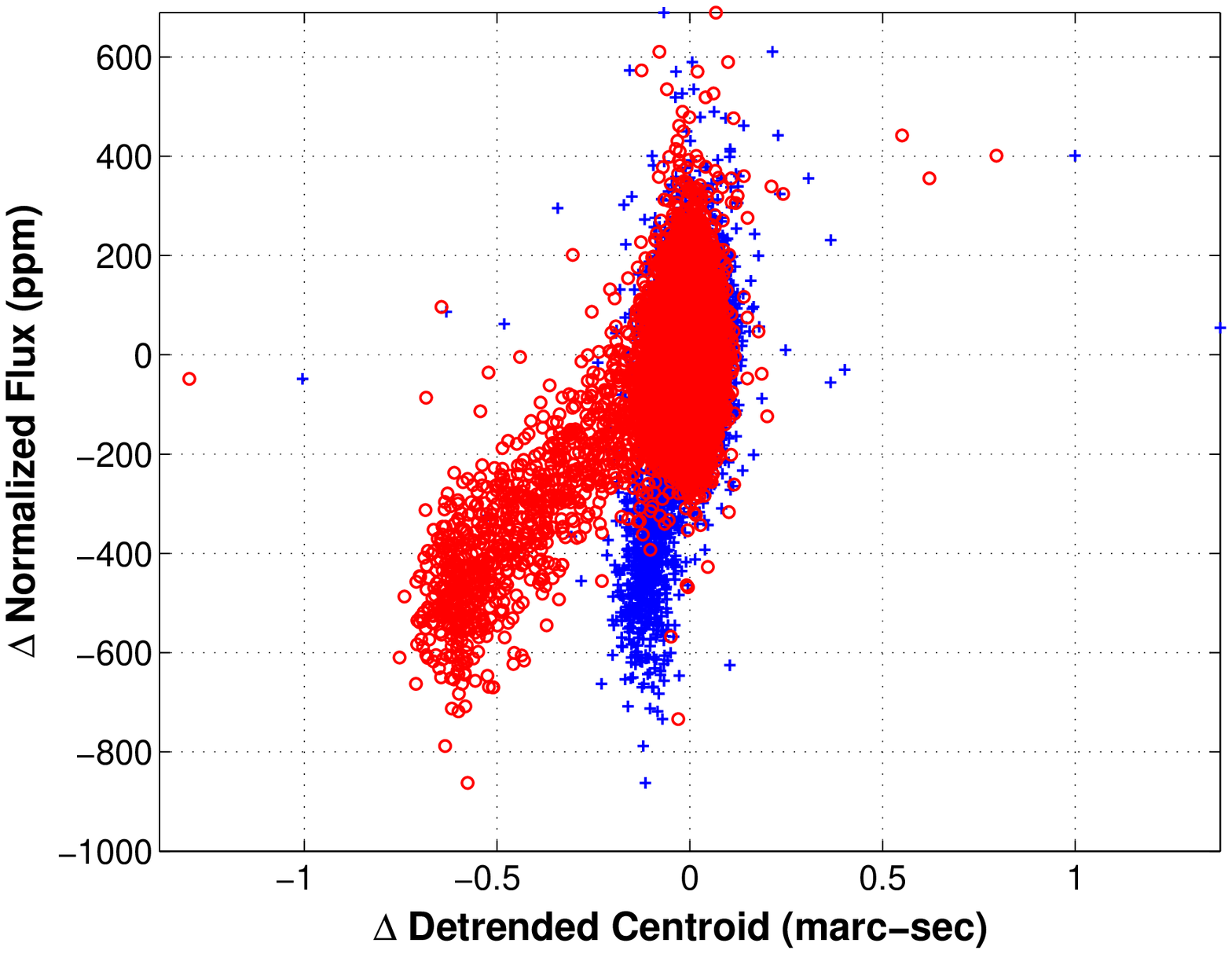}
\caption{Example cloud plots where the normalized residual flux ($y$-axis) is
plotted against the centroid shift ($x$-axis).  Each point 
plots the normalized, median-detrended flux value against the
median-detrended RA (blue crosses) or Dec (red circles) centroid time series in a single long cadence.  
In both figures most points are from out-of-transit cadences and form
a cloud around (0,0).
Left: When the transit is on an isolated target star  (KOI-221 in this example), the centroid does not shift
when in transit, so in-transit points are directly below the out-of-transit points.
Right: When the transit is on an object offset from the target (KOI-109 in this example), the in-transit
centroids are shifted relative to the out-of-transit centroids and appear below to one side, indicating 
a strong possibility of a background false positive.  In this example the
Dec centroid components show a shift while the RA components to not, indicating that the  
transit source is offset in the Dec direction. }
\label{fig:rainPlots}
\end{center}
\end{figure}

A graphical method showing the correlation between the centroid shift and the transit signal is to plot the median-detrended centroid time series
against the normalized, median-detrended light curve flux value.  The results is a {\it cloud plot}, 
shown in Figure~\ref{fig:rainPlots}.  
Most points in a cloud plot are out-of-transit cadences and form a cluster around (0,0).  The size of the cloud
reflects the sensitivity of the photometric
centroid computation to noise in the pixel values.  
When there is no centroid shift associated with transits, 
the points in transit (with negative normalized flux) fall directly below the out-of-transit points.
When there is a centroid shift associated with the transit, points in transit will fall to the side of the out-of-transit cloud.
Seeing sideways motion of the in-transit points as shown in the right panel of Figure~\ref{fig:rainPlots} 
indicates a centroid shift associated with the transit.
This suggests that the transit source may be offset from the target star.
As explained above, care must be taken 
when interpreting cloud plots because there may be a non-trivial centroid shift correlated with the transit 
even when the target star is the transit source.  

\subsection{Correlating Centroid Motion with the Transit Model} \label{section:centroid_correlation}

The centroid time series is sensitive to photometric noise, so quantitatively measuring the correlation of the centroid shift 
with the photometric transit signal can be difficult, particularly for low SNR transits.  A simple approach is to
identify all in- and out-of-transit cadences, and compute the average (or median) in- and out-of-transit centroid values.
The average centroid shift is then given by the difference of the in- and out-of-transit average centroid locations.
This method encounters many difficulties, however: quarter-to-quarter differences in aperture shape will introduce 
systematic errors, and non-transit related variability will degrade these averages as measures of transit-related shifts.
A better method is to fit a transit model computed during data validation \citep{Wu_DV}
to the centroid time series.  This will provide a more robust measurement of $\Delta C$.

\begin{figure}[htbp]
\begin{center}
\includegraphics[trim = 2cm 1.5cm 0 0, clip, scale=0.7]{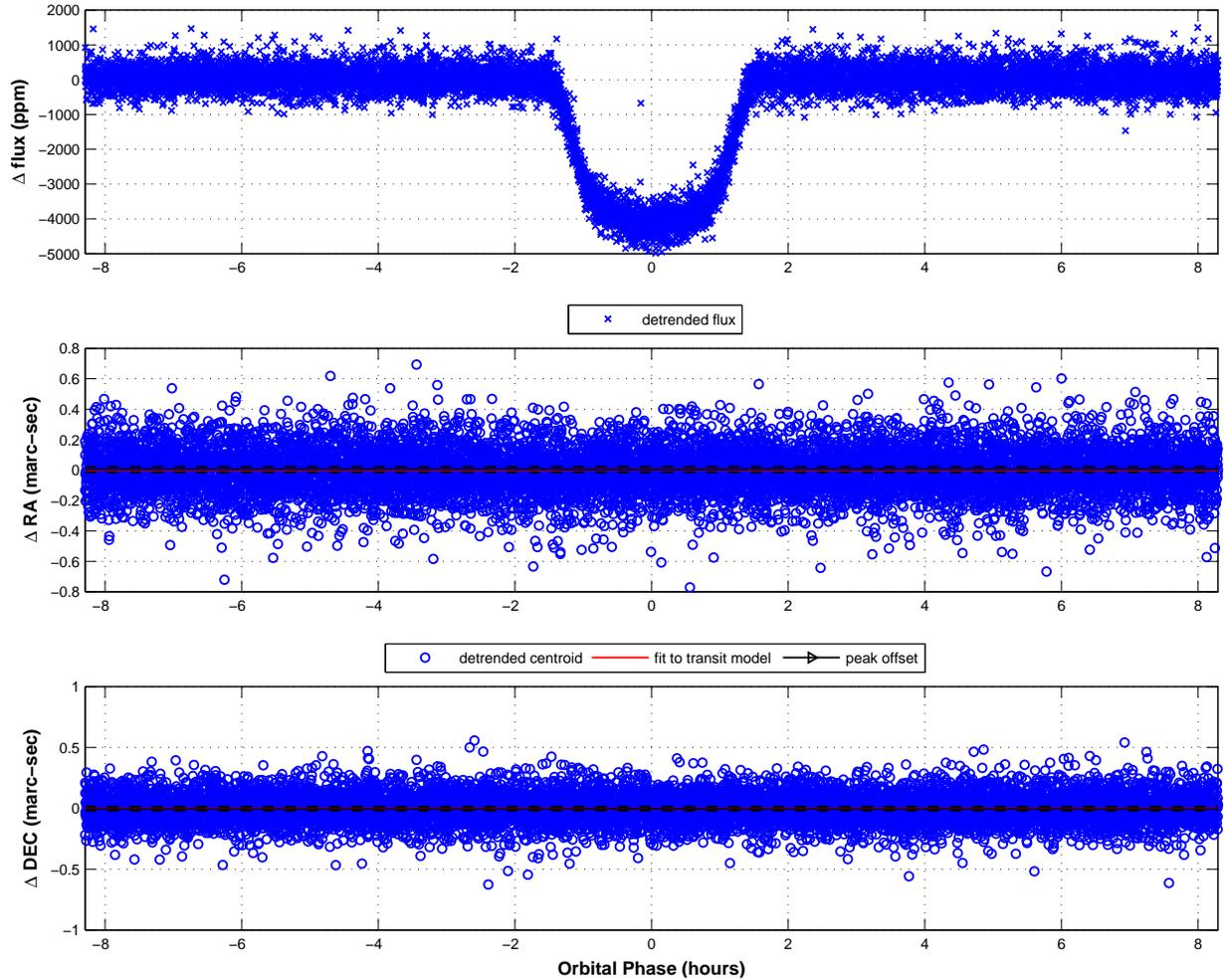} 
\caption{An example of a fit of the centroid time series to the transit model for a case when the 
transit source is at the same location as the target star (KOI-221).  Top: the
detrended flux light curve over all quarters folded on the transit period, with a closeup on the transit.
Middle and Bottom: the RA and Dec detrended centroid shifts $\Delta C$ 
for the same cadences in milli-arc seconds.  There is no apparent change in the centroid positions at the
time of the transit.}
\label{fig:centroidFit221}
\end{center}
\end{figure}

\begin{figure}[htbp]
\begin{center}
\includegraphics[trim = 2cm 1.5cm 0 0, clip, scale=0.7]{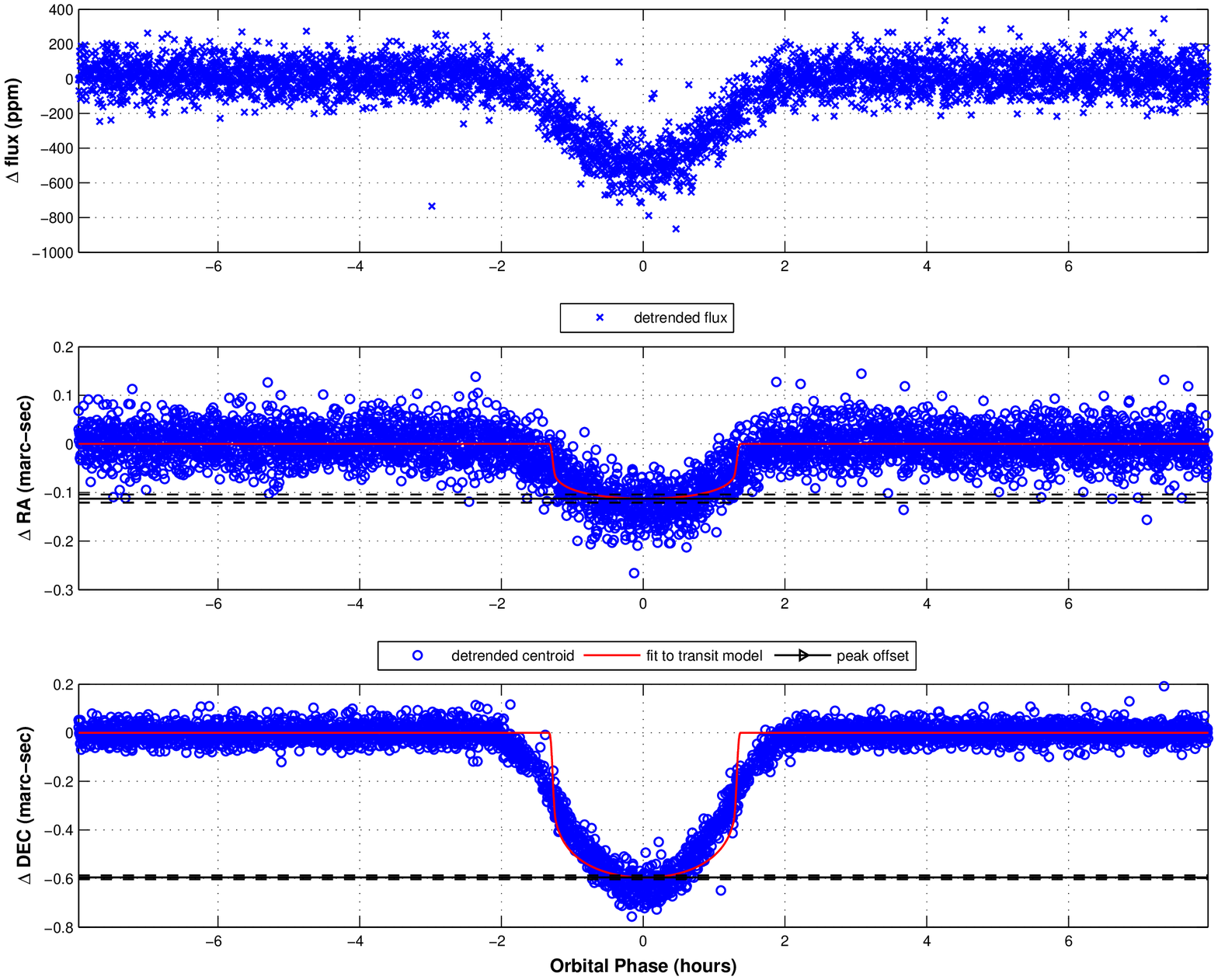}
\caption{An example of a fit of the centroid time series to the transit model for a case where the 
transit source is offset from the target star (KOI-109).  Top: the 
detrended flux light curve over all quarters folded on the transit period, with a closeup on the transit.
Middle and Bottom: the RA and Dec centroid shifts $\Delta C$ 
for the same cadences in milli-arc seconds.  There is a readily apparent change in the centroid shifts at the
time of the transit, particularly in Dec.  The transit model that best fits the flux light curve is superimposed on each
centroid shift plot, scaled by the coefficient $\gamma$ in Equation (\ref{eqn:centroid_model_fit_solution}).   
The value of $\Delta C = \gamma$ in declination is about 0.1 milli-second of arc. The poor model fit is due to 
the fact that the transit source for KOI-109 is in fact a deep eclipsing binary while the model assumes a
planetary transit.}  
\label{fig:centroidFit109}
\end{center}
\end{figure}

In this section we define the centroid shift time series $\Delta C_n = C_n - C^{\mathrm{out}}$ where $C^{\mathrm{out}}$ 
is the average out-of-transit centroid and $n$ labels the cadence.  In this section 
We assume that the transit model has been whitened to remove secular variations such as those due to pointing drift
and stellar variability \citep{Wu_DV}, in which case the centroid shift time series $\Delta C_n$ must be whitened in the same way.  
We compute a least-squares fit of the centroid shift time series $\Delta C_n$ to the 
transit model $M_n$ multiplied by a constant $\gamma$, weighted by the centroid uncertainties.
This fit is most easily done by requiring that the transit model and the centroid shift time series both have
zero mean when the transit is not occurring.  This implies that the transit model $M_n = 0$ for out-of-transit
cadences.
When this is the case we minimize 
\begin{equation}
\chi^2 = \sum_{n=1}^N \frac{1}{\left( \sigma_{\Delta C_n} \right)^2}\left(\Delta C_n  - \gamma M_n \right)^2 \label{eqn:centroid_model_fit}.
\end{equation} 
This least-squares minimization problem has the solution
\begin{equation}
\gamma = \frac{\sum_{n=1}^N \frac{\Delta C_n  M_n}{\left( \sigma_{\Delta C_n} \right)^2}  }
	{\sum_{n=1}^N \frac{M_n^2}{\left( \sigma_{\Delta C_n} \right)^2} }.
\label{eqn:centroid_model_fit_solution}
\end{equation} 
Examples of this fit are given in Figures~\ref{fig:centroidFit221} and \ref{fig:centroidFit109}.

Assuming that the centroid and transit model uncertainties are uncorrelated over time, and neglecting uncertainties in the 
transit model values, the uncertainty in $\gamma$ is 
\begin{equation}
\sigma_\gamma = \left( \sum_{n=1}^N \frac{M_n^2}{\left( \sigma_{\Delta C_n} \right)^2} \right)^{-\frac{1}{2}}.
\label{eqn:centroid_model_fit_sigma}
\end{equation}
Only in-transit cadences contribute to the computation of $\gamma$ and $\sigma_\gamma$ because $M_n = 0$ for out-of-transit
cadences.
Because $M_n$ is fit to the whitened and normalized flux light curve, it has unit variance, so $\gamma$ is in the same units as $\Delta C_n$ and 
directly gives an estimate of the in- vs. out-of-transit shift: $\Delta C \approx \gamma$.  
When the centroids shifts are in RA and Dec coordinates, all quarters of 
data can be simultaneously fit.  
From Equation (\ref{eqn:centroid_model_fit_sigma}) we see a $\sqrt{N^{\mathrm{in}}}$ reduction in the
uncertainty, where $N^{\mathrm{in}}$ is the total number of in-transit cadences,
so combining many quarters increases the precision of the estimate of $\Delta C$ in each coordinate.  

Once the shift is estimated in RA and Dec (in seconds of arc), the shift distance is simply 
\begin{equation}
D = \sqrt{\Delta C^2_{\mathrm{RA}} + \Delta C^2_{\mathrm{Dec}}},
\end{equation}
with uncertainty
\begin{equation}
\sigma_{D} = \frac{\sqrt{ \Delta C^2_{\mathrm{RA}} \sigma^2_{\Delta C_{\mathrm{RA}}}
	+ \Delta C^2_{\mathrm{Dec}} \sigma^2_{\Delta C_{\mathrm{Dec}}}}}{D}.
\end{equation}

A high-level detection statistic indicating whether a detected shift is statistically significant is also computed.
This statistic measures the probability that the detected shift is 
due to an actual signal rather than a statistical fluctuation in white noise by subtracting 
the residual $\chi^2$ from the signal $\chi^2$.  From this statistic a significance metric is constructed
that is normalized to the range $\left[ 0, 1 \right]$, where 1 means that there is no detected shift and 0 means
that the shift is highly significant.
This is equivalent to Equation (4) of \citet{Wu_DV}, which in our notation is given by 
\begin{eqnarray}
l = \frac{\sum_{n=1}^N \Delta C_n  M_n  }
	{\sigma_{\Delta C} \sqrt{\sum_{n=1}^N M_n^2} }.
\end{eqnarray}

\subsubsection{The impact of crowding and variability on the centroid shift estimate} \label{section:centroid_shift_error}

\begin{figure}[htbp]
\begin{center}
\includegraphics[scale=0.6]{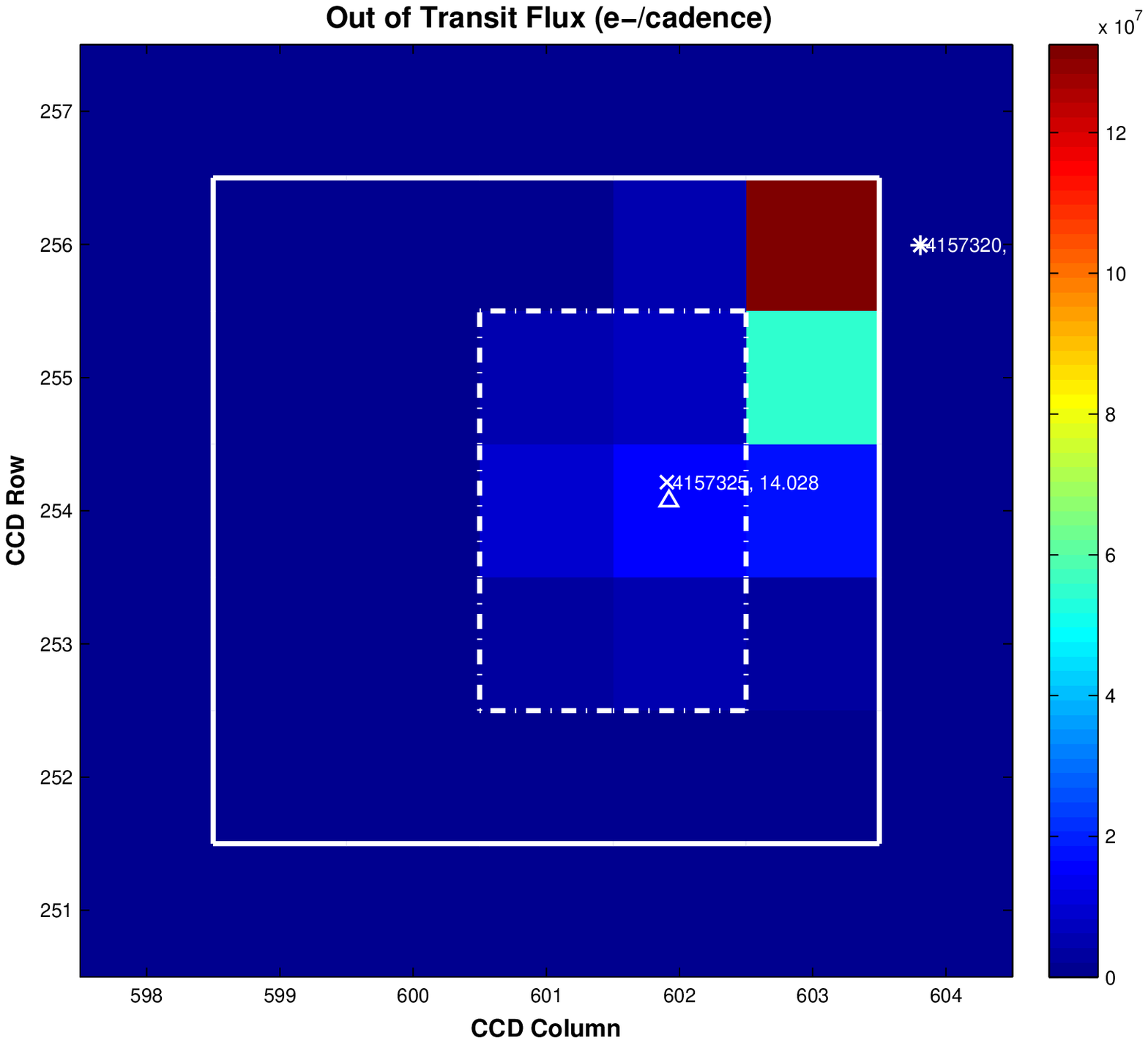} 
\caption{The pixels collected for KOI-1860 in quarter 10.  The pixels are dominated by the field star KIC 4157320
which is 2.7 magnitudes brighter than the target star.  KIC 4157320 has strong variability.  In addition, because it is only 
partially captured in the pixels, spacecraft pointing variations are apparent in the pixel flux light curves.}
\label{fig:koi1860DirectImage}
\end{center}
\end{figure}

\begin{figure}[htbp]
\begin{center}
\includegraphics[scale=0.9]{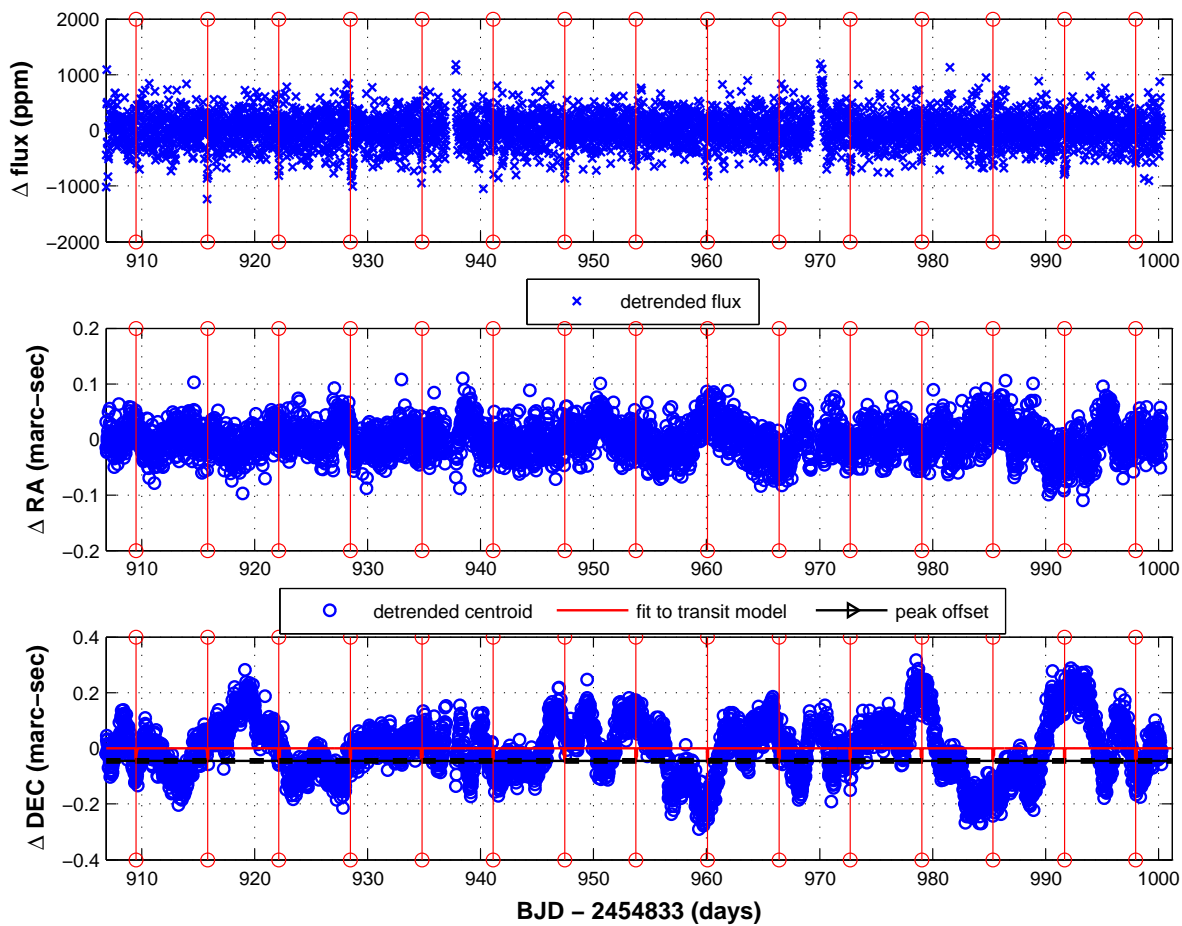} 
\caption{The (not folded) flux and photometric centroid time series for KOI-1860 in quarter 10.  The vertical red 
lines indicate times of transit.  The bright field star
at the edge of the aperture (see Figure~\ref{fig:koi1860DirectImage}) causes strong variations in the 
centroid time series due to the intrinsic variability of that star combined with spacecraft pointing
jitter, which is exacerbated by that star being only partially captured in the pixels.  These variations cause a spurious 
centroid shift that is correlated with the transit signal.}
\label{fig:koi1860Q10CentroidTimeSeries}
\end{center}
\end{figure}

The computation of the in-transit centroid shift assumes that the transiting object is the only source of time varying flux that is
correlated with the transit signal in the target star's pixels.  While this is usually a reasonable assumption, it is sometimes violated, introducing 
systematic error into the centroid shift estimate.
A dramatic example is KOI-1860, whose pixels are shown in Figure~\ref{fig:koi1860DirectImage}.  In this case there is a field star
that is 2.7 magnitudes brighter than the target star at the edge of the collected pixels.  Examination of the pixel flux time series shows that
this bright star has moderately high variability on short time scales.  In addition, because this bright star is at the edge of the collected
pixels and is only partially captured, there are strong variations in flux due to spacecraft pointing jitter.  The effect of these variations
on the centroid time series are shown in Figure~\ref{fig:koi1860Q10CentroidTimeSeries}.  These variations are on a time scale that 
occasionally correlates with the transit signal, leading to a small spurious measured centroid shift in the fit (\ref{eqn:centroid_model_fit_solution}).
The reconstructed transit source location using this spurious shift measurement, 
described in \S\ref{section:fw_source_estimate}, indicates a transit source separated from the 
target star by about 4 arcseconds.   As we will see in \S\ref{section:multiq_averaging}, however, the PRF-fit technique provides strong evidence that 
the transit source is only about a third of an arcsecond from the target star.  

\subsection{Estimating the Transit Source Location from Centroid Motion} \label{section:fw_source_estimate}

Photometric centroids are the weighted average of all flux in the target star's pixels, so they
do not provide direct information about the location of the target star or the
transit source.  In particular, as explained in \S\ref{section:fw_centroids}, 
a statistically significant shift does not necessarily 
imply that the transit source is offset from the target star.  In Appendix \ref{appendix_a} we derive a 
formula approximating the location of the transit source from the observed transit depth (based on the light curve
created by summing the pixels used for centroiding), the out-of-transit centroid location $C$ 
and the centroid shift $\Delta C$.
Remarkably, this formula applies in the presence an arbitrary background signal, including any number of 
field stars in or near the aperture, and
does not depend on the brightness of those stars.  This formula only assumes that the
flux from the transit source is the only time-varying signal in the aperture, so no other 
stars or the background flux vary in brightness.  These assumptions are never exactly true, 
but in many cases they are very nearly true and in these cases we can estimate the
transit source location.  We can then compare the transit source location to the 
catalog location of the target star to estimate the offset of the transit source from the target star.
We assume that the centroids are provided in RA and Dec coordinates, denoted $\left( \alpha, \delta \right)$.  

\begin{figure}[htbp]
\begin{center}
\includegraphics[scale=1.2]{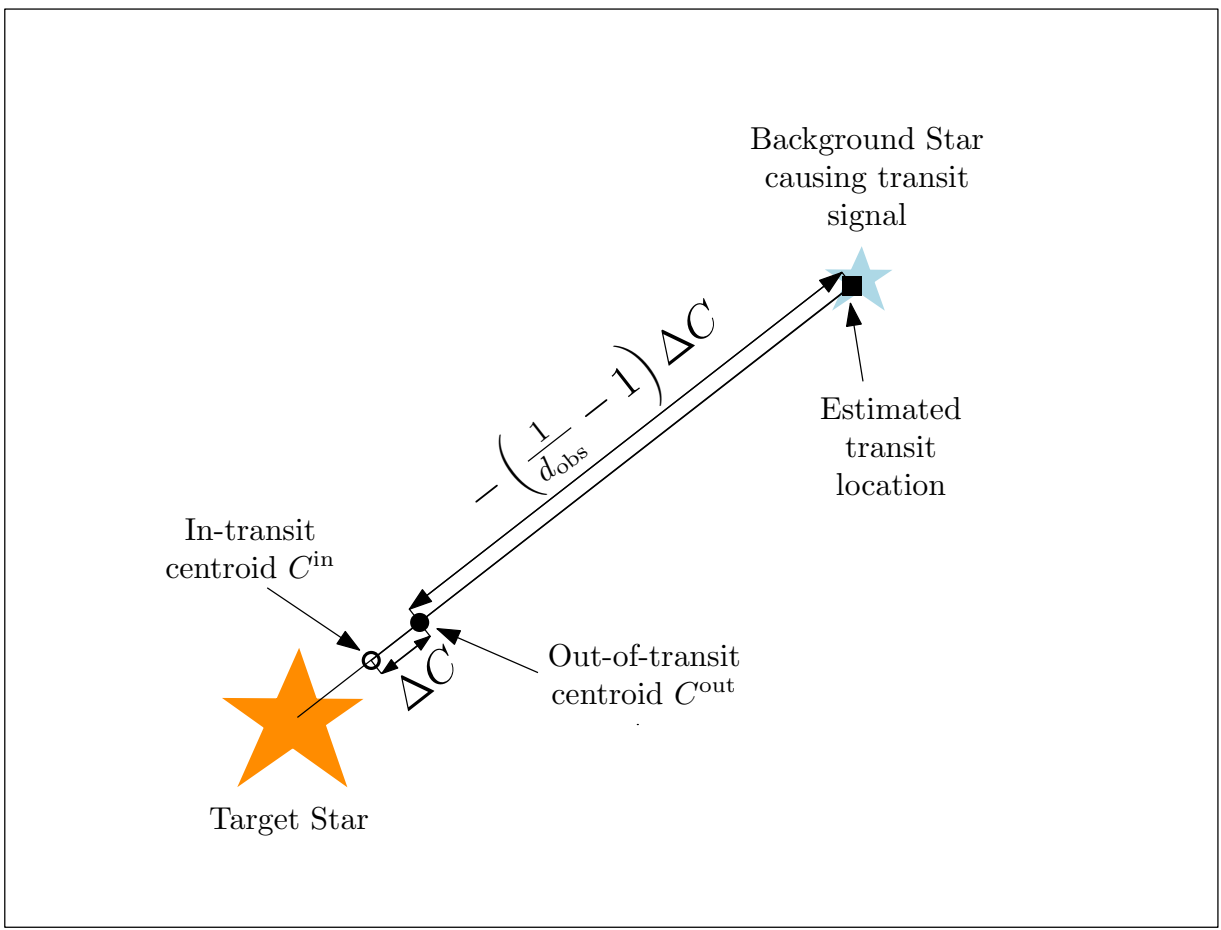}
\caption{An illustration of the relationship between centroids, centroid shifts, the background eclipsing 
binary causing the transit signal, and the target star in Equation (\ref{eqn:source_location}) for an otherwise empty aperture.  The
photometric centroid when a transit is not occurring is given by $C^{\mathrm{out}}$ (filled circle).  
If the transit is due to an eclipse on the background star, during the eclipse the
centroid will shift towards the target star to $C^{\mathrm{in}}$ (open circle).  The resulting transit shift is
$\Delta C = C^{\mathrm{in}} - C^{\mathrm{out}}$. Applying Equation (\ref{eqn:source_location})
gives an estimate of the transit source location (filled square), which in an idealized case will correspond to
the location of the transit source.
}
\label{fig:centroidSourceLocation}
\end{center}
\end{figure}

We denote the RA and Dec components of the average out-of-transit centroid as $\left(C^{\mathrm{out}}_\alpha, C^{\mathrm{out}}_\delta \right)$, and 
the centroid shift measured as described in \S\ref{section:centroid_correlation} 
as ($\Delta C_\alpha, \Delta C_\delta$).  If  
the observed transit depth is $d_{\mathrm{obs}}$, then as shown in Appendix~\ref{appendix_a} the centroid of the flux from the transit source that
falls in the aperture is at RA and Dec
\begin{equation}
  \alpha_{\mathrm{transit}} = C^{\mathrm{out}}_\alpha - \left( \frac{1}{d_{\mathrm{obs}}} - 1 \right) \frac{\Delta C_\alpha}{\cos \delta}, \qquad 
  \delta_{\mathrm{transit}} = C^{\mathrm{out}}_\delta - \left( \frac{1}{d_{\mathrm{obs}}} - 1 \right) \Delta C_\delta
  \label{eqn:source_location}
\end{equation}
(see Figure~\ref{fig:centroidSourceLocation}).  
When all flux from the transit source is captured in the aperture, then this centroid gives the location of the transit source.

The formal uncertainty in the source position is given in terms of the centroid uncertainty
$\sigma_{C_\alpha}$ and depth uncertainty $\sigma_{d_{\mathrm{obs}}}$ by 
\begin{eqnarray}
  \sigma_{\alpha_{\mathrm{transit}}} & = & \sqrt{ \sigma_{C_\alpha}^2 + \left( \frac{1}{d_{\mathrm{obs}}} - 1
  \right)^2\frac{ \sigma^2_{C^{\mathrm{out}}_\alpha} + \sigma^2_{C^{\mathrm{in}}_\alpha} }{\cos^2 \delta} + \frac{\Delta C_\alpha^2}{\cos^2 \delta}
  \frac{\sigma_{d_{\mathrm{obs}}}^2}{d_{\mathrm{obs}}^4} }\\
  \sigma_{\delta_{\mathrm{transit}}} & = & \sqrt{  \sigma_{C_\delta}^2 + \left( \frac{1}{d_{\mathrm{obs}}} - 1
  \right)^2 \left( \sigma^2_{C^{\mathrm{out}}_\delta} + \sigma^2_{C^{\mathrm{in}}_\delta} \right) + \Delta C_\delta^2
  \frac{\sigma_{d_{\mathrm{obs}}}^2}{d_{\mathrm{obs}}^4} }.
\end{eqnarray}
These uncertainties do not account for systematic error due to other sources of varying flux. 

For $d_{\mathrm{obs}} \ll 1$ Equation (\ref{eqn:source_location}) reduces to 
\begin{equation}
  \alpha_{\mathrm{transit}} \simeq C_\alpha - \frac{\Delta C_\alpha}{d_{\mathrm{obs}} \cos \delta}, \qquad 
  \delta_{\mathrm{transit}} \simeq C_\delta - \frac{\Delta C_\delta}{d_{\mathrm{obs}}},
\end{equation}
the approximation given in Equation (2) of \citet{Wu_DV}.  The uncertainties are similarly approximated
by replacing $\left( 1/d_{\mathrm{obs}} - 1 \right)$ by $1/d_{\mathrm{obs}}$. This approximation has an error 
that is proportional to $d_{\mathrm{obs}}$, which is very small for most {\it Kepler} planetary candidates.

Once we have the centroid source location from Equation (\ref{eqn:source_location}), we compare 
it with the target location to determine the source offset.  The target star location cannot, however, be
reliably determined from the centroid time series, so we take the target star position from the Kepler Input Catalog.
This choice potentially introduces new sources of systematic error, particularly due to unknown proper motion.

Given the target star's catalog location $\left( \alpha_{\mathrm{target}}, \delta_{\mathrm{target}} \right)$, we can 
compute the target offset and uncertainty from the offset components 
$\Delta \alpha = \left( \alpha_{\mathrm{transit}} - \alpha_{\mathrm{target}} \right) \cos \delta$
and $\Delta \delta = \delta_{\mathrm{transit}} - \delta_{\mathrm{target}}$ as 
\begin{equation}
D = \sqrt{ \Delta \alpha^2 + \Delta \delta^2}, \qquad
	\sigma_{D} = \frac{\sqrt{ \Delta \alpha^2 \sigma^2_{\Delta \alpha}
	+  \Delta \delta^2 \sigma^2_{ \Delta \delta}}}{D}
\label{eqn:offset_distance}
\end{equation}
where $ \sigma_{\Delta \alpha} = \sqrt{ \sigma_{\alpha_{\mathrm{transit}}}^2 
+ \sigma_{\alpha_{\mathrm{target}}}^2} \cos \delta$ and
$ \sigma_{\Delta \delta} = \sqrt{\sigma_{\delta_{\mathrm{transit}}}^2 
+ \sigma_{\delta_{\mathrm{target}}}^2}$.

We can now determine if the transit source is statistically significantly offset from the target star 
by observing whether $D > 3 \sigma_{D}$.  

\subsubsection{Systematic errors in the source position estimate} \label{section:fw_error}

\begin{figure}[htbp]
\begin{center}
\includegraphics[scale=0.6]{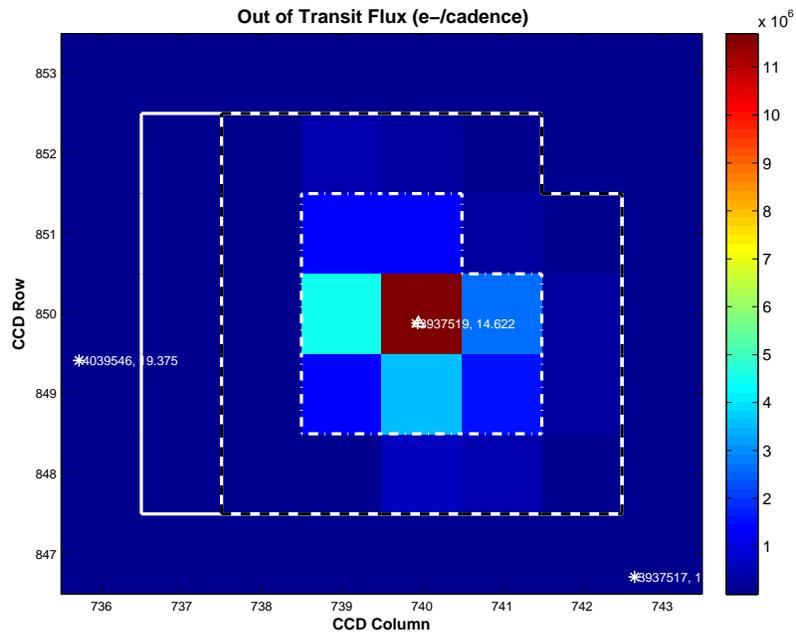} 
\caption{The optimal aperture compared with the pixels used for photometric centroiding.  The optimal aperture pixels are 
outlined by the dot-dashed line, while the pixels used for photometric centroiding are outlined by the dashed line.}
\label{fig:pixelsUsedForCentroiding}
\end{center}
\end{figure}

As discussed in Appendix \ref{appendix_a}, the above analysis does not describe the current implementation 
in the {\it Kepler} pipeline.  The {\it Kepler} pipeline uses the photometrically optimal aperture \citep{bryson_TAD}
to compute the transit depth and the optimal aperture plus one ring of surrounding pixels to compute
the centroid (see Figure~\ref{fig:pixelsUsedForCentroiding}).  
This use of different pixel apertures to compute the depth and centroid invalidates
the above analysis when significant flux from the transit source falls outside the optimal aperture.
Because optimal apertures are as small as a single pixel,
such overshoot is possible when the transit source and target star are separated by more than
one {\it Kepler} pixel (3.98 arcseconds).  

In the typical background false positive case when the transit source is associated with a field star that is significantly dimmer
than the target star, the observed depth in the optimal aperture (the depth computed by the {\it Kepler} 
pipeline) will be smaller than the depth that would have been observed using the centroided pixels.  
This will result in an overestimate of the distance of the transit source from the out-of-transit photometric centroid $C^{\mathrm{out}}$
in Equation (\ref{eqn:source_location}).  Occasionally the field star associated with the transit source will
be brighter than the target star so the flux from the target star dominates the centroids.  
In this case the observed depth in both apertures will be similar, resulting in less of an overshoot.  This behavior is observed in \S\ref{section:performance_accuracy}.
See Appendix \ref{appendix_a} for details.

\begin{figure}[htbp]
\begin{center}
\includegraphics[scale=0.6]{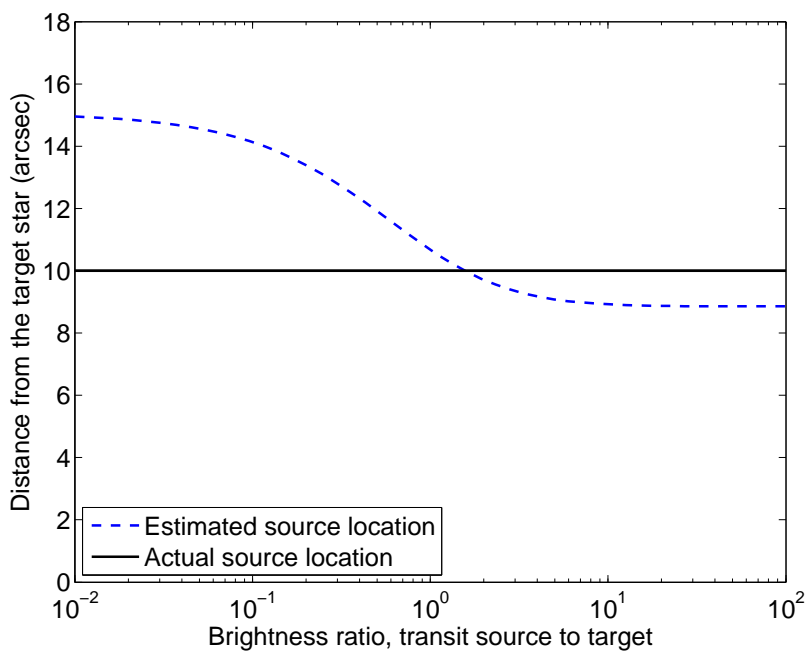} 
\caption{The photometric-based transit source offset as a function of the ratio of the background source
brightness to the target star brightness.  The example shown here is for a 0.1\% transit on a background star 
that is 10 arcseconds from the target star.  The optimal aperture in this case is $2 \times 2$ {\it Kepler} 
pixels ($7.96 \times 7.96$ arcseconds), so the background star is outside the optimal aperture in the halo pixels.
Because significant flux from the background star falls outside the captured pixels, the source position
estimate (Equation (\ref{eqn:source_location}))
underestimates the actual position of the background star.}
\label{fig:overshootVsBrightness}
\end{center}
\end{figure}

The dependence of the source offset estimate on the ratio of the brightness of the background star to that of the target star is
shown in Figure~\ref{fig:overshootVsBrightness}.  This example is similar to that in Figure~\ref{fig:koi1860DirectImage},
where the background star causing the transit signal is outside
the optimal aperture and mostly, but not completely, captured in the centroided pixels.  When the background star is dim,
the estimated transit source overshoots the correct offset.  When the background star is significantly brighter than the target star 
then the flux from the background star dominates the depth estimate, so the depth based on the centroided pixels is about the same
as the depth based on the optimal apertures.  But because the background star is close to the edge of the centroided pixels not all
flux from the background star is captured.  Therefore the source offset estimate in Equation (\ref{eqn:source_location}) gives
the centroid of the flux in the pixels from the background star, which is closer to the target star than the background star itself.

\section{Difference Imaging} \label{pixel differencing}

The \emph{difference imaging technique} computes the difference between average in
and out of transit pixel values.  These pixel differences provide an image of the transit source at
its true location.  A centroid of this \emph{difference image}
provides the location of the transit source.  To measure this centroid we fit the {\it Kepler Pixel Response
Function} (PRF), looking for the PRF position that best matches the difference pixels.  We compare this position 
to the PRF fit to the out-of-transit position, which provides the target star position when it is not crowded by
field stars.  The difference of these centroids gives us the offset of the transit signal from the target star.  
This method is more robust against photometric variability than the photometric centroid method, but is sensitive
to scene crowding.   


\subsection{The Concept of Difference Imaging} \label{section:difference_imaging}

The difference image technique is based on the insight that subtracting the in-transit pixel values
from the out-of-transit pixel values give an image that shows only those pixels that have changed
during the transits.  Further, if the changes during transits are due to a change in brightness of a star
(as is the case for a planetary transit or an eclipsing binary) then the bright pixels in the difference 
image will be those of that star with flux given by the fractional transit depth times the flux of that
star.  

More precisely, consider a set of pixels that contain flux from $M$ stars, labeled by the index $j$, 
at locations $\left( \alpha_j, \delta_j \right)$ with flux $b_j$ (we neglect background flux
in this simple analysis).  The PSF will distribute the flux from each of these stars over
several pixels. 
We express the flux on the pixel at row $r$ and column $c$ due to star $j$ by 
the unit flux function $f \left( \alpha_j, \delta_j, r, c \right)$
(so the sum over all pixels of $f \left( \alpha_j, \delta_j, r, c \right) = 1$).
Then the out-of-transit pixel values due to all stars will be given by
$F^{\mathrm{out}}\left(r, c \right) = \sum_{j=1}^M b_j f \left( \alpha_j, \delta_j, r, c \right)$.  
If star $k$ has a transit of depth
$d_{\mathrm{back}}$ then during mid transit the pixel values would be given by 
$F^{\mathrm{in}}\left(r, c \right) = \sum_{j=1, j \neq k}^M b_j f \left( \alpha_j, \delta_j, r, c \right)
+ \left(1 - d_{\mathrm{back}} \right) b_k f \left( \alpha_k, \delta_k, r, c \right)$.
In the ideal case where the only flux change is in star $k$, the difference image will be 
$F^{\mathrm{out}}\left(r, c \right) - F^{\mathrm{in}}\left(r, c \right) = d_{\mathrm{back}} b_k f \left( \alpha_k, \delta_k, r, c \right)$, 
which is exactly the image of star $k$ with flux $d_{\mathrm{back}} b_k$.

Difference images provide direct information about the location of the transit source, as opposed
to the use of photometric centroids in \S\ref{section:fw_centroids}, where the source location
is inferred.

\begin{figure}[htbp]
\begin{center}
\includegraphics[scale=0.9]{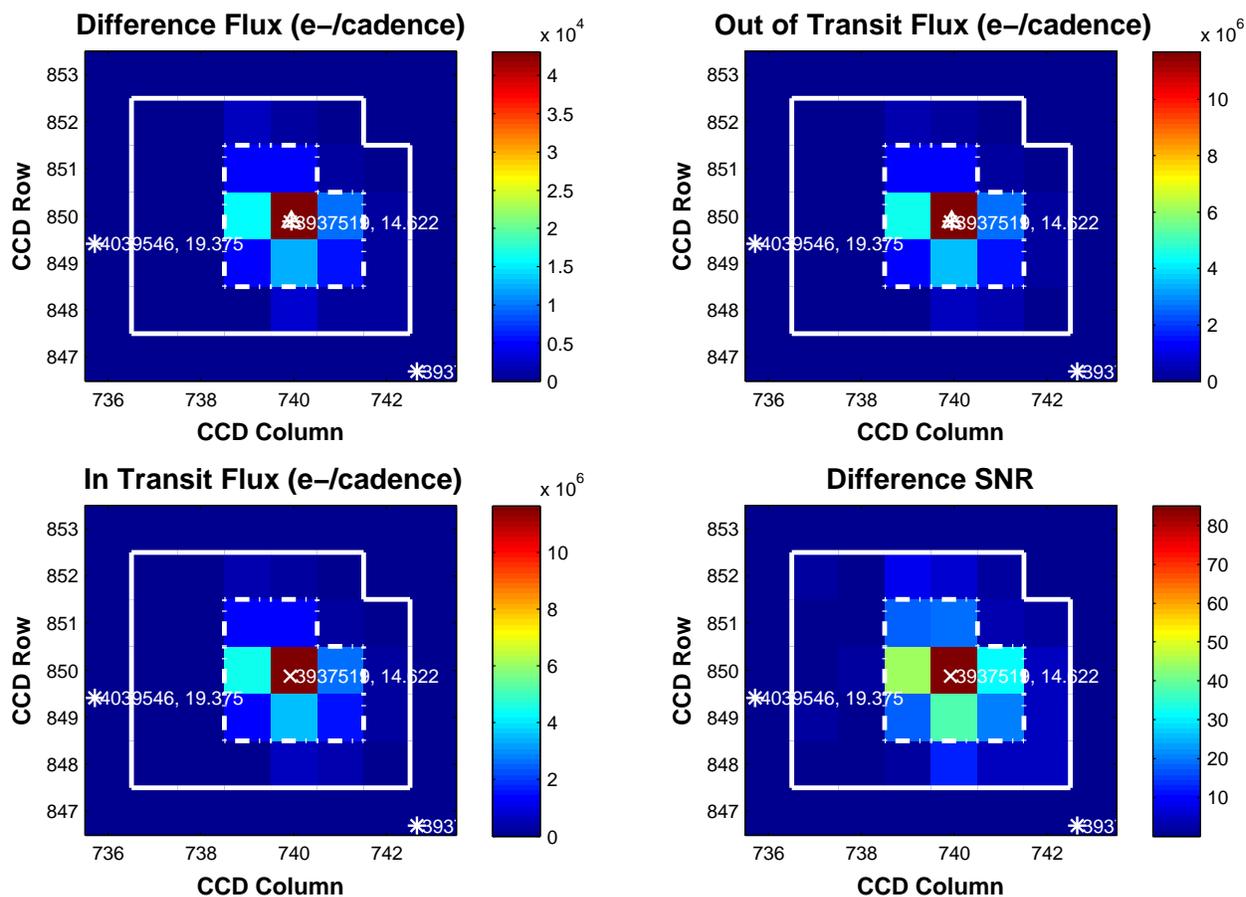}
\caption{Example pixel images for KOI-221 in quarter 7, which shows no indication that the transit is
not on the target star. In all figures, the dotted white
line borders the pixels of the optimal aperture, while the solid white line borders all pixels collected
for this target.  Known stars are shown as white asterisks, with each star's KIC catalog number and {\it Kepler} 
magnitude.  Upper Right: the averaged out-of-transit pixel image.  Lower Left: the in-transit pixel image.
Upper Left: the difference image = out-of-transit pixel image - in-transit pixel image.  Lower right: the difference
image normalized by pixel value uncertainty.  In this case the difference image appears identical to the in- and
out-of-transit images, which indicates that the transit source is coincident with the target star.}
\label{fig:KOI_221_diff_image}
\end{center}
\end{figure}

\begin{figure}[htbp]
\begin{center}
\includegraphics[scale=0.9]{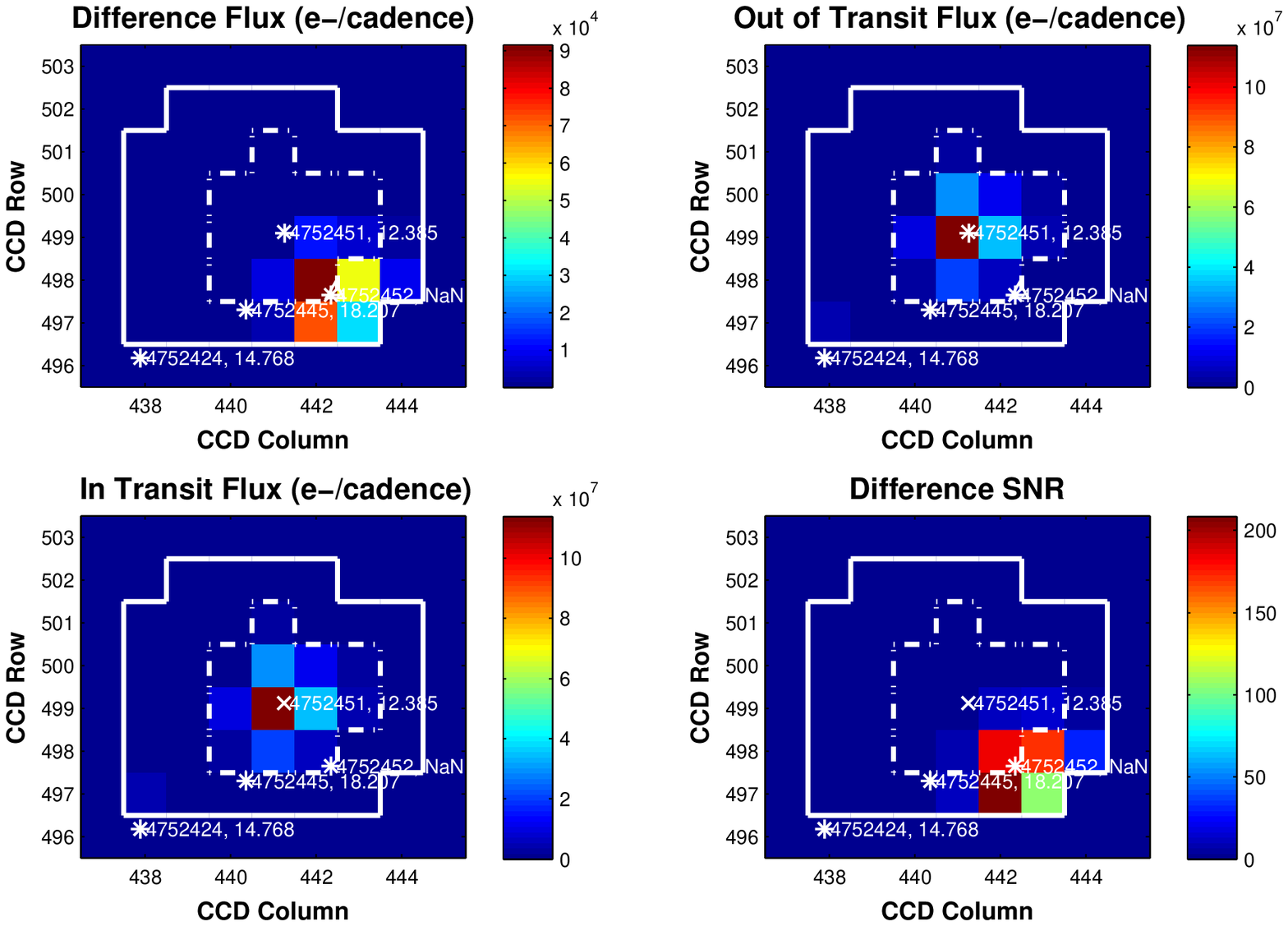}
\caption{Example pixel images for KOI-109 in quarter 4, which shows indications that the transit is
not on the target star. Upper Right: the averaged out-of-transit pixel image.  Lower Left: the in-transit pixel image.
Upper Left: the difference image = out-of-transit pixel image - in-transit pixel image.  Lower right: the difference
image normalized by pixel value uncertainty.  In this case the difference image appears to be very different from the in- and
out-of-transit images, which indicates that the transit source is coincident with the star KIC 4752452.}
\label{fig:KOI_109_diff_image}
\end{center}
\end{figure}

Example pixel images are shown in Figures \ref{fig:KOI_221_diff_image} and \ref{fig:KOI_109_diff_image}.
In Figure \ref{fig:KOI_221_diff_image} we see an example of a star (KOI-221) for which there is no apparent 
offset between the target star and the transit source.  In this case the difference image looks much like the 
in- and out-of-transit images, likely because the target star is itself the source of the transit (and there are no other
stars of comparable brightness in the out-of-transit image).  Therefore the
only difference between the difference image and the out-of-transit image is the flux level in the pixels.
Figure \ref{fig:KOI_109_diff_image} shows a case (KOI-109) where the difference image is dramatically different
from the out-of-transit image, and appears as a star image coincident with the dim unclassified star KIC 4752452.
Because KIC 4752452 is unclassified, it does not have a {\it Kepler} magnitude.  In this case the pixel data show that 
the transit source is clearly not on the target star.


\begin{figure}[htbp]
\begin{center}
\includegraphics[trim = 1.7cm 1cm 0 0, clip, scale=0.7]{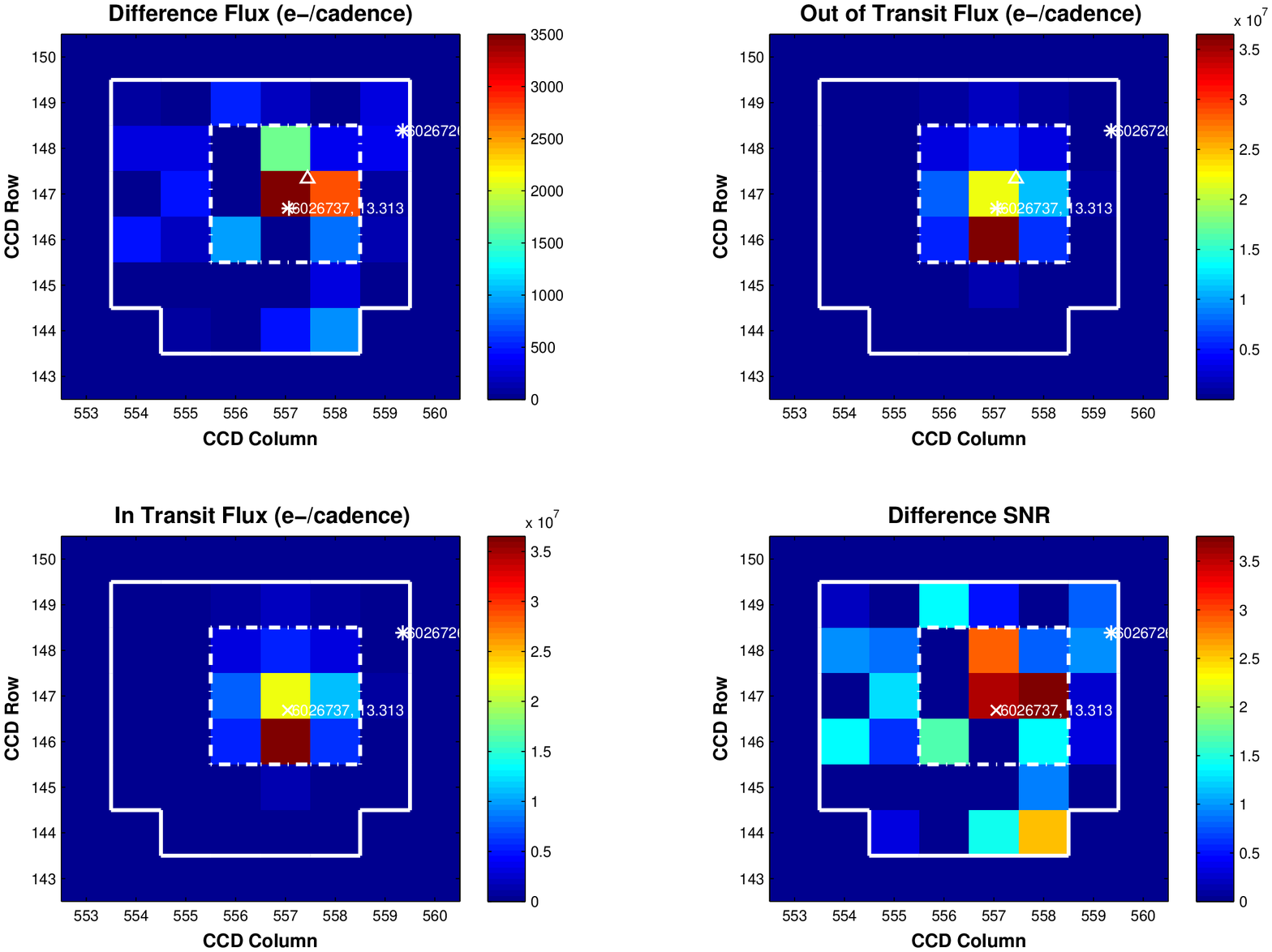}
\caption{Pixel images for a low SNR transit on KOI-2949 with an SNR of 11.  The  difference image
appears significantly different from the out-of-transit image in this quarter, indicating that the
transit source is not on the target star.  But other quarters show the transit source in other 
locations including on the target star.  This situation is typical for low SNR transits, and more reliable 
measurement of the transit source location
can be attained by combining 
the quarters as described in \S\ref{section:combining_quarters}. 
In this example the combined quarter result indicates that the transit location is statistically consistent with the
target star.}
\label{fig:KOI_2949_diff_image_low_snr}
\end{center}
\end{figure}

When the transit SNR is high the pixel images appear as in Figures \ref{fig:KOI_221_diff_image} (SNR = 378) and \ref{fig:KOI_109_diff_image}
(SNR = 101), with very well defined star-like difference images.  When the SNR is high and the transit is on the target
star, as in Figure \ref{fig:KOI_221_diff_image}, we expect the difference image to look like the out-of-transit image.  
Figure \ref{fig:KOI_2949_diff_image_low_snr} shows an example
of a low SNR transit on KOI-2949 with an SNR of 11.  In this figure the difference image looks significantly different from the out-of-transit 
image, so a cursory inspection of only this quarter's out-of-transit and difference images would indicate a significant offset.  
But examination of other quarters finds offsets
in other directions in some quarters and much smaller offsets in other quarters.  When the SNR is low, the difference image is subject
to pixel-level systematics that can pollute the difference image.  As we will see in \S\ref{section:combining_quarters}, combining
quarters puts the transit source statistically close to the target.  When the SNR is very low, the difference image is dominated
by noise because the transit does not have sufficient signal in individual quarters.


When the offset is as dramatic as that in Figure \ref{fig:KOI_109_diff_image}, cursory visual inspection 
is sufficient to determine that the transit signal does not occur on the target star.  We are interested, however,
in measuring smaller offsets that may not be so visually obvious.  
In addition we wish to have the ability to automatically measure and detect such transit-source offsets for thousands of transit 
signals.  This can be done by measuring the centroid of
the difference image and comparing with estimates of the target star position.  This approach encounters
several difficulties:
\begin{itemize}
\item Difference images can be noisy, particularly for low SNR transits.  This is particularly a problem 
for transits near spacecraft thermal events and in multiple planet systems, where the transit signals from 
multiple planets can interfere with each other.
\item Determination of the location of the target star should use the same method as the difference
image to minimize the impact of systematic measurement errors. 
\item The structure of the background signal for the target star due to crowding will be very different
from the difference image background signal because non-variable background stars will cancel out in the difference image.
\item In different quarters stars fall in different places on different pixels and pixel apertures vary from quarter to 
quarter.  Therefore the offsets measured in different quarters can be different.
\end{itemize}
We address these difficulties through the following strategies:
\begin{itemize}
\item Careful construction of the in- and out-of-transit images, described in \S\ref{section:average_images},
so the difference image is as clean as possible.
\item Determining the location of stars in the difference or 
out-of-transit image via PSF-type fitting to the pixel data using the {\it Kepler} Pixel Response Function (PRF), 
described in \S\ref{section:prf_fitting}, which is  
more robust against noise than photometric centroids.
\item Either carefully averaging the quarterly offsets (\S\ref{section:multiq_averaging}), 
or performing a joint multi-quarter fit (\S\ref{section:joint_multiq_fit}).
\end{itemize}

\subsection{Construction of in- and out-of-transit and difference pixel images} \label{section:average_images}


Our goal is to measure the location of the change in the flux due to the transit signal. Therefore we 
want to create a difference image by subtracting pixel flux in transit from pixel flux near transit.  
We want to avoid pixel flux away from the transit
so changes due to stellar variability are less likely to enter into the difference image.  We also want to avoid changes in flux
that are not related to the transit under examination, such as spacecraft thermal or pointing events or transits 
due to other planets orbiting the target star in multiple systems.
We minimize noise by averaging as many in- and out-of-transit measurements as possible subject to these
constraints.  

In each quarter, {\it Kepler} collects about 4300 long cadences, 
from which in- and out-of-transit exposures 
need to be identified.  We use the (unwhitened) transit model $M_n$ constructed in Data Validation \citep{Wu_DV} to select 
these cadences.  

In-transit cadences 
are defined as those cadences where the model is less than a threshold proportional to the model transit depth.
The current threshold is 3/4 of the transit depth: when the model is normalized so that
$M_n$ = 0 for out-of-transit cadences, in-transit cadences are those for which the model values $M_n <  -\frac{3}{4} d$,
where $d$ is the modeled fractional transit depth. 

\begin{figure}[htbp]
\begin{center}
\includegraphics[scale=0.9]{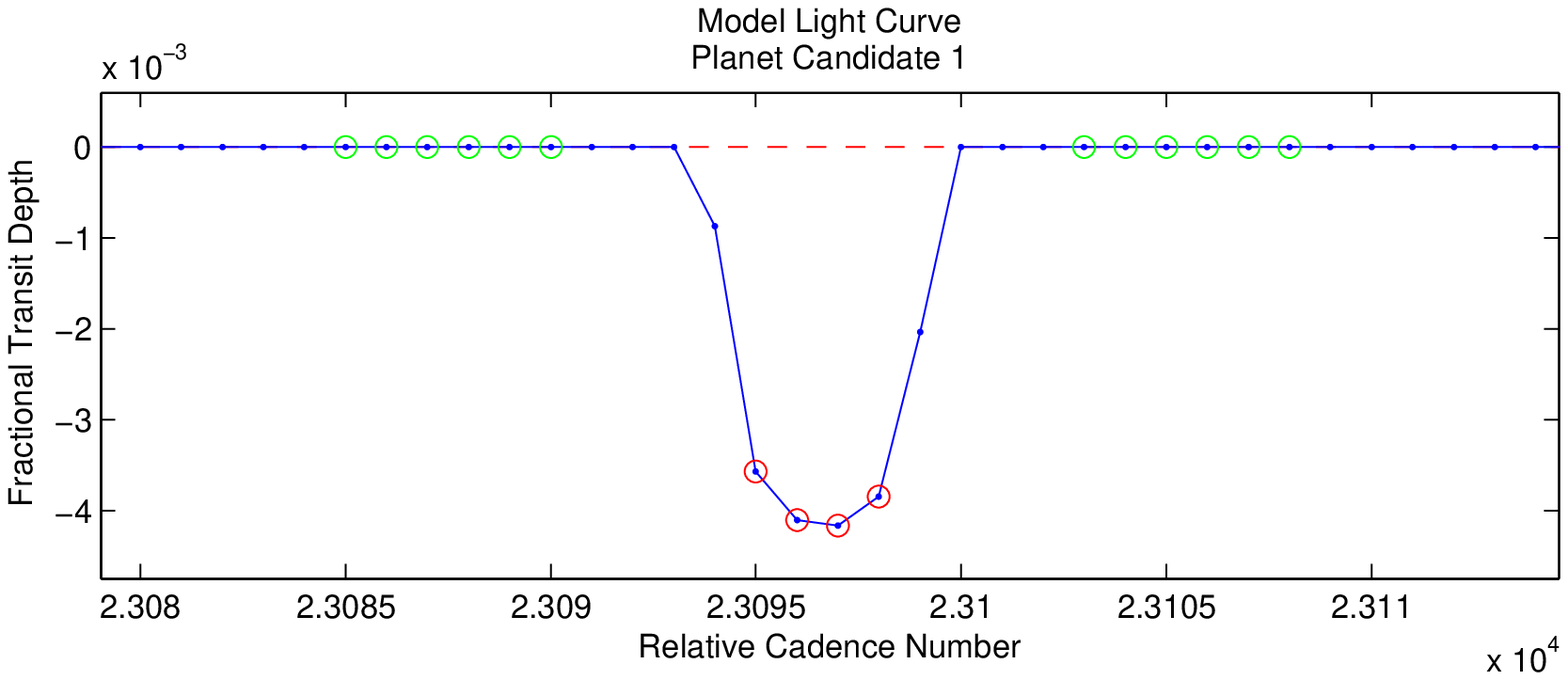} \\
\includegraphics[scale=0.9]{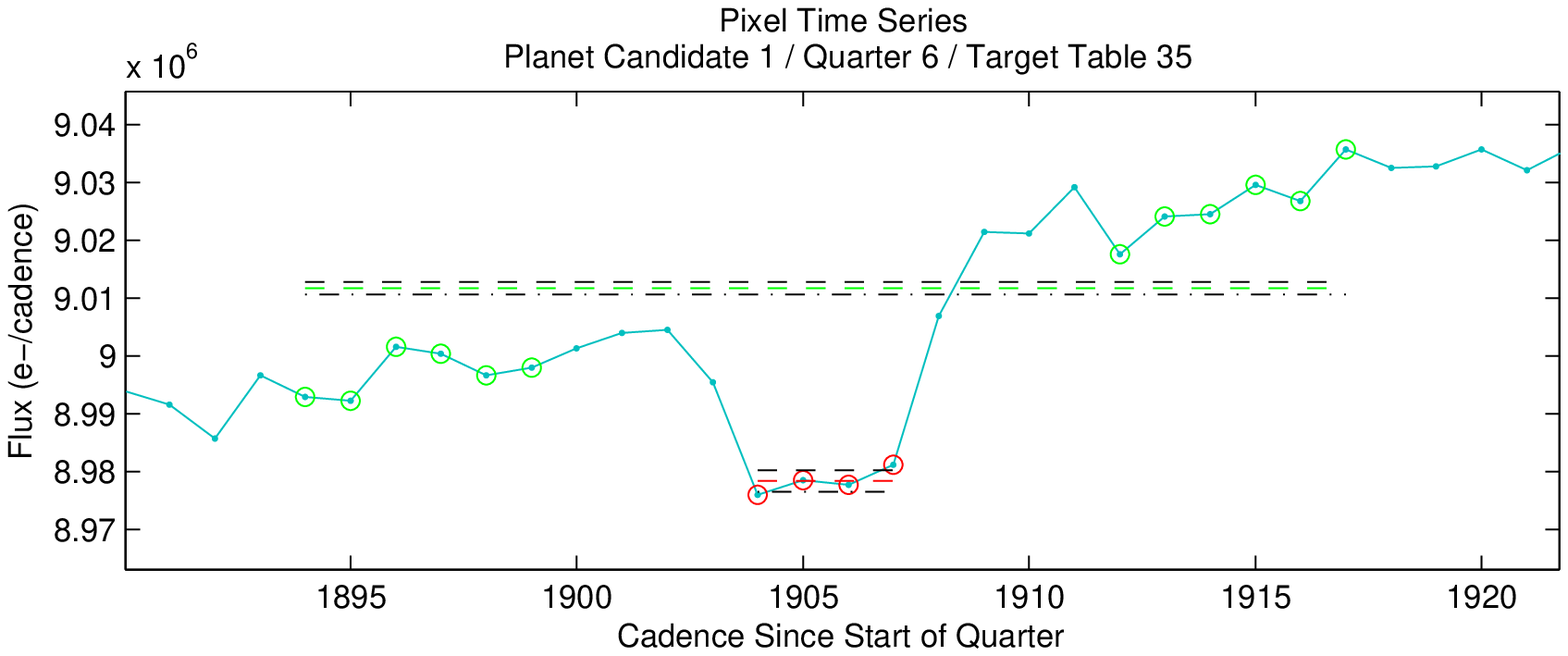}
\caption{An example of in- and out-of-transit cadence selection (KOI-221).  
Top: the transit model $M_n$ for a selected cadence range in quarter 6.  The $x$-axis shows the
cadences since the beginning of the {\it Kepler} science operations. 
The circles at the bottom of the transit show the cadences that were chosen for the in-transit image.
$N_{\mathrm{in}} = 4$ cadences were chosen in the transit because they are below the threshold described in the text.
The circles outside the transit show the cadences chosen for the out-of-transit image.  
The full transit is six cadences wide so $N_{\mathrm{out}} = 6$ cadences were chosen on both
sides of the transit.  The out-of-transit cadences are $N_{\mathrm{buffer}} = 3$ cadences from the transit.
Bottom: the actual transit in one of the brighter pixels.  The $x$-axis shows the
cadences since the beginning of quarter 6.}
\label{fig:diff_image_cadences}
\end{center}
\end{figure}

The out-of-transit cadences are chosen near each transit under the following criteria:
\begin{itemize}
\item Out-of-transit cadences are chosen on both sides of the transit so that an average of these out-of-transit 
cadences removes any locally linear secular trends.
\item Not too many cadences are chosen so that nonlinear variability on time scales longer than the transit are small.
\item Out-of-transit cadences should not be too close to the transit.
\end{itemize}
The number of out-of-transit cadences $N_{\mathrm{out}}$ is chosen as the number of cadences that occur during the
entire transit duration where $M_n < 0$.  
This is generally not the same as $N_{\mathrm{in}}$.  The 
out-of-transit cadences are chosen to lie more than $N_{\mathrm{buffer}}$ cadences from the cadences 
for which $M_n < 0$.  
Fig.~\ref{fig:diff_image_cadences} shows an example of selected cadences for a typical transit.


\begin{figure}[htbp]
\begin{center}
\includegraphics[scale=0.9]{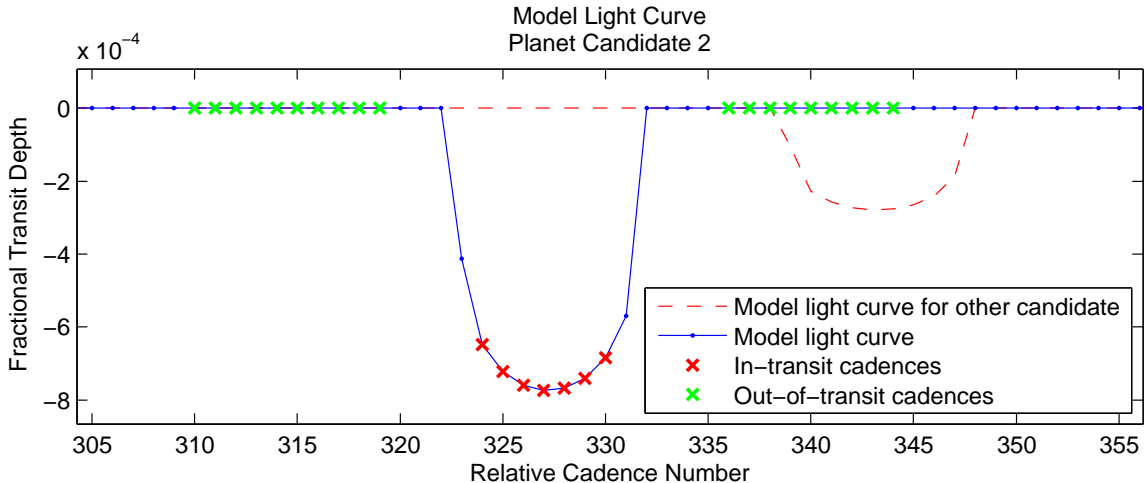}
\caption{An example of the interference with cadences chosen for a transit in the Kepler-11 system.
Seven out-of-transit points to the right of the transit are excluded because of the interfering transit by the other
planet candidate, which causes the entire transit to be excluded from the construction of the average pixel images.}
\label{fig:transit_interference}
\end{center}
\end{figure}

After in- and out-of-transit cadences are chosen they are excluded if they are associated with any of the following events:
\begin{itemize}
\item Data gaps such as Earth points and safe modes.
\item Cadences within a day after major spacecraft thermal events, such as recovery from Earth points and safe modes that
significantly change the temperature distribution of the spacecraft and require many hours to return to 
thermal equilibrium.
\item Pointing anomalies such as attitude tweaks, and loss of fine-point events.
\item Interference by transits from other planets in multiple planet systems.  An example of such interference is shown
in Figure \ref{fig:transit_interference}
\end{itemize}

If more than a small number of cadences associated with a transit are excluded then the entire transit is excluded from the 
construction of the difference image.  This threshold is currently set to zero, so if any cadences are excluded then the entire
transit is excluded.  As {\it Kepler} detects longer-period transits, so fewer transits will be available, this threshold will be relaxed to 
one or two excluded cadences per transit.

Once the final set of transits and their in- and out-of-transit cadences are identified, the in-transit pixel values are averaged to produce the in-transit image
and the out-of-transit cadences are averaged to produce the out-of-transit image.  The pixel values are not whitened or otherwise detrended:
we rely on the averaging described in this section to remove local secular trends.  First the average pixel values 
are computed for each transit, then each transit's averaged pixels in a quarter are averaged together to produce the final in- and out-of-transit
average pixel images for that quarter.
The difference image for the quarter is then the out-of-transit pixel image minus 
the in-transit pixel image.

\subsection{Fitting the Pixel Response Function} \label{section:prf_fitting}

In this section we describe how the {\it Kepler} pixel response function (PRF) \citep{bryson10}
is used to provide a robust, high-precision estimate of the target star and transit locations using the average out-of-transit
and difference images constructed as described in \S\ref{section:average_images}.
This technique requires that the target star is several magnitudes brighter than other stars in the out-of-transit pixels,
and that the transit signal is sufficiently strong in the difference image.  In \S\ref{section:prf_fit_quality} we describe
a quantitative measure of whether the average images for a given target star have the required properties.
\S\ref{section:prf_error} 
describes various ways in which this method can be compromised and discuss mitigation strategies.

The PRF gives the long-cadence brightness of a pixel due to a star at 
a specified location.  The PRF can be thought of as the convolution of the optical PSF with the 
effects of pointing, sub-pixel response and system electronics.  In this section we write the PRF as a 
unit flux function $f \left( \alpha, \delta, r_i, c_i \right)$
so $\sum_{i=1}^{P_{\mathrm{total}}} f \left( \alpha, \delta, r_i, c_i \right) = 1$,
where $P_{\mathrm{total}}$ is the 
number of all pixels that contain
flux from a star at sky coordinates $\left( \alpha, \delta \right)$, and $r_i$ and $c_i$ are those pixels' row
and column coordinates.  If the star has flux
$b$, then the value of a pixel at row $r_i$ and column $c_i$ due to that star
will be $p_i = b f \left( \alpha, \delta, r_i, c_i \right)$, and the sum of all pixels containing flux from that star 
is $\sum_{i=1}^{P_{\mathrm{total}}} p_{r_i, c_i} = b$.  (In \citet{bryson10} the star location is defined in pixel
coordinates rather than sky coordinates.  In this paper we include the projection from sky coordinates to
pixel coordinates in the PRF function $f$).

Assume we are given a set of $P$ pixel values $p_i$ with rows $r_i$ and columns $c_i$ 
that form a pixel image.  The $P$ pixels need not contain all the flux from the target star,
so $P$ may be less than $P_{\mathrm{total}}$. 
A PRF fit to these pixels is the determination of 
sky coordinates $\left( \alpha_{\mathrm{fit}}, \delta_{\mathrm{fit}} \right)$
and flux $b_{\mathrm{fit}}$  that minimize the function 
\begin{equation}
\chi^2 = \sum_{i=1}^P \frac{1}{\sigma_{p_i}^2}
	\left(p_i  -  b f \left( \alpha, \delta, r_i, c_i \right) \right)^2 \label{eqn:prf_fit}
\end{equation} 
where $\sigma_{p_i}$ is the uncertainty in the pixel value $p_i$.  
This fit is performed iteratively via the non-linear Levenberg-Marquardt algorithm \citep{levenberg,marquardt}.  Formally this is a
three dimensional fitting problem in the parameters $\alpha, \delta$ and $b$.  The fit to $b$, however,
can be reduced to a linear problem once the position is known, so this problem can be treated
as a much faster two-dimensional non-linear fit in $\alpha$ and $\delta$.  In each iteration of the Levenberg-Marquardt algorithm
the pixels $p_i$ at $\left(r_i, c_i \right)$ and the fit parameters $\alpha$ and $\delta$ are provided to the model function.  We first evaluate 
the uncertainty-normalized {\it Kepler} PRF at $\alpha$ and $\delta$, computing $\hat{p}_i = f \left( \alpha, \delta, r_i, c_i \right)/ \sigma_i$ for each pixel.
The flux $b$ is the linear least-squares fit of the input pixel values $p_i$ to the model $b \hat{p}_i$, given by
\begin{equation}
b = \frac{\sum_{i=1}^P p_i \hat{p}_i}{\sum_{i=1}^P \hat{p}_i^2}.
\end{equation}
The product $b \hat{p}_i$ is then returned by the model function.  The Levenberg-Marquardt 
algorithm seeks the $\alpha$ and $\delta$ that minimizes $\sum_{i=1}^P \left(p_i - b \hat{p}_i / \sigma_i \right)^2$
after several iterations.  (In the {\it Kepler} pipeline this is implemented as 
a model function passed to the MATLAB function {\it nlinfit}.)  Once the iteration has converged, providing 
$\left( \alpha_{\mathrm{fit}}, \delta_{\mathrm{fit}} \right)$, the final estimate of 
$b$ can be computed as $b_{\mathrm{fit}} = \left(\sum_{i=1}^P p_i \hat{p}_i \right)/\left(\sum_{i=1}^P \hat{p}_i^2 \right)$, where
now $\hat{p}_i = f \left( \alpha_{\mathrm{fit}}, \delta_{\mathrm{fit}}, r_i, c_i \right)/ \sigma_i$.

The typical implementation of the Levenberg-Marquardt algorithm returns the Jacobian $J$, which contains the 
derivatives of the model function with respect to position.
To estimate the uncertainty of the fit location we need the Jacobian of the position with respect to the pixel
values given by the model function.  We obtain this by inverting $J$, using the pseudo-inverse, to 
give the transformation $T = \left( J^T J \right)^{-1} J^T$.  T is a $P \times 2$ matrix, and the columns of T are normalized 
by the pixel uncertainties: $T_{ij} \rightarrow T_{ij}/\sigma_i$ for $j = 1,2$.
Then the PRF fit location covariance matrix is ${\cal C} = T^T {\cal C}_{\mathrm{pixel}} T$, where ${\cal C}_{\mathrm{pixel}}$ is the pixel covariance,
and the fit location uncertainties are 
the square root of the diagonal of $\cal C$: $\sigma_\alpha = \sqrt{{\cal C}_{1,1}}$  
and $\sigma_\delta = \sqrt{{\cal C}_{2,2}}$.

The PRF is fit separately to the difference image and the out-of-transit image.  Because the fit to the difference image
$\left( \alpha_{\mathrm{diff}}, \delta_{\mathrm{diff}} \right)$ measures
the position of the transiting source and the fit to the out-of-transit image 
$\left( \alpha_{\mathrm{OOT}}, \delta_{\mathrm{OOT}} \right)$ measures  the position of the target star, the offset
of the transit source from the target is simply $\left( \Delta \alpha, \Delta \delta \right) 
= \left(\left(\alpha_{\mathrm{diff}} - \alpha_{\mathrm{OOT}} \right) \cos \delta_{\mathrm{OOT}}, \delta_{\mathrm{diff}} - \delta_{\mathrm{OOT}} \right)$.
Then the offset distance and uncertainty are computed as in Equation~\ref{eqn:offset_distance}.

In- and out-of-transit pixel images, and therefore difference images, can only be constructed on a quarter-by-quarter basis.  
Images cannot be combined across quarters in a useful way because 
\begin{itemize}
\item The same star will fall on slightly different pixel 
locations in each quarter due to pointing differences and small asymmetries in the construction of the {\it Kepler} focal plane.
\item The {\it Kepler} PRF at the star's location can have large changes from quarter to quarter.
\item The pixel aperture generally varies in both size and shape from quarter to quarter.
\end{itemize}
Two approaches to combining 
quarters will be described in \S\ref{section:combining_quarters}.

\subsubsection{Systematic PRF fit error} \label{section:prf_error}

Systematic error in the PRF fit arises from primarily from two classes of sources: error in the PRF model being fit and crowding.  
These errors cause biases in the offset vector $\left( \Delta \alpha, \Delta \delta \right)$.
There are various ways to control systematic PRF fit errors, so we examine these
errors in detail.

\subsubsubsection{Sources of PRF fit error}

\paragraph{PRF Model Error}
The PRF model contains various sources of error \citep{bryson10} which lead to {\it a priori} unpredictable
bias in the PRF-fit centroid.  
Because the target star falls on different parts of the {\it Kepler} field of view in different quarters, variation of the PRF across the focal plane
causes the PRF error bias to vary from quarter to quarter.
 
\begin{figure}[htbp]
\begin{center}
\includegraphics[trim = 1.7cm 1cm 0 0, clip, scale=0.7]{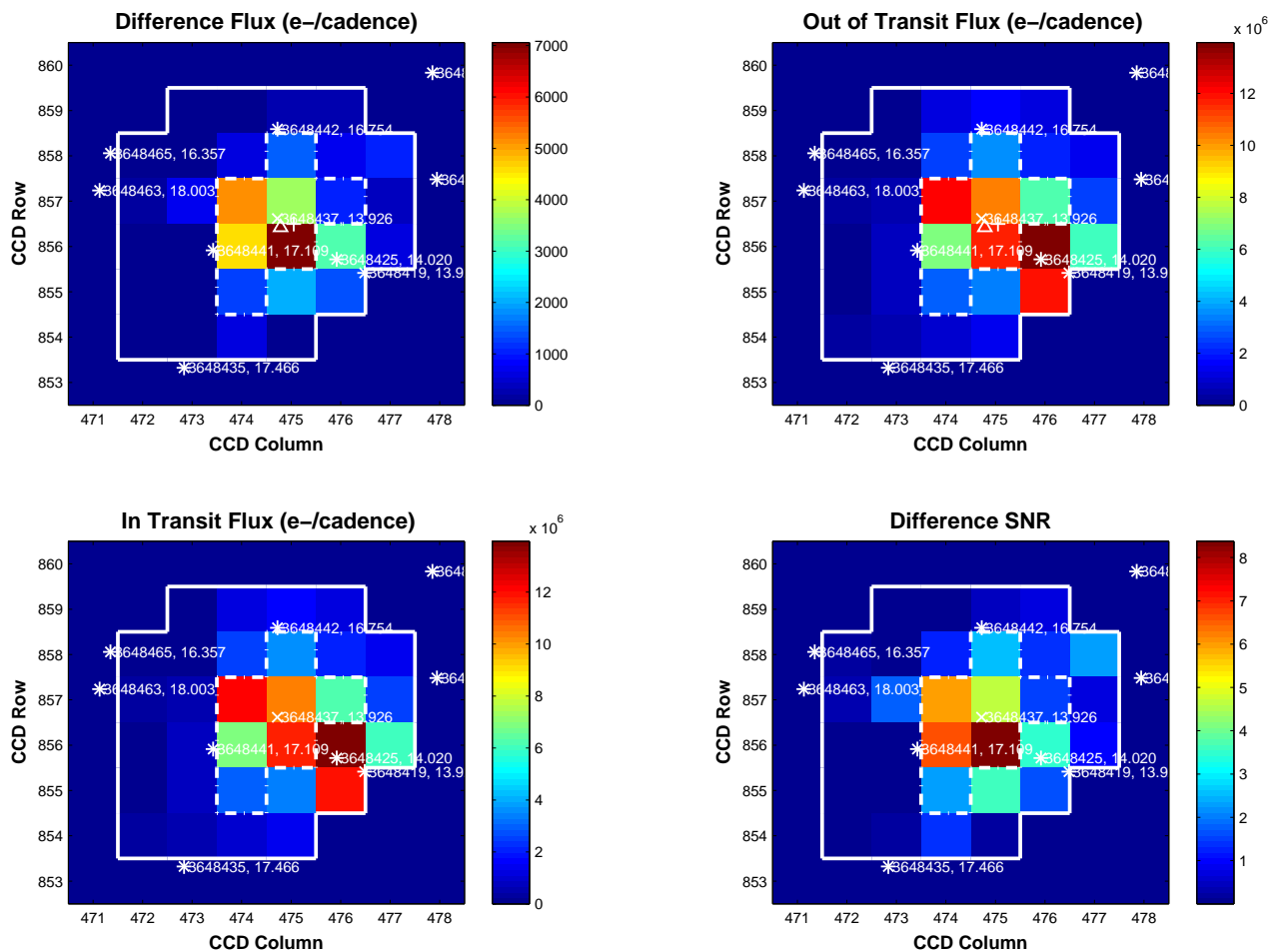}
\caption{An example of a target with large amounts of crowding (KOI-1861).  The in- and out-of-transit images
do not appear as a typical star, and the fact that this is due to crowding is indicated by the large number of asterisks on
the image indicating many relatively bright background stars.  The difference image, on the other hand, looks much more like
a star because most of the background stars in the image have cancelled out, though there is still some 
residual background contamination.  In this case the fit to the out-of-transit image will have a large bias 
relative to the target star, while the bias in the difference image fit will be much smaller.  This results in a
biased offset measurement of the transit source relative to the target star.  Visual inspection of the
difference image, however, indicates that the transit source is closer to the target star than the biased 
measurement would indicate.}
\label{koi-1861_crowded_diff_image}
\end{center}
\end{figure}

\paragraph{Crowding Bias}
The PRF fit is a single-star fit, and therefore assumes that the target star in the out-of-transit image and the transit signal in the 
difference image are the only stars present in the pixels.  This is rarely the case in the out-of-transit image and sometimes not the case in 
the difference image due to variability of field stars.  Unlike the case of photometric centroids described in \S\ref{centroiding},
the effect of crowding on the PRF fit
is difficult to predict.  Because field stars mostly cancel in the difference image, the crowding signal in the out-of-transit and 
difference images can be very different.  Therefore the PRF fit to the out-of-transit and difference images can have 
very different biases, which leads to errors in the offset vector $\left( \Delta \alpha, \Delta \delta \right)$.  An example of a target 
with a large amount of crowding is shown in Figure~\ref{koi-1861_crowded_diff_image}.


\begin{figure}[htbp]
\begin{center}
\includegraphics[trim = 1.7cm 1cm 0 0, clip, scale=0.7]{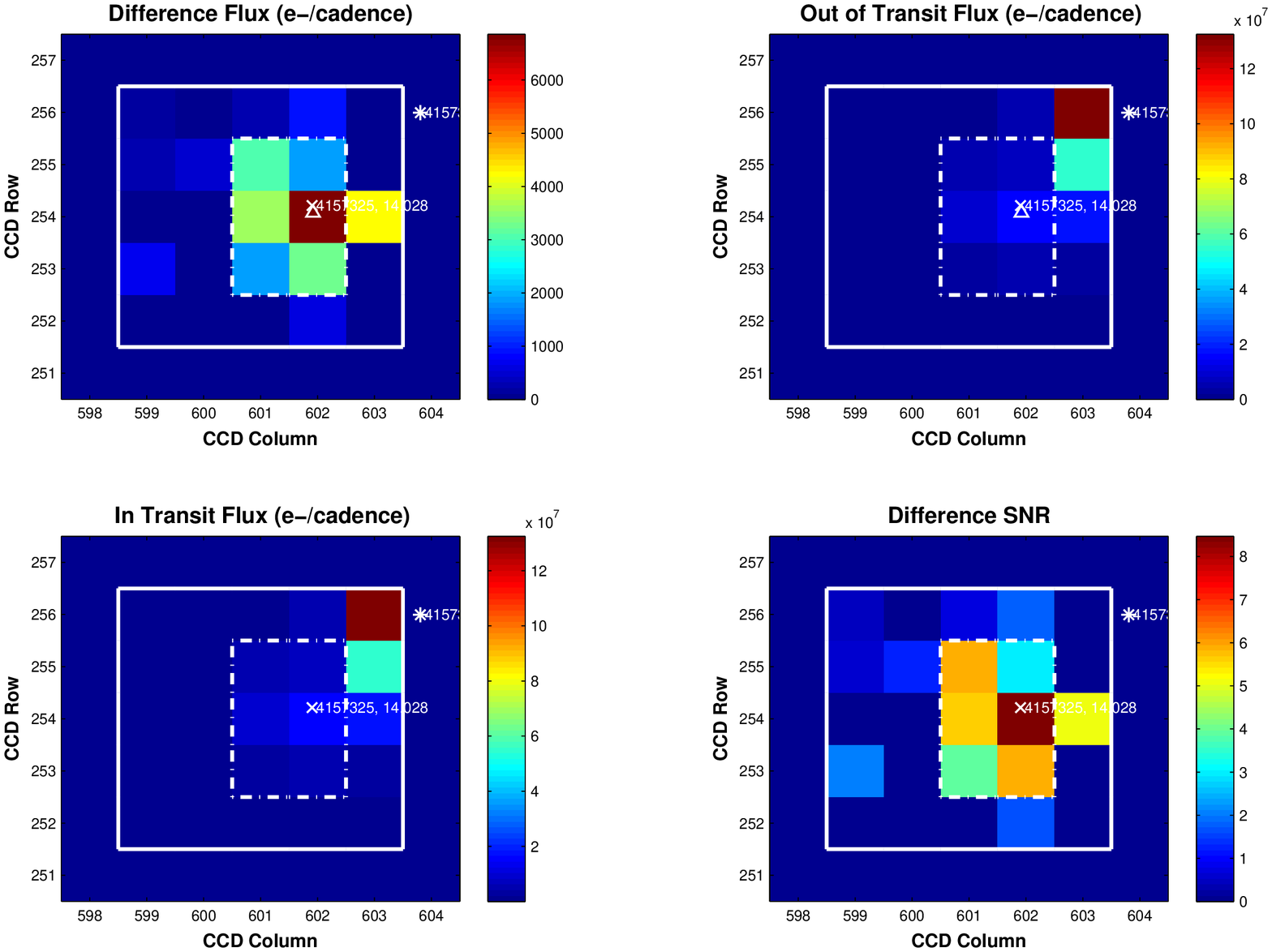}
\caption{An example of a target with bright field star that captures the out-of-transit PRF fit (KOI-1860).  
The out-of-transit image is dominated by the bright star in the upper right corner, so this field star position will
be returned by the PRF fit to the out-of-transit image.  The difference image, however, shows a nicely star-shaped
pattern at the location of the target star, so the target star position will be returned by the PRF fit to the difference image.
The resulting offset vector measures the distance of the transit source (target star in this case) to the bright field
star rather than the distance of the transit source to the target star.  In this case blindly using the offset values
would lead to the erroneous identification of a background false positive. 
}
\label{koi-1860_spoofed_diff_image}
\end{center}
\end{figure}

In the worst case there is a 
field star in the out-of-transit image brighter than the target star, so the PRF fit to the out-of-transit image
returns the centroid of the field star rather than the target star.
When this bright field star cancels in the difference image, so the difference image is dominated by a transit on the target star, 
the offset vector $\left( \Delta \alpha, \Delta \delta \right)$ gives the distance of the transit signal from the field star rather than the target star.
The result is an incorrect measurement of a significant offset of the transit source from the target star.  
An example of this situation, KOI-1860 (discussed in \S\ref{section:centroid_shift_error}),
is shown in Figure~\ref{koi-1860_spoofed_diff_image}.

\subsubsubsection{Mitigation of the impact of PRF fit error within a quarter}

Average out-of-transit and difference images are computed for each quarter, and these are fit by the PRF to 
estimate the offset of the transit source from the target star.  PRF model error and crowding 
contribute systematic errors in this estimate.  Here we discuss ways
to mitigate these systematic errors within each quarter.  In \S\ref{section:multiq_averaging} we discuss
ways the possibility of averaging out these systematics across quarters.  

The {\it Kepler} PRF for nearby stars will be very nearly the same, so the PRF model error for those stars will be similar.  
Assuming low crowding, the PRF fit of the out-of-transit image and the fit to the difference image will have similar biases
due to PRF model error.  When forming the offset vector $\left( \Delta \alpha, \Delta \delta \right)$ as the difference between these 
two fits, these biases should approximately cancel.  We therefore prefer the offset vector computed as the difference
between the two out-of-transit fits when the target star is not highly crowded.

When the target star is highly crowded, crowding bias will dominate the out-of-transit PRF fit but rarely the difference image PRF fit. 
This bias is usually due to an error in the measurement of the target star position.  As an alternative we compute 
the transit source offset relative to the target star's catalog position.  We define 
$\left( \Delta \alpha, \Delta \delta \right)_{\mathrm{catalog}}
= \left(\left(\alpha_{\mathrm{diff}} - \alpha_{\mathrm{catalog}} \right) \cos \delta_{\mathrm{catalog}}, 
\delta_{\mathrm{diff}} - \delta_{\mathrm{catalog}} \right)$,
where
$\left( \alpha_{\mathrm{catalog}}, \delta_{\mathrm{catalog}} \right)$ is the catalog position of the target star
(usually from the {\it Kepler} input catalog).  When $\left( \Delta \alpha, \Delta \delta \right)$ 
differs from $\left( \Delta \alpha, \Delta \delta \right)_{\mathrm{catalog}}$ by more than a {\it Kepler} pixel (3.98 arcseconds), the 
out-of-transit measurement of the target star position $\left( \alpha_{\mathrm{OOT}}, \delta_{\mathrm{OOT}} \right)$
likely contains large errors and the offset vector $\left( \Delta \alpha, \Delta \delta \right)$ should be considered unreliable.
The catalog-based offset error $\left( \Delta \alpha, \Delta \delta \right)_{\mathrm{catalog}}$ can be used instead,
but is itself subject to error because
a) it does not mitigate fit error due to PRF error and b) is subject to catalog errors due to, for example, unknown proper motion 
of the target star.  In this case the PRF fit results should be considered qualitative and to have lower accuracy than non-crowded 
targets, regardless of the formal propagated uncertainty.  
In the example in Figure~\ref{koi-1860_spoofed_diff_image} the magnitude of the offset vector in that 
quarter is about 11 arcseconds, while the magnitude of the offset from the catalog position is about 0.6 arcseconds.

A work in preparation \citep{bryson_Morton13} will describe the use of modeling to identify and mitigate bias due to crowding. 

In the majority of cases the bias will be due to a mix of crowding and PRF model error, with comparably small contributions from each.  
In this case we reduce the overall bias by taking advantage 
of the variation in bias across quarters via averaging as described in \S\ref{section:combining_quarters}.

\subsubsection{PRF Fit Quality} \label{section:prf_fit_quality}

The quarterly out-of-transit and difference images can be polluted by various types of contamination.  For example the 
out-of-transit image may have bright stars in addition to the target star.  The difference image may have more than
one stellar image due to the variability of a field star, or the transit may have low SNR, causing the difference image 
to be poorly formed as in Figure~\ref{fig:KOI_2949_diff_image_low_snr}.  These cases will degrade the 
reliability of the PRF-fit source offset measurement.  
The quality of the PRF fit can be determined by evaluating the PRF at the fit position,
creating a synthetic pixel image containing only one star at that position, and compare this to 
the observed average pixel image.  This synthetic image
will have the pixel values $\tilde{p_i} = b_{\mathrm{fit}} f \left( \alpha_{\mathrm{fit}}, \delta_{\mathrm{fit}}, r_i, c_i \right)$
($ = b_{\mathrm{fit}} \hat{p_i}$), where the subscript ``fit'' refers to ``diff'' or ``OOT'' as appropriate.  
These can be compared to the actual pixel values $p_i$ to determine if the fitted
PRF reproduces the observed pixels.  One simple comparison is to compute the correlation between $\tilde{p_i}$ and $p_i$, 
and declare the fit good if this correlation is above some threshold.  For the difference image fit quality we set the threshold 
to 0.7.  When the correlation is below this threshold, then the difference image is likely dominated by noise,
typically because the transit has a very low SNR.  When the correlation is below threshold for the out-of-transit fit,
then it is likely that there is more than one bright star in the image, which compromises the fit due to crowding.  In both cases 
the source offset measurement is likely to be unreliable.

\subsection{Combining Quarterly Results} \label{section:combining_quarters}


A comparison of PRF-fit star positions with their catalog RA and Dec show that the combination of crowding and PRF error bias has an
approximately Gaussian distribution
with a median of ~1 millipixel (0.004 arcsec) and a median absolute deviation of ~22 millipixels (0.09 arcsec) \citep{bryson10}.  
While the quarter-to-quarter variation in the PRF fit of a particular star can have larger spreads, we find that for most stars this 
quarter-to-quarter variation is approximately zero-mean on average.  We therefore combine the quarterly offsets to 
improve the precision of the PRF-fit centroid offset vector.  

\subsubsection{Multi-Quarter Averaging} \label{section:multiq_averaging}

We denote the single-quarter PRF fit offset vectors by $\left( \Delta \alpha_q, \Delta \delta_q \right)$, where $q$ labels the quarter.  
A simple average of $Q$ quarters, $\frac{1}{Q} \sum_{q=1}^Q \left( \Delta \alpha_q, \Delta \delta_q \right)$ with its uncertainties 
\linebreak $\frac{1}{Q} \sqrt{  \sum_{q=1}^Q \left( \sigma_{\Delta \alpha_q}^2,  \sigma_{\Delta \delta_q}^2 \right)}$ can be used
but this has the weakness that the uncertainties do not reflect scatter in the quarterly averages.  For example a set of points 
on a large circle with some uncertainty will have the same average and average uncertainty as a set of points with  
the same uncertainty that all lie at the center of the circle.  We would like the uncertainty to reflect the scatter of the quarterly
offsets.

We accomplish this by treating the quarterly offset vectors and their uncertainties as a time series, and compute
the average offset $\left(\overline{\Delta \alpha}, \overline{\Delta \delta} \right)$ by robustly fitting this time 
series with a constant.  In other words we compute a least-squares robust 
fit of a 0th-order polynomial to the quarterly data, minimizing
\begin{equation}
\sum_{q=1}^Q \frac{1}{\left( \sigma_{\Delta \alpha_q} \right)^2}\left(\Delta \alpha_q  - \overline{\Delta \alpha} \right)^2,
\qquad  \sum_{q=1}^Q \frac{1}{\left( \sigma_{\Delta \delta_q} \right)^2}\left(\Delta \delta_q  - \overline{\Delta \delta} \right)^2.
\label{eqn:multiq_average}
\end{equation} 
We compute a robust fit to suppress statistical outliers in the belief that these are due to transient biases resulting from systematic
events such as pointing or thermal anomalies.  The uncertainties in the above fit are typically returned by the robust fit algorithm
used to compute $\left(\overline{\Delta \alpha}, \overline{\Delta \delta} \right)$.  Care must be taken when estimating these 
uncertainties {\it a priori} from the quarterly data because every fourth quarter the spacecraft orientation is strongly correlated.

The above estimate of the average uncertainty assumes Gaussian statistics.  While PRF fit biases appear nearly Gaussian
in the statistical sense, they may not be Gaussian for individual targets.  We therefore compute an alternative uncertainty 
via bootstrap analysis, which provides a more general estimate of the uncertainty.  
We use a resample-with-replacement strategy, creating an ensemble
of $Q^2$ simple multi-quarter averages.  
Specifically, given the set of $Q$ measured offsets $\left( \Delta \alpha_1, \Delta \alpha_2, \ldots, \Delta \alpha_Q \right)$, 
$Q^2$ realizations are created, where in each realization we replace each element with an offset randomly chosen from the measured set.
Examples of these realizations when $Q = 5$ include 
$\left( \Delta \alpha_3, \Delta \alpha_1, \Delta \alpha_5, \Delta \alpha_4, \Delta \alpha_2 \right)$ and 
$\left( \Delta \alpha_2, \Delta \alpha_4, \Delta \alpha_1, \Delta \alpha_4, \Delta \alpha_1 \right)$.
Averages are computed for each of these realizations, and the standard deviation of the resulting ensemble of $Q^2$ averages 
provides the bootstrap uncertainty estimate.  The bootstrap uncertainty is typically very similar to the uncertainty returned by the
robust fit described above, but can be significantly different for specific targets.  
We choose the larger of the two uncertainty estimates as the final uncertainty estimate for the multi-quarter average $\sigma_{\Delta \alpha}$.
A similar analysis applies to $\sigma_{\Delta \delta}$.

\begin{figure}[htbp]
\begin{center}
\includegraphics[scale=0.9]{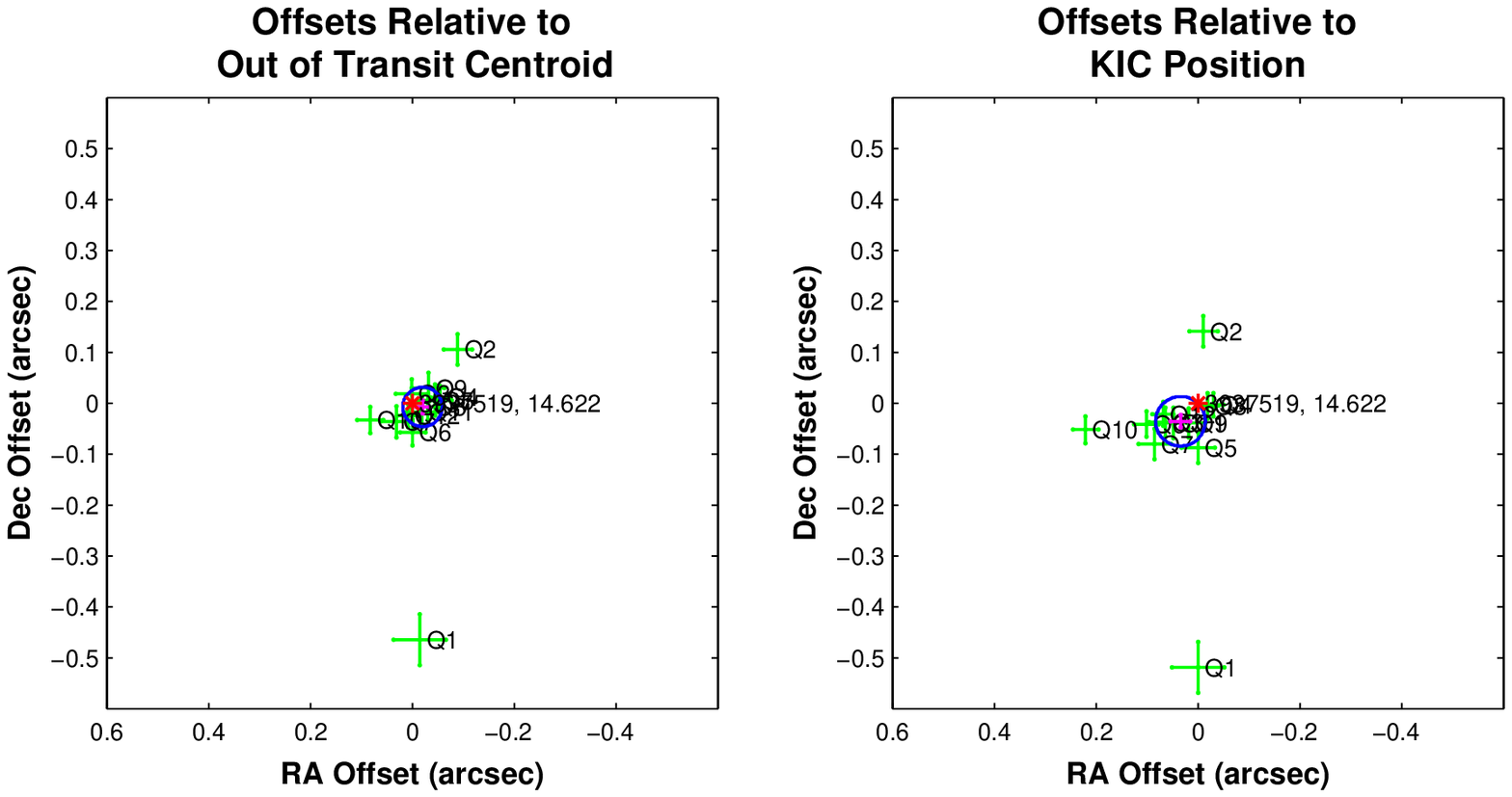} 
\caption{An example of multi-quarter offset analysis when the transit signal seems to be on the target star (KOI-221). 
In both figures the $x$- and $y$-axes give the offsets $\Delta \alpha$ and $\Delta \delta$, with $\left( 0,0 \right)$ being
the catalog location of the target star.  The green crosses show the individual quarter offsets labeled by quarter, and the length of the crosses are 
equal to the uncertainties $\sigma_{\Delta \alpha}$ and $\sigma_{\Delta \delta}$.  The location of the multi-quarter average 
$\left(\overline{\Delta \alpha}, \overline{\Delta \delta} \right)$ is shown as a magenta cross 
(obscured by the tight cluster of green crosses).  The blue circle has radius equal to three times the 
uncertainty in the magnitude of $\left(\overline{\Delta \alpha}, \overline{\Delta \delta} \right)$.  Star locations relative to
the target star are shown
as asterisks, with the target star in red (there happen to be no other stars in this figure).  The KIC catalog number
and {\it Kepler} magnitudes are shown next to each star.  We see that most offsets are tightly clustered
within 0.1 arcseconds of the target star with Q1 and Q2 as outliers.  Left: the offsets 
$\left( \Delta \alpha, \Delta \delta \right)$
relative to the PRF fit to the out-of-transit 
centroid.  Right: the offsets $\left( \Delta \alpha, \Delta \delta \right)_{\mathrm{catalog}}$ relative to the catalog position of the target star.
The difference between the left and right plots is not a simple translation because the two plots have different biases due
to PRF error and crowding (see \S\ref{section:prf_error}).
}
\label{fig:koi_221_multiq_plot}
\end{center}
\end{figure}

\begin{figure}[htbp]
\begin{center}
\includegraphics[scale=0.9]{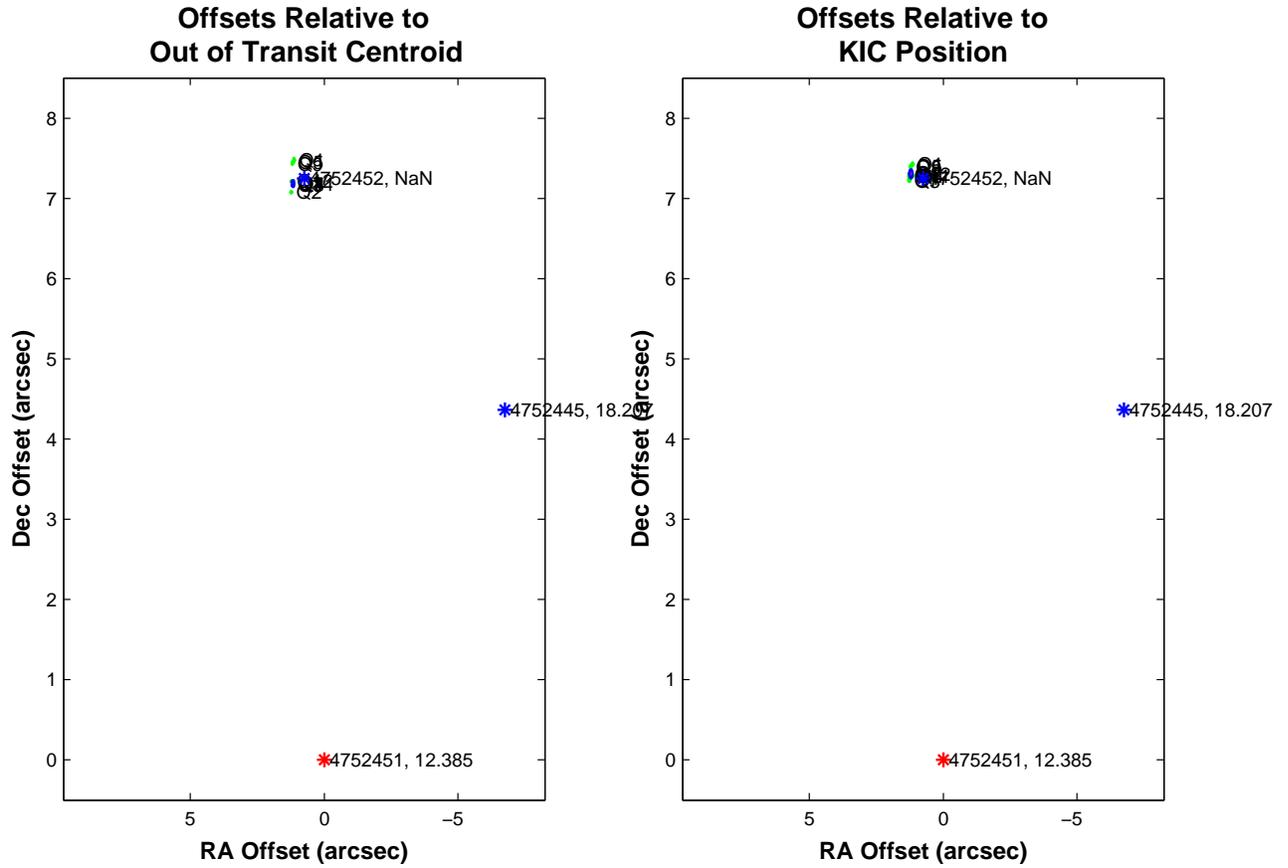} 
\caption{An example of multi-quarter offset analysis when the transit signal seems to be on
a different star than the target star (KOI-109).  The quarterly offsets are tightly clustered 
around the star KIC 4752452, indicating that this star is the 
source of the transit. See the caption to Figure~\ref{fig:koi_221_multiq_plot} for a description
of these plots.  
}
\label{fig:koi_109_multiq_plot}
\end{center}
\end{figure}

\begin{figure}[htbp]
\begin{center}
\includegraphics[scale=0.9]{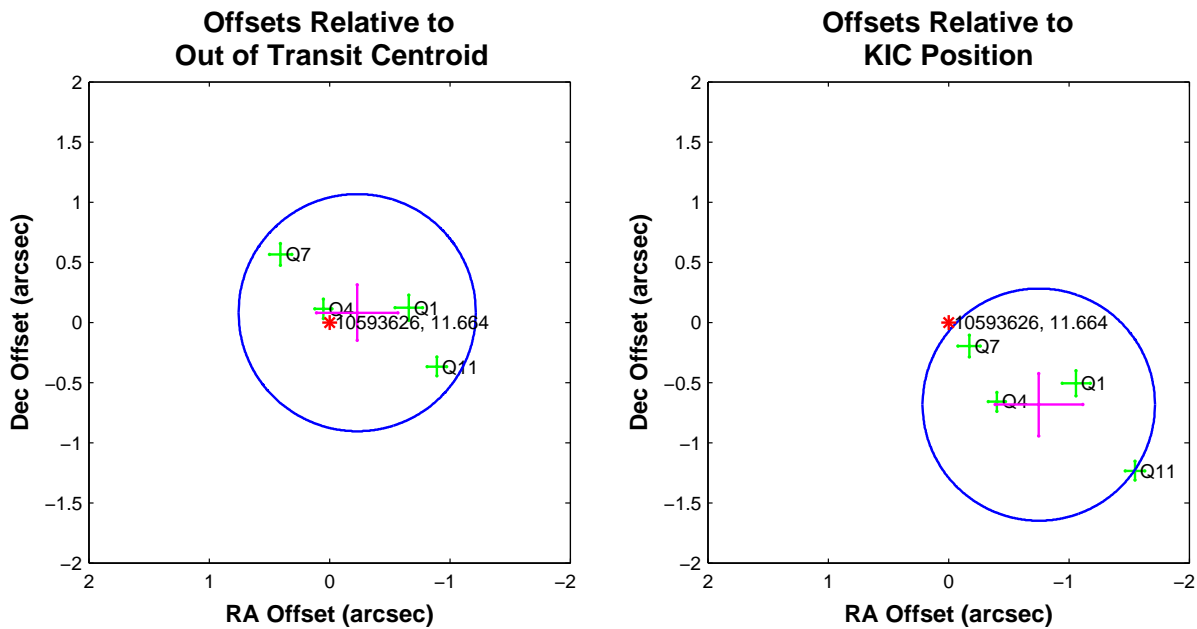} 
\caption{An example of multi-quarter offset analysis for a confirmed planet signal (Kepler-22b) with a very long
period orbit, so only four quarters show transits.  
The result is a larger scatter and higher average uncertainty compared to the case where there are transits present in 
every quarter.  Also there is a significant difference in the offsets relative to the out-of-transit centroid in the 
left panel and relative to the target star's catalog position in the right panel.  
This is likely due to a combination of not-fully-averaged PRF bias and catalog error.  If this planet were not confirmed by other
methods \citep{bor12_Kepler22} we would have only moderate confidence that the transit signal is on the target star.
See the caption to Figure~\ref{fig:koi_221_multiq_plot} for a description
of these plots.  
}
\label{fig:kepler-22_multiq_plot}
\end{center}
\end{figure}

\begin{figure}[htbp]
\begin{center}
\includegraphics[scale=0.9]{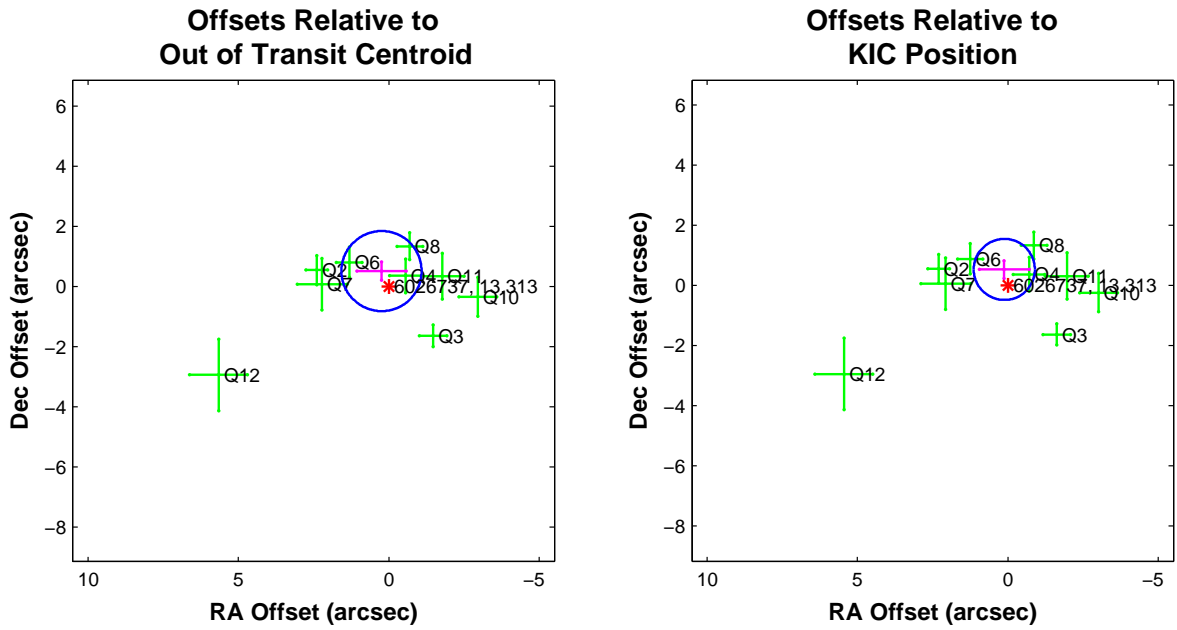} 
\caption{An example of multi-quarter offset analysis for a low SNR transit signal (KOI-2949) with SNR = 11.  
In this case the quarterly offsets have a large scatter measured in arcseconds, but the average across quarters 
is within 3 standard deviations of the target star.
See the caption to Figure~\ref{fig:koi_221_multiq_plot} for a description
of these plots.  
}
\label{fig:koi_2949_multiq_plot}
\end{center}
\end{figure}

\begin{figure}[htbp]
\begin{center}
\includegraphics[trim = 0 3cm 0 0, clip, scale=0.9]{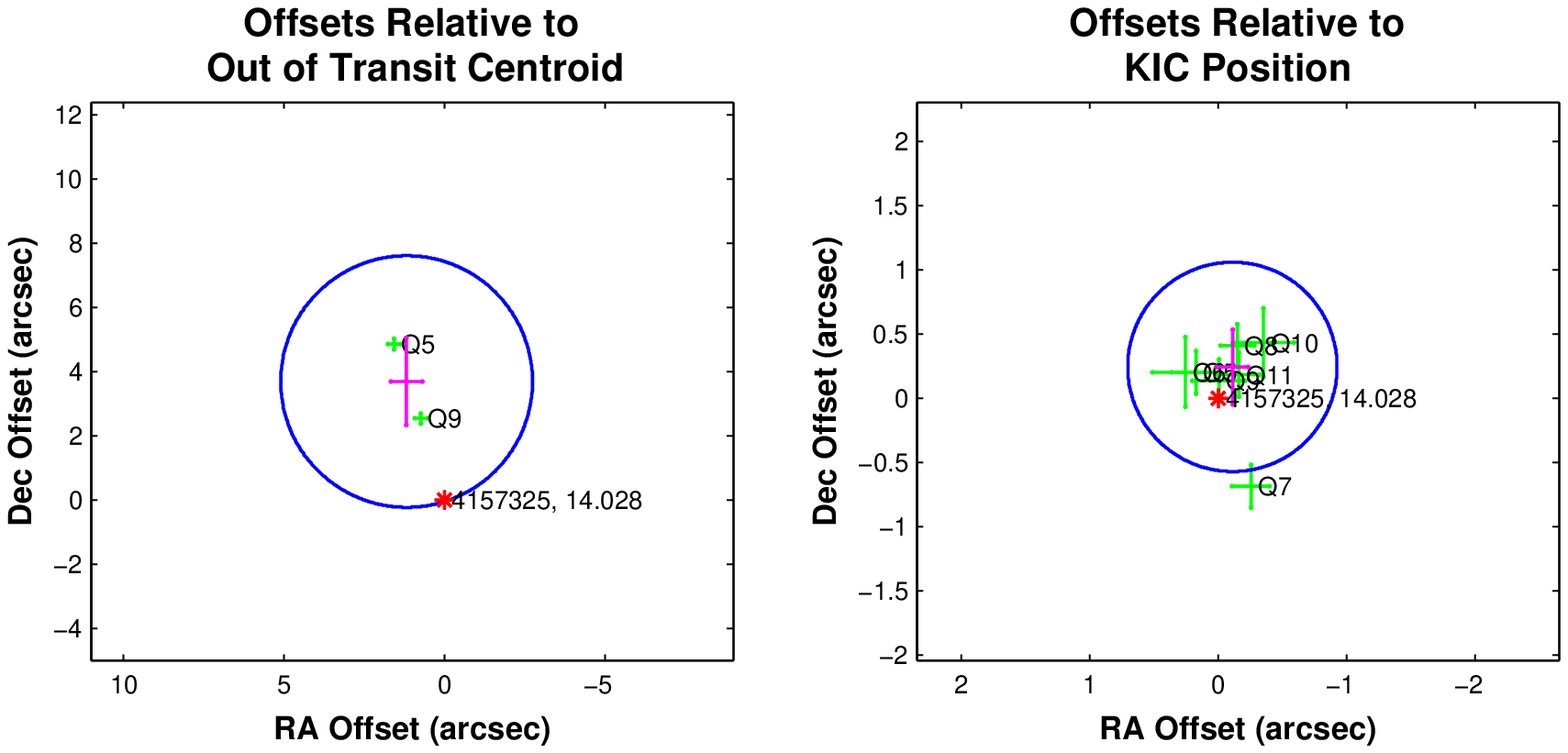} 
\caption{An example of multi-quarter offset analysis for a target star (KOI-1860, also discussed in \S\ref{section:centroid_shift_error})
whose pixels contain a brighter field star (see Figure~\ref{koi-1860_spoofed_diff_image}).  The offsets relative to the out-of-transit centriod
are large because the bright star captured the out-of-transit PRF fit.  The out-of-transit PRF fit also failed
in many quarters because the bright star is at the edge of the pixel aperture.  The offsets relative to the 
target star's catalog position are, however, well clustered around the target star indicating that the offset of the transit is 
not statistically significant.  We therefore conclude that the large offset relative to the out-of-transit centroid is due to systematic
effects from the bright field star in the pixels.  
See the caption to Figure~\ref{fig:koi_221_multiq_plot} for a description
of these plots.  
}
\label{fig:koi_1860_multiq_plot}
\end{center}
\end{figure}

Examples of this multi-quarter averaging technique are shown in Figures~\ref{fig:koi_221_multiq_plot} through \ref{fig:koi_1860_multiq_plot}.
Figure \ref{fig:koi_221_multiq_plot} shows
a case with no significant offset while Figure~\ref{fig:koi_109_multiq_plot} 
shows a case with a significant offset, indicating that the transit signal is on a
background star.  For long-period transiting planets, where there are few quarters that contain transits, the 
benefits of multi-quarter averaging will diminish.
In such cases, however, multi-quarter averaging can often provide good results, an example of which 
is shown in Figure~\ref{fig:kepler-22_multiq_plot}.   Figure~\ref{fig:koi_2949_multiq_plot} shows the low SNR example discussed in
\S\ref{section:difference_imaging}, where we see that there is a large scatter in the quarterly measurements, but the multi-quarter average
is within three standard deviations of the target star.  

The case of KOI-1860, where a bright field star at the
edge of the captured pixels introduces large systematic error, is examined in Figure~\ref{fig:koi_1860_multiq_plot}.  The offset
relative to the out-of-transit 
centroid is measured to be about 4 arcseconds, which is a statistically significant $4 \sigma$.  For most quarters, particularly
those which would show a larger offset, the PRF fit to the out-of-transit image failed because the bright star falls very close to the edge of
the captured pixels.  The offset relative to the catalog position, however, is much smaller, with a mult-quarter average of about 0.3 arcseconds
or $1 \sigma$.  Because we are aware of the bright star crowding for KOI-1860, we defer to the offset relative to the catalog position, which is
not statistically significant.  


\begin{figure}[htbp]
\begin{center}
\includegraphics[scale=0.45]{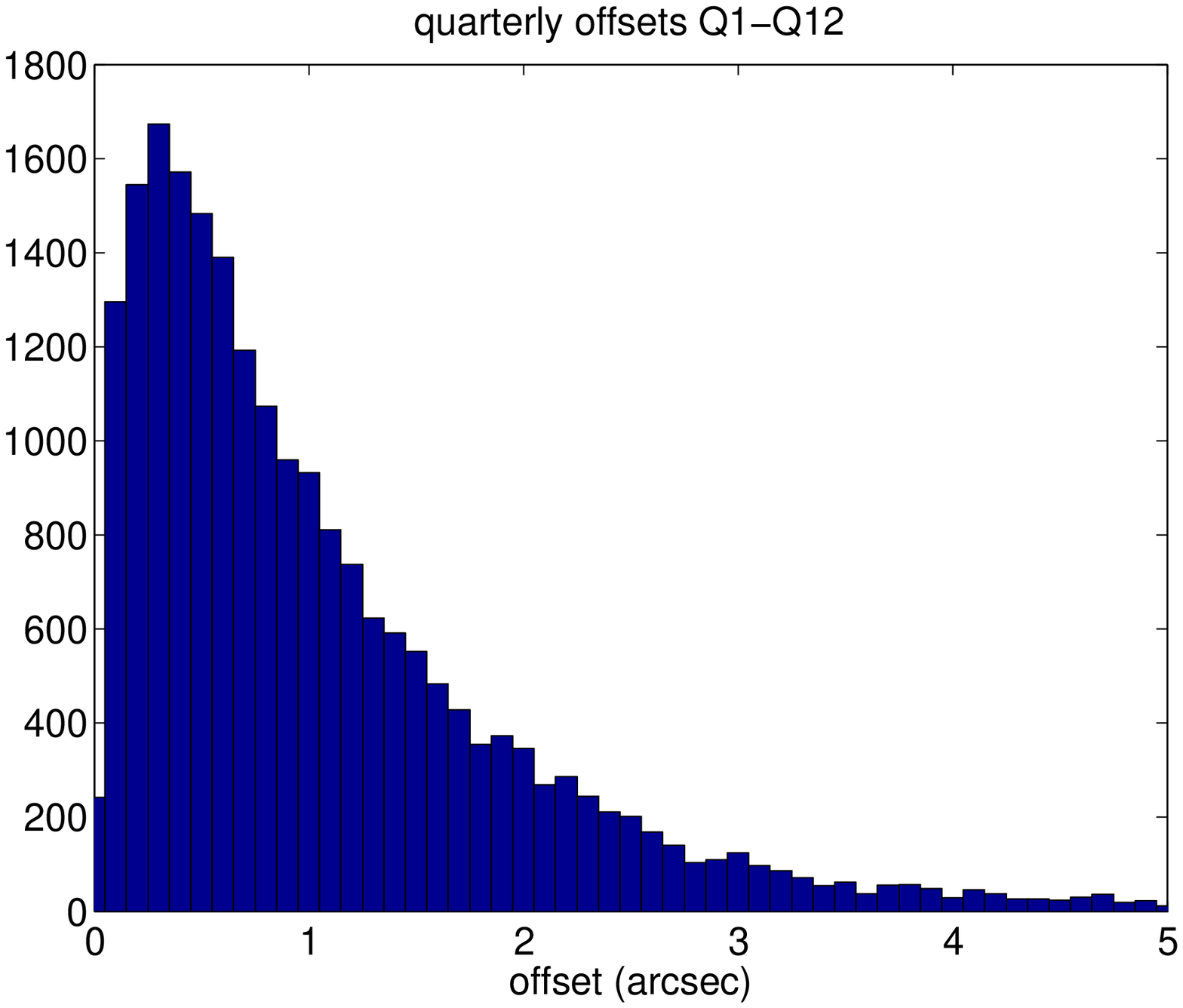} 
\includegraphics[scale=0.45]{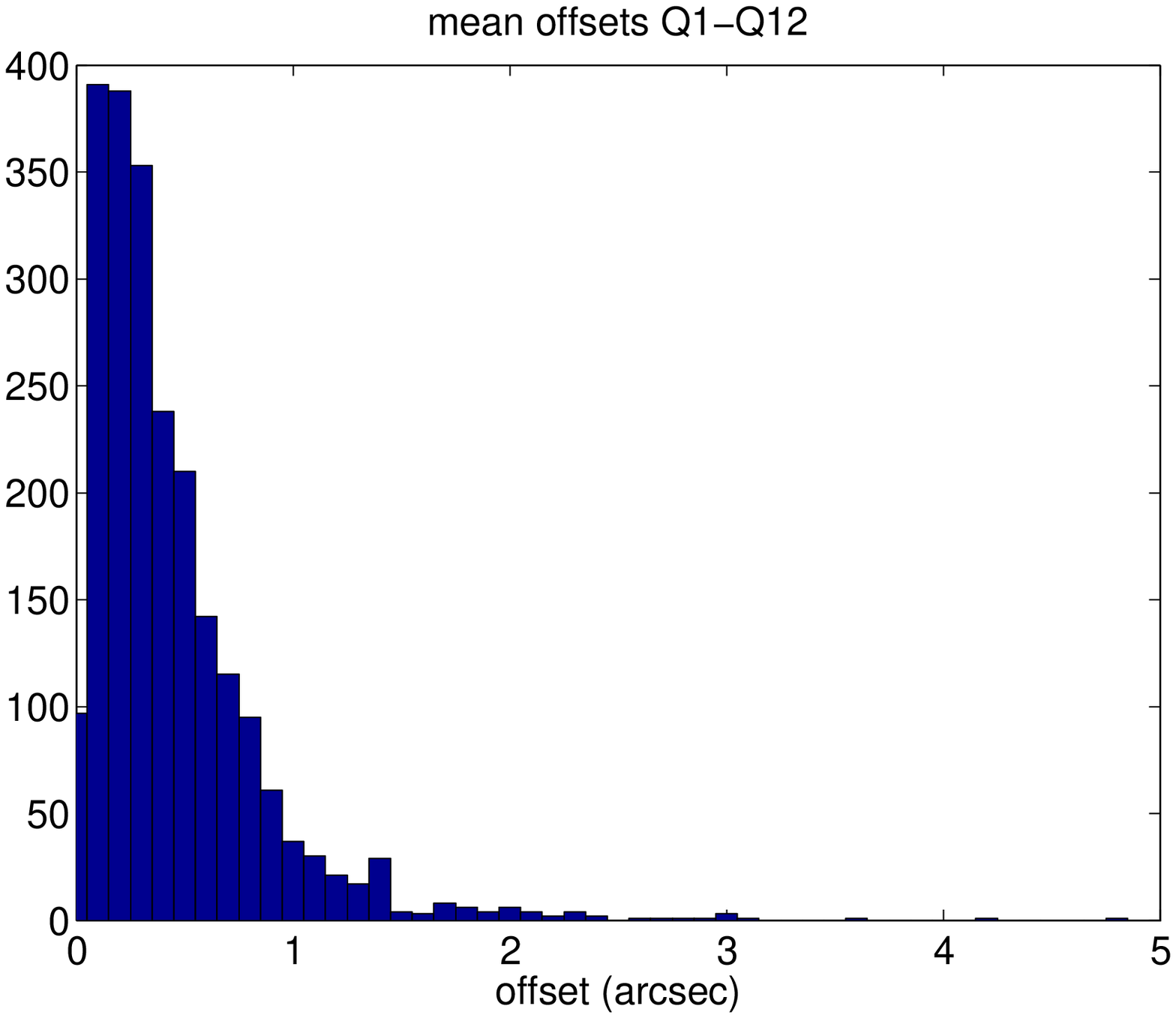} 
\caption{Distributions of the PRF-fit offset from the target catalog position for 2,278 KOIs whose quarterly 
averaged offsets are less than $3 \sigma$ and whose offsets from the target are $< 5$ arcseconds.  Left: the distribution of
individual quarter offsets.  Right: the distribution of the multi-quarter averages.}
\label{fig:mean_dist}
\end{center}
\end{figure}

We demonstrate the increased precision of the multi-quarter average in Figure~\ref{fig:mean_dist}.  The offset distance 
from the target catalog position is shown for 
both individual quarter PRF fits and their quarterly average.  This analysis uses 2,278 KOIs whose quarterly 
averaged offsets are less than $3 \sigma$ and whose offsets from the target are $< 5$ arcseconds in the Q1-Q12 data.  The left panel shows the 
21,401 individual quarter offsets, while the right panel shows the offset of the average over all quarters for each target.  The
individual quarter offsets have a standard deviation of 0.90 arcseconds, while the multi-quarter averages over 12 quarters have a standard deviation
of 0.41 arcseconds.  Strong year-to-year correlations prevent the standard deviation from scaling as $1/\sqrt{Q}$, but do not prevent 
an improvement as $Q$ increases.  

\begin{figure}[htbp]
\begin{center}
\includegraphics[scale=0.9]{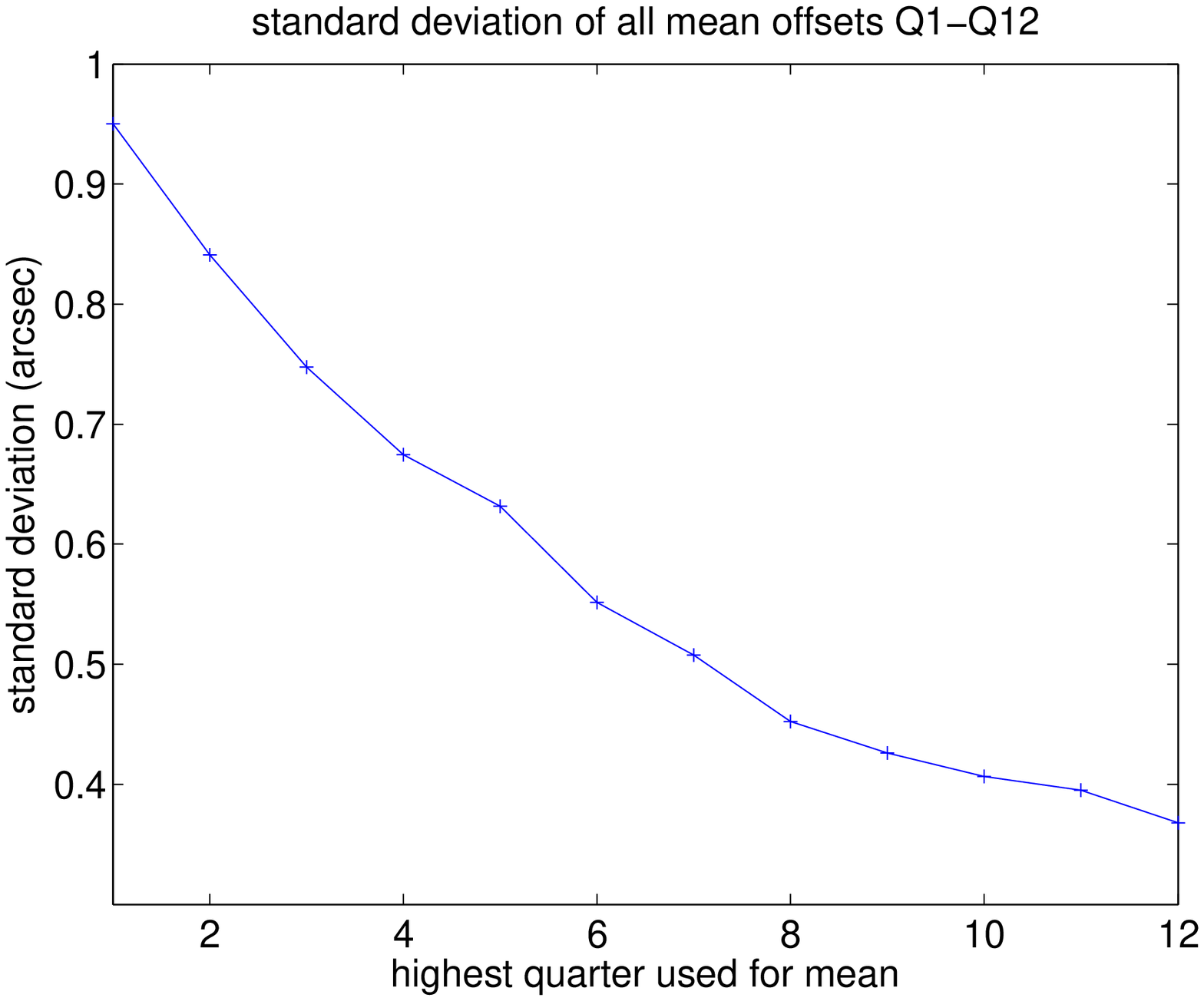} 
\caption{The standard deviation of the multi-quarter average as a function of the number of 
quarters used in the average.  The $x$-axis shows quarters used, where for each point 
the average is taken for the transits found in quarters 1 through the $x$-axis value. }
\label{fig:mean_vs_quarter}
\end{center}
\end{figure}

Figure~\ref{fig:mean_vs_quarter} shows how the standard deviation depends on the number of quarters averaged.  
We see that adding a quarter always statistically increases the precision of the multi-quarter average, though this may not
be the case for every individual target.


\subsubsection{Joint Multi-Quarter PRF Fit} \label{section:joint_multiq_fit}

When the transit SNR is very low, there may not be enough signal in each quarterly difference image to support per-quarter PRF fitting.
In this case we perform a joint multi-quarter fit, where the pixel images for all quarters are supplied to the PRF fitter, and the single
RA and Dec (and quarter-specific PRF amplitude) is found that minimizes the pixel-level difference between the pixel images 
and PRF-reconstructed pixels over all quarters.  In other words, the joint multi-quarter fit finds the single sky position 
$\left( \alpha, \delta \right)$ that minimizes the function
\begin{equation}
\chi^2 = \sum_{q=1}^Q \sum_{i=1}^P \frac{1}{\sigma_{p_{i,q}}^2}
	\left(p_{i,q}  -  b_q f_q \left( \alpha, \delta, r_{i,q}, c_{i,q} \right) \right)^2 \label{eqn:prf_join_fit}
\end{equation} 
where the subscript $q$ means the quarter-specific values of each quantity.  
So in each quarter the flux-normalized PRF $b_q f_q$ for that quarter is evaluated at $\left( \alpha, \delta \right)$ (which is common to
all quarters) for that quarter's pixels $\left( r_{i,q}, c_{i,q} \right)$.  
These PRF-based pixel values are subtracted from the observed pixel values $p_{i,q}$ for each quarter.  
The square of this difference normalized by the uncertainty is summed over all the pixels in that quarter, 
and finally summed over all quarters producing the test $\chi^2$ value.  The sky position is  
varied until the $\left( \alpha, \delta \right)$ that minimize $\chi^2$ is found.
The details of the computation in each quarter are similar to the single-quarter fit in \S\ref{section:prf_fitting}.  

The propagated uncertainty in this fit does not account for scatter across quarters due to systematic error,
so it dramatically underestimates the actual uncertainty in this fit.  We compute a more accurate uncertainty via a bootstrap approach
much like that for the multi-quarter averages described in \S\ref{section:multiq_averaging}, except the data consist of  
pixel images rather than offsets and each element of the ensemble is a joint PRF fit.
Specifically, the multi-quarter PRF fit takes as input the set of pixel images $\left( I_1, I_2, \ldots, I_Q \right)$ constructed 
in \S\ref{section:average_images}, where $I_q$ is the pixel image for each quarter.
The bootstrap approach creates an ensemble of resamplings-with-replacement sets of pixel images, for example $\left( I_4, I_5, I_3, I_2, I_2 \right)$ 
if $Q=5$.  The multi-quarter fit is performed on each element of the ensemble, computing a best fit $\left( \alpha, \delta \right)$ for each one.  
Each element of the ensemble is fit with the parameters from the quarter for that component.  For example if the
first element of the ensemble is $I_4$, then the PRF from quarter 4 is applied to those quarter 4 pixels.  
The uncertainty in the joint multi-quarter fit is then set to the standard deviation of the ensemble of fit positions.

The size of the resampled ensemble needs to be chosen with care.  The time to compute the joint multi-quarter fit scales 
with the number of quarters $Q$.  If the usual choice of $Q^2$ were chosen for the size of this ensemble, the full computation 
of the joint fit and its uncertainties would scale as $Q^3$.  In the {\it Kepler} pipeline, a bootstrap joint fit of 8 quarters took about 20 minutes, 
which indicates that a 16-quarter fit would take almost three hours.  It is prohibitive to run this on all 15,000 to 20,000 
threshold crossing events identified by the pipeline.  The joint PRF fit is therefore not routinely run in the {\it Kepler} pipeline, but is reserved for
low SNR transits for which the multi-quarter average does not provide a sufficiently precise result.  The possible use of a smaller 
resampled ensemble is under investigation.

\section{Pixel Correlation Images} \label{pixel correlation}

The \emph{pixel correlation method} computes the degree to which the transit signal over time appears in each pixel.
This information is used to create a pixel image, where the value of each pixel is the degree of correlation
between the pixel flux and the transit signal.  This image is centroided via PRF fitting similar to the difference image method.
This method has a different response to non-transit photometric variability from the photometric and difference image methods, 
so it can be useful for resolving cases when the other methods provide ambiguous results.

\begin{figure}[htbp]
\begin{center}
\includegraphics[trim = 3.1cm 0 0 0, clip, scale=0.62]{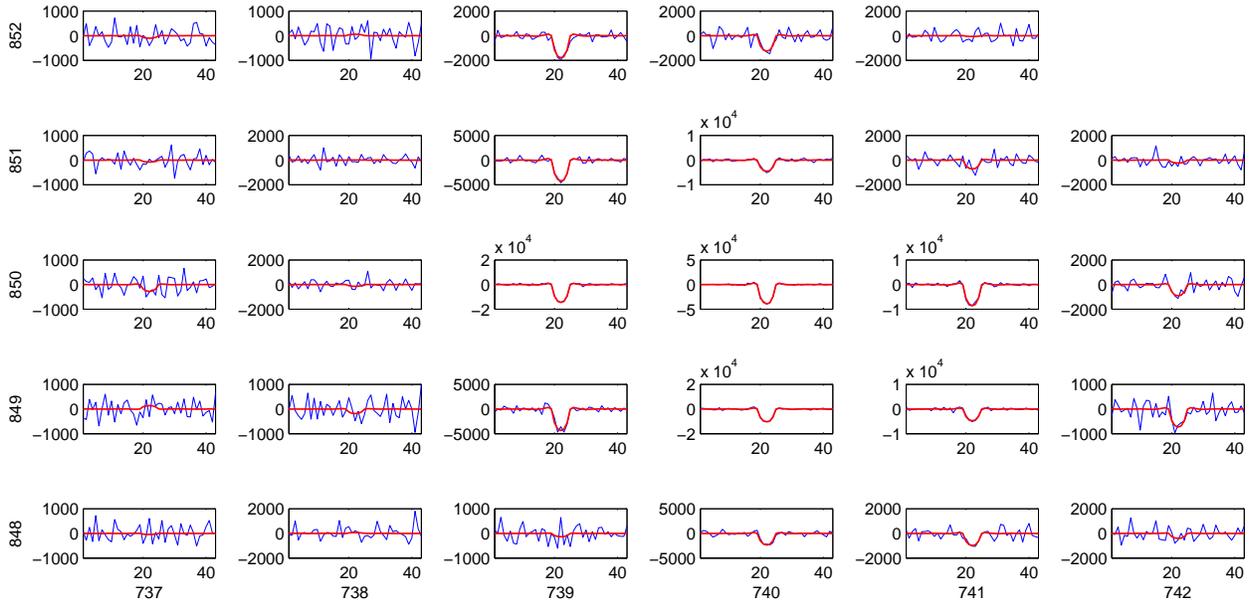} 
\caption{Fits of the transit model to individual pixel flux time series for 
KOI-221 in quarter 7.  The pixel flux time series is shown in blue and transit model 
is in red.  Each pixel flux time series is detrended and folded on the
transit period.  A closeup of the transit event is shown, with the same time interval 
on all $x$ axes.  The $y$-axes show the pixel values and are scaled to show the variation in each pixel time series.
The pixel rows are shown along the left, and pixel columns along the bottom.
The pixels that strongly contain the transit signal indicate the location of the transit source. }
\label{fig:koi_221_pixel_model_fits}
\end{center}
\end{figure}
\begin{figure}[htbp]
\begin{center}
\includegraphics[scale=0.45]{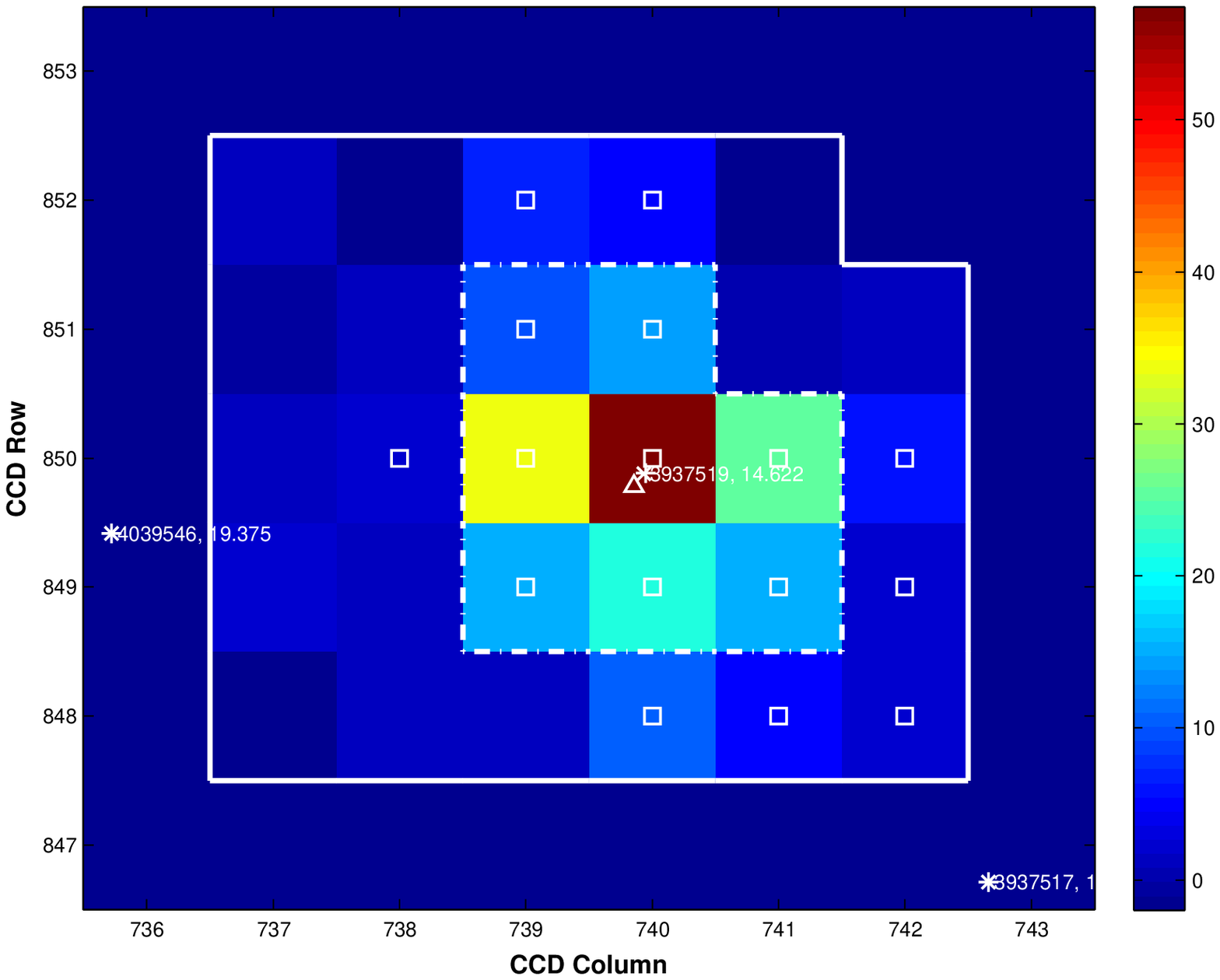} 
\includegraphics[scale=0.45]{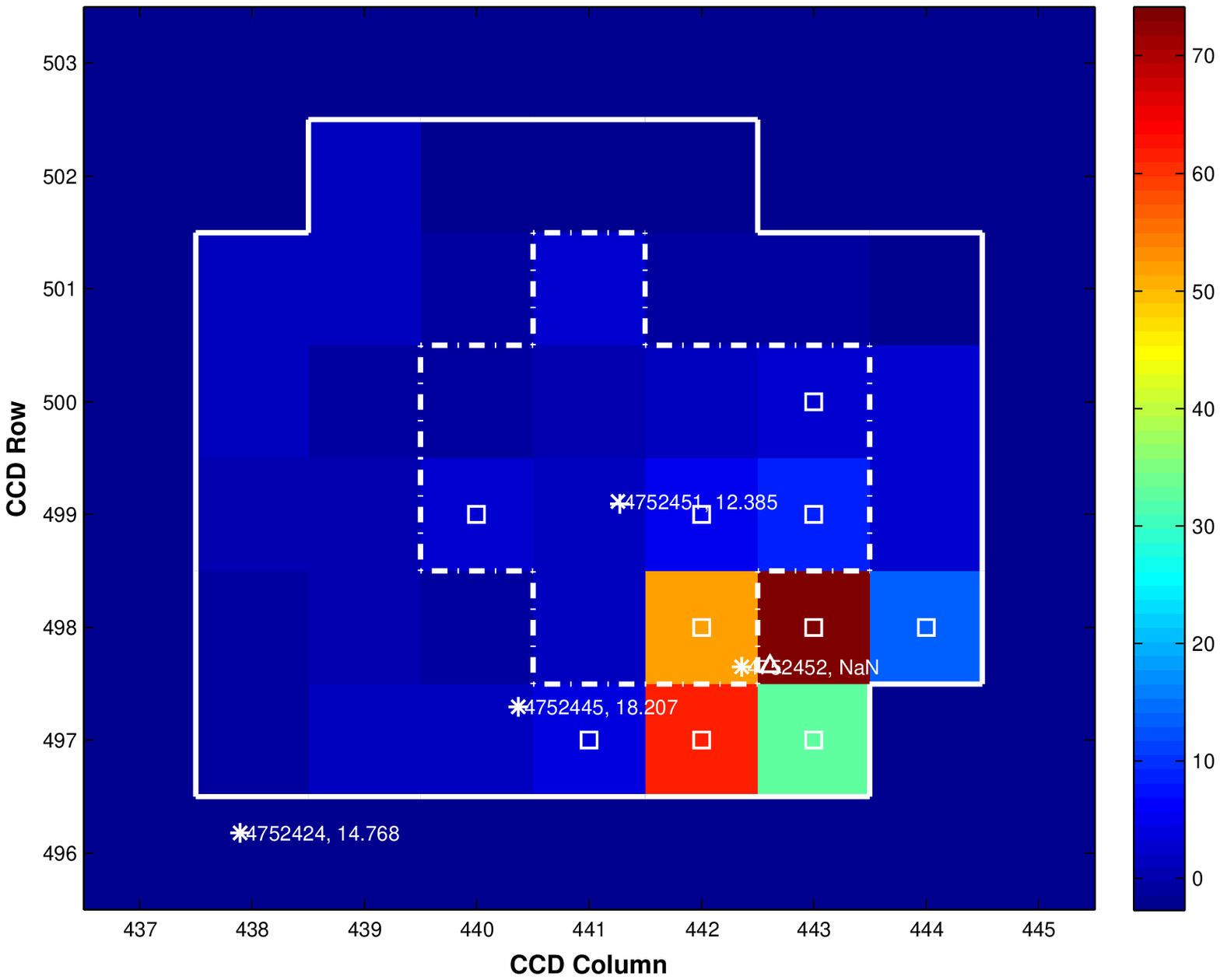} 
\caption{Correlation images, created by assigning each pixel the 
scale factor that multiplies the transit model to best fit that pixel's flux time series.
Left: the example from Figure~\ref{fig:koi_221_pixel_model_fits} of the transit
signal being coincident with the target star (KOI-221).  Right: an example with the transit
signal significantly offset from the target star (KOI-109).   In these figures the small white squares
indicate pixels for which the fit scaling is above a threshold.}
\label{fig:koi_good_correlation_image}
\end{center}
\end{figure}

The correlation between the pixel-level flux and the transit signal over time is computed via a 
fit of the transit model to the individual pixel flux time series.  This uses the same fitting
method described in \S\ref{section:centroid_correlation}, with the centroid time series 
replaced by the pixel flux time series.  In this case the fit constant $\gamma$ is a measure
of the presence of the transit signal in each individual pixel.   An example of these fits
is shown in Figure~\ref{fig:koi_221_pixel_model_fits}.  A 
{\it pixel correlation image} can be constructed by setting the value of each 
pixel to its model fit value $\gamma$.  When this is done for the example in 
Figure~\ref{fig:koi_221_pixel_model_fits}, we get the pixel image in the left
panel of Figure~\ref{fig:koi_good_correlation_image}.  The right panel of 
Figure~\ref{fig:koi_good_correlation_image} shows an example where the
transit signal is offset from the target star.  For such high SNR targets, the transit 
signal is readily apparent in the pixels, and the correlation image has a star-like appearance.
In these cases the photometric or PRF centroiding can be applied to quantitatively and 
automatically compute the location of the transit, which can be compared to the catalog position
of the target star or the target star location from the PRF fit to the difference image.

\begin{figure}[htbp]
\begin{center}
\includegraphics[scale=0.45]{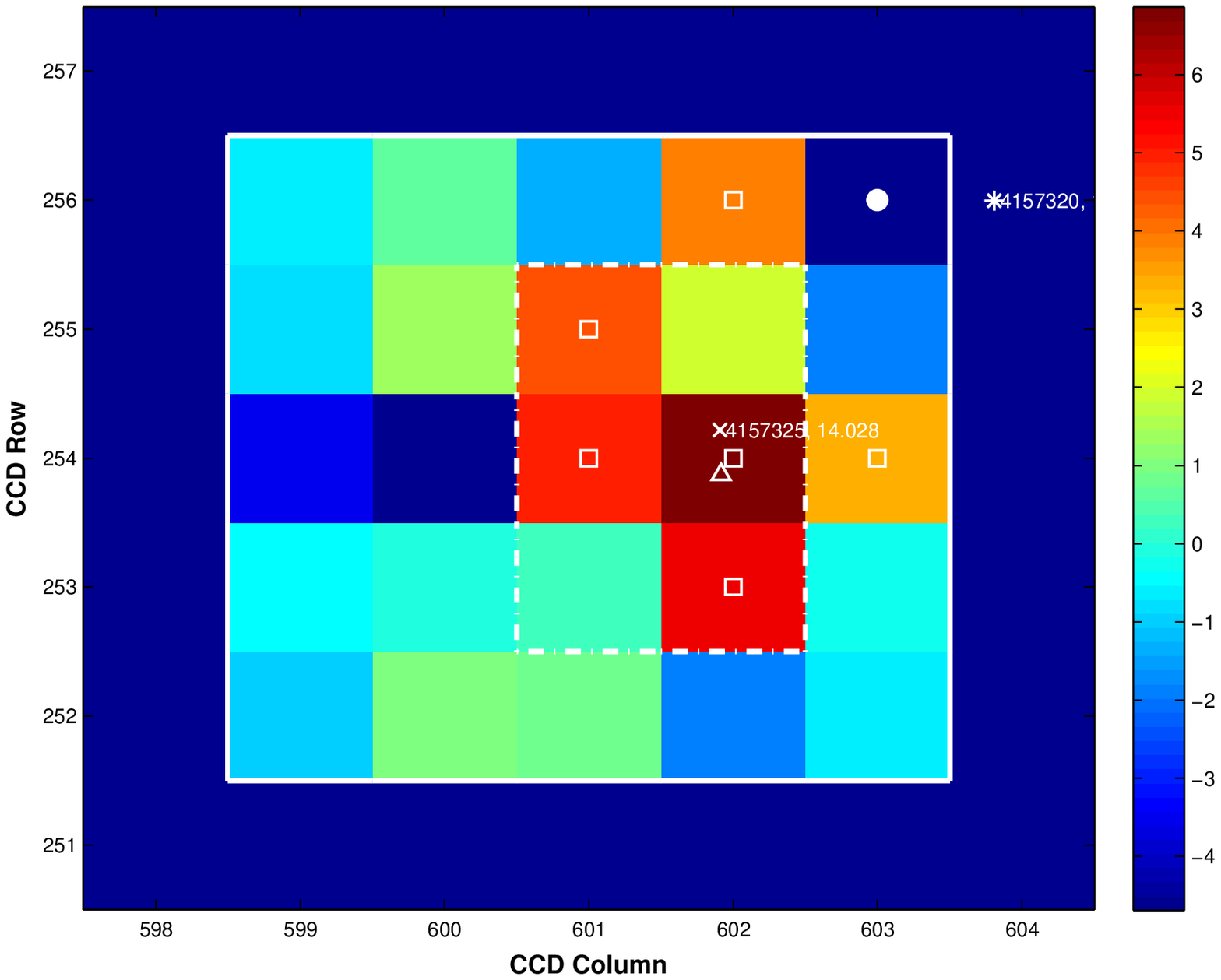} 
\includegraphics[scale=0.45]{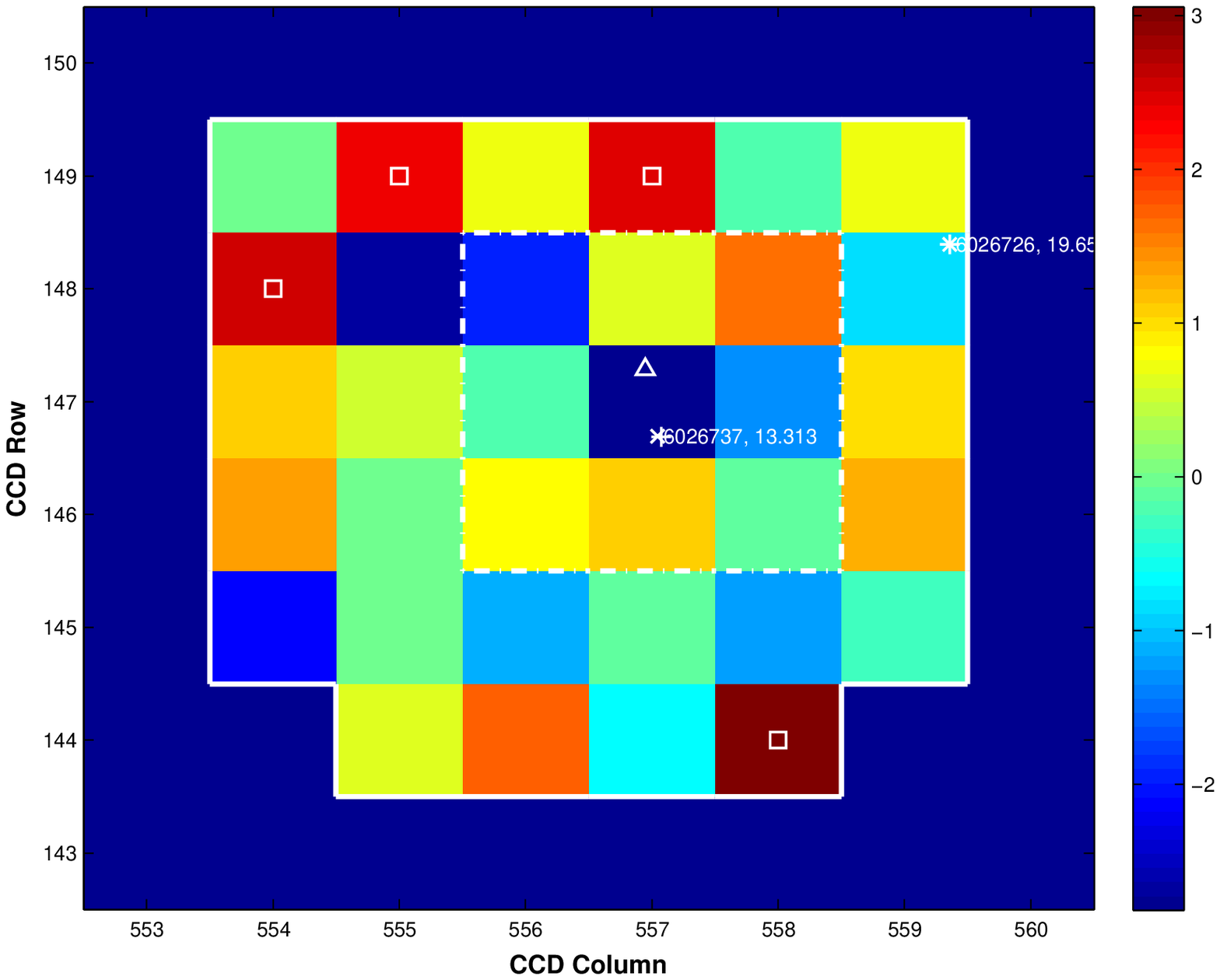} 
\caption{Correlation images for more problematic transits.  Left: an 
example where there is a field star in the aperture brighter than the
target star (KOI-1860).  Variability of the bright star pollutes the correlation image,
but the transit signal is still apparent.  Right: a low SNR example (KOI-2949) with
SNR = 11.  For such low SNR transits, the transit signal is barely discernable in the 
individual pixel time series, which causes the correlation image to be dominated
by background variability and pixel-level systematics.}
\label{fig:koi_bad_correlation_image}
\end{center}
\end{figure}

When the transit has low SNR or the pixels have significant flux from other sources, 
the pixel correlation image can be of much lower quality.  Two examples of this situation
are shown in Figure~\ref{fig:koi_bad_correlation_image}.

Because the correlation image is degraded by background flux and can have poor behavior 
at low SNR, it is not generally used for false positive identification.  There are circumstances, however,
where the correlation image can be used in combination with the other methods to make a
determination.  For example, some low SNR targets have marginal difference and correlation images, but if they 
show the transit signal in the same pixel location then we have increased 
confidence that the transit signal in those pixels is real.


\section{Saturated Targets} \label{saturated targets}


Target stars with {\it Kepler} magnitudes brighter than $\sim11.5$ can exhibit saturation, where the flux in a pixel exceeds that pixel's full well and spills up
and down the pixel columns \citep{instrum}.  The result is that the pixel image of the star can be highly distorted, invalidating
all of the centroid methods described in this paper.  Saturation can be highly asymmetric, so even photometric centroids 
are of limited use.  Visual inspection of the difference image can, however, reveal large, 
multi-pixel offsets indicating that the transit is not on the saturating star.  

\begin{figure}[htbp]
\begin{center}
\includegraphics[trim = 1.7cm 1cm 0 0, clip, scale=0.7]{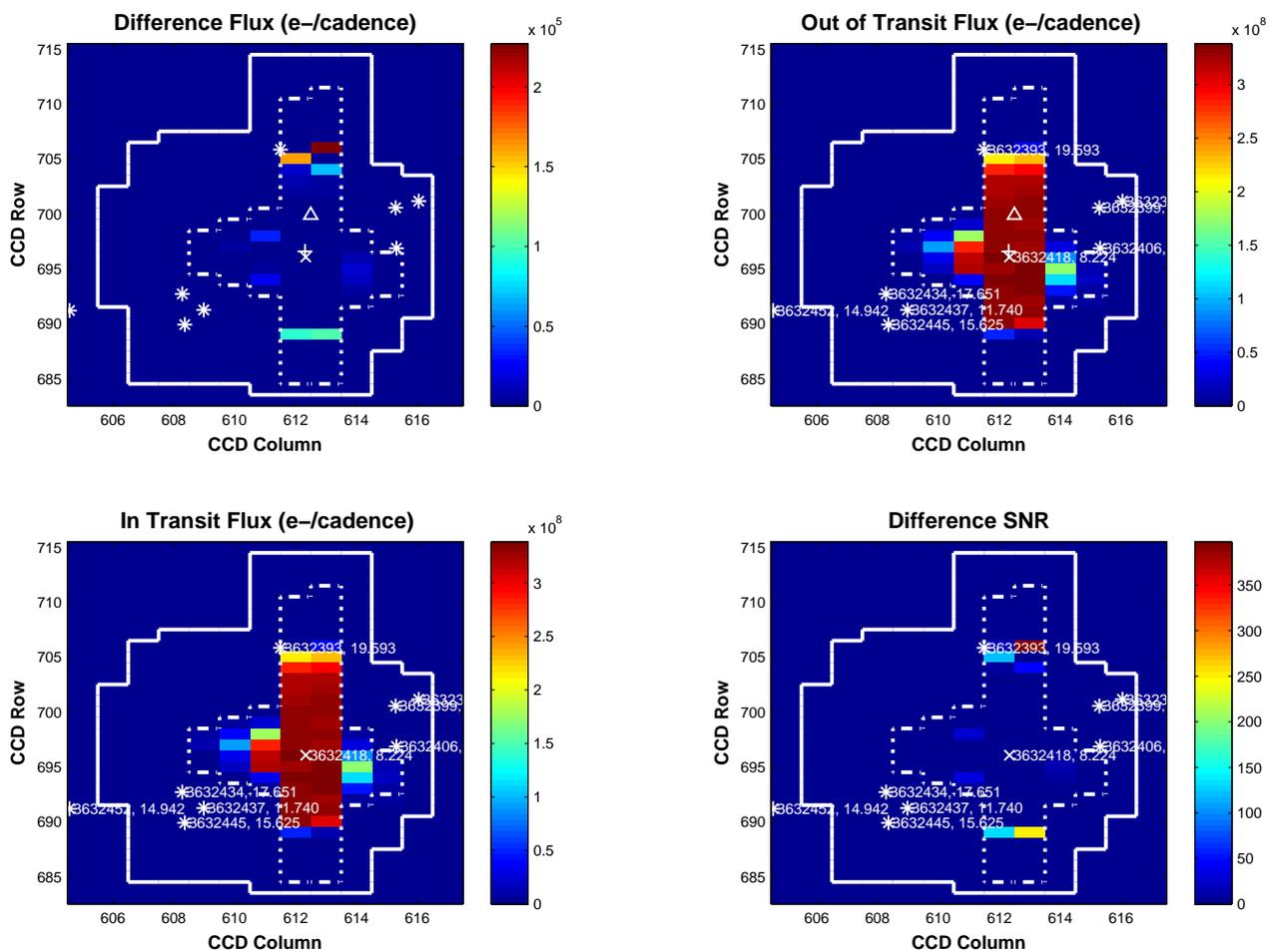} 
\caption{An example of a transit signal on a saturated star for the confirmed planet 
Kepler-21b \citep{howell12}.  The host star has  {\it Kepler} magnitude = 8.4 and is highly saturated.  In the difference
image the transit is apparent in the pixels at the end of the saturation in columns 612 and 613 (the star 
labels have been removed from the difference image for clarity).  The target star
is near the boundary between these two columns, which is why there is about equal saturation in both columns.
Note the strong asymmetry in the saturation for this quarter, with the saturation going up the columns
significantly further than down.}
\label{fig:koi_975_diff_image}
\end{center}
\end{figure}

When the saturated star is the transit source, the difference image will have a distinctive, non-star-like, pattern.
Because the saturation spills along columns and the amount of spill is approximately proportional to the flux of the star,
a transit signal on a saturated star will appear in the difference image as changes at the ends of the saturated columns.  
An example is shown in Figure~\ref{fig:koi_975_diff_image}.  This is a characteristic pattern in the difference images
for saturated targets.  All that can be said in this case is that the transiting source is in approximately the same 
column position as the target star, between the ends of the saturation.  If the transit were due to a field star that
is not in the saturated pixels, the difference image would show that star and not the signal from the saturated pixels.  

Special investigation of saturated targets can sometimes refine the location of the transit signal.  
The appearance of the transit at the end of the saturated columns is sensitive to the column position of the
transiting source.  If the 
transit SNR is high enough, the wings of the transits can be subject to a PRF fit while masking out the saturated columns.
These techniques have been applied with some success, identifying the location of the transit signal to within 4 arcseconds,
for Kepler-21b \citep{howell12}.  We refer the reader to that publication for details.  


\section{Performance and Comparison of Techniques} \label{results}


In this section we examine the performance of our transit-source location estimation via photometric and PRF-fit centroids.  
We focus on offset distances because that is the high-level metric used in initial false positive identification.
We examine three populations of targets: 
\begin{itemize}
\item all Kepler objects of interest (KOIs) dimmer than {\it Kepler} magnitude 11.5 
(to avoid saturated targets \citep{instrum}), which have well-defined transit-like signals
of sufficient quality to pass vetting and produce an ephemeris and valid PRF fits (4,049 KOIs).  Many of these KOIs are in multiple systems.
\item unsaturated KOIs that have been identified as being due to
transit sources that are unlikely to be on the target, called Active Pixel Offsets (APOs), that have valid PRF fits 
as of July 2012 (178 KOIs).
\item a small number of APO KOIs whose transit signals have been identified with stars in the {\it Kepler} input catalog (16 KOIs).  
\end{itemize}
In this section we focus on the following questions: 
\begin{itemize}
\item How well do the methods identify the location of these sources?
\item Is there evidence that the source locations correspond to a uniform distribution of background sources?
\item How do these methods compare with one another with respect to accuracy and precision?
\end{itemize}
We also address an issue that arises with high-transit-SNR targets, where offsets can be very small but the formal
uncertainty can be much smaller.  In this situation we encounter residual bias that is not accounted for in the uncertainty,
which causes offsets to incorrectly seem statistically significant.  

\subsection{Accuracy} \label{section:performance_accuracy}

\begin{figure}[htbp]
\begin{center}
\includegraphics[scale=0.44]{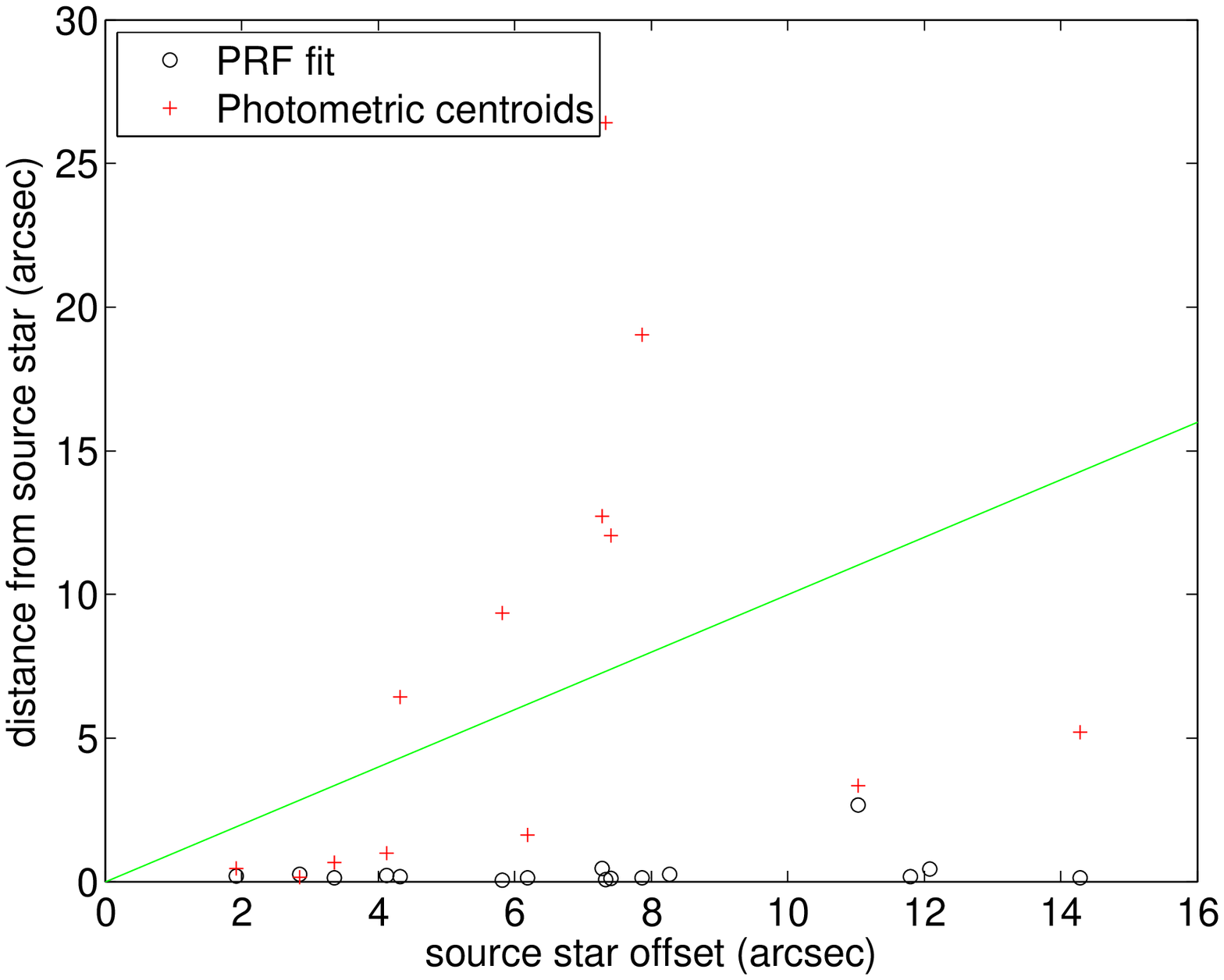} 
\includegraphics[scale=0.44]{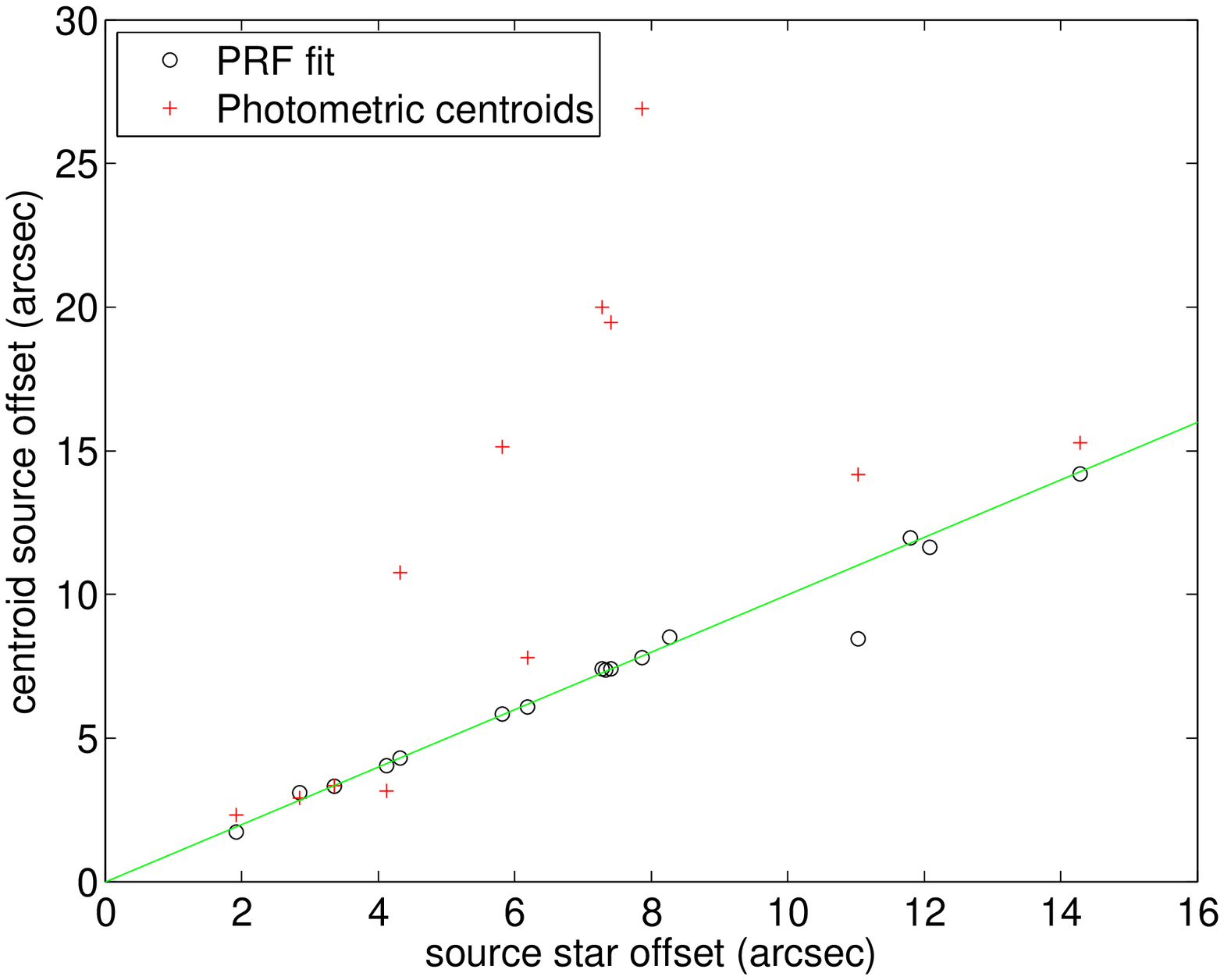} 
\caption{Left: The distance of the PRF-fit and photometric centroids from known stars that are likely to be the source of confirmed APO transit signals ($y$-axis) 
vs. the distance of the known star from the target star ($x$-axis).   Right: the same stars, showing the offset of the centroid from the target star 
($y$-axis).  
The PRF offsets are relative to the target star catalog location for consistency with the photometric offsets.}
\label{fig:known_star}
\end{center}
\end{figure}

We use APO targets whose transit signals have been associated with known stars to measure how accurately
our two primary methods of photometric and PRF-fit centroids identify the source location.  This association 
is determined by manual investigation of the difference images independently of the offset computations.  
We see in Figure~\ref{fig:known_star}
that the PRF estimate of the transit source offset is close to the star identified as the transit signal source.  
For APOs with small offsets ($< 4$ arcseconds) the photometric centroids also have good accuracy.  For APOs with larger
offsets, however, photometric centroids show large errors.  This behavior is expected because the {\it Kepler} pipeline uses 
one set of pixels to estimate the depth of the transit signal and a larger set of pixels to compute the photometric centroid.
As described in \S\ref{section:fw_error}, when the transit source has significant flux that falls outside the pixels used for
the depth estimate, which is the case when the source is more than 4 arcseconds from the target star,
there can be significant error in the transit source location inferred from the photometric centroids.

\begin{figure}[htbp]
\begin{center}
\includegraphics[scale=0.44]{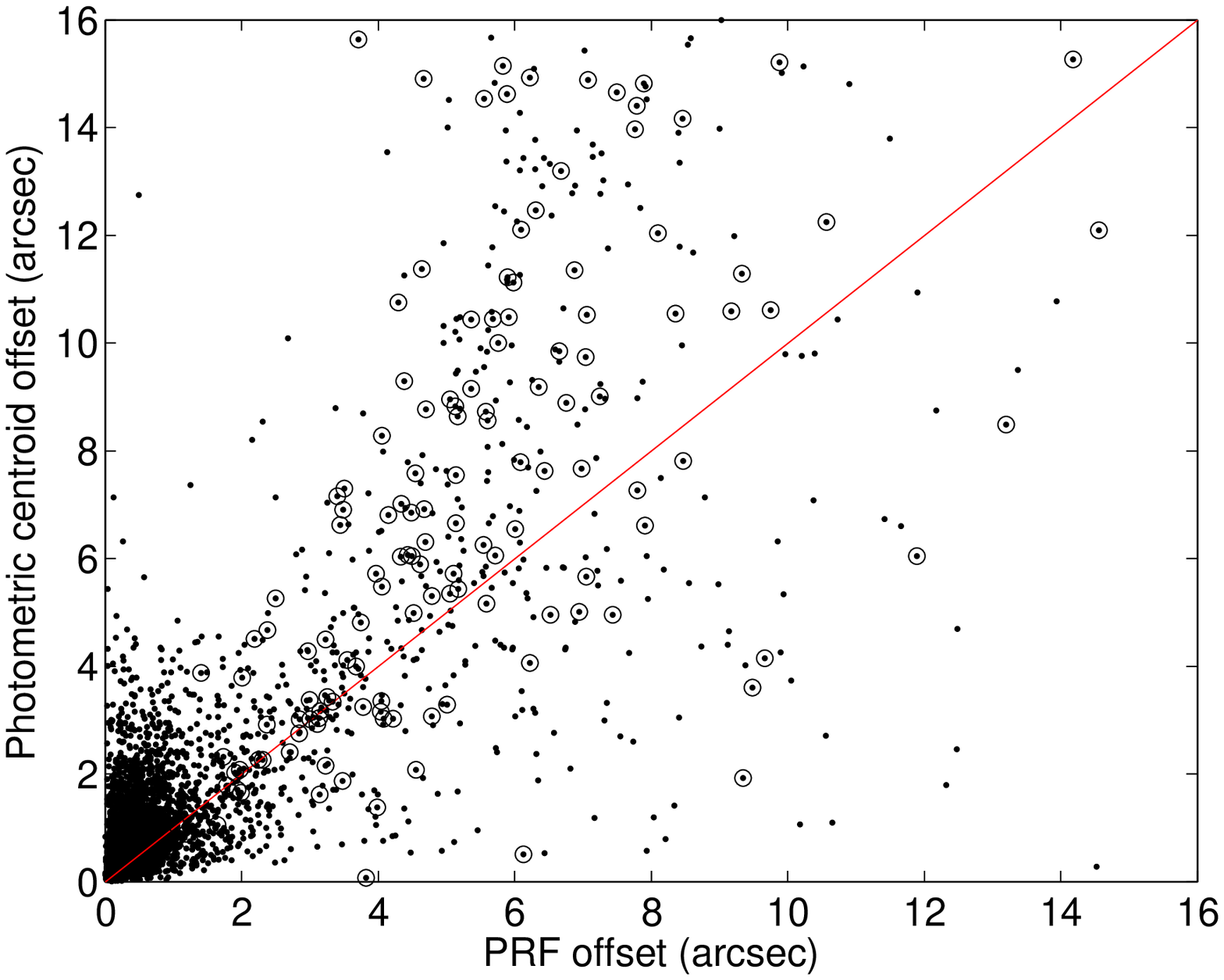} 
\includegraphics[scale=0.44]{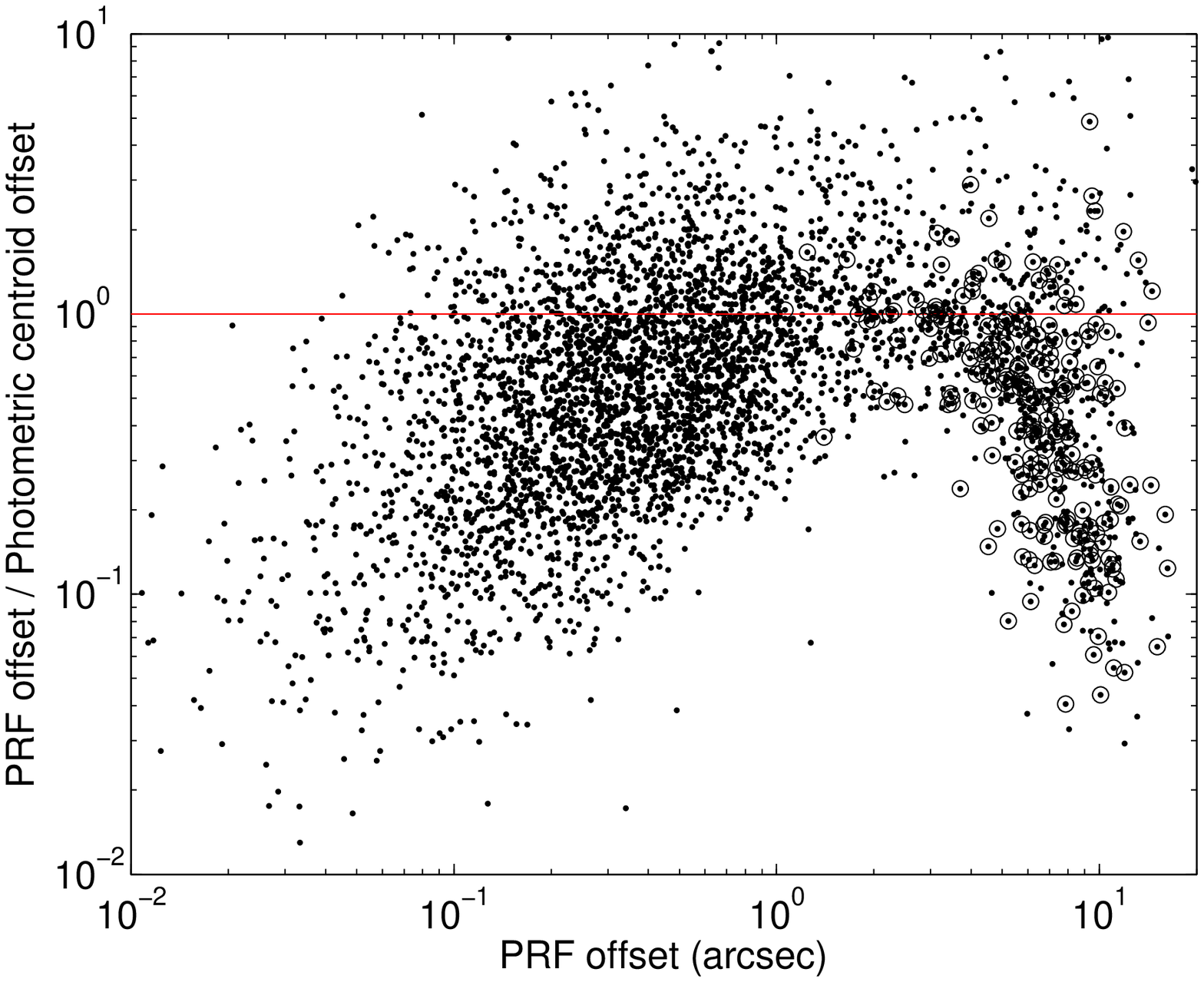} 
\caption{Left: A comparison between the PRF-fit offsets ($x$-axis) 
and the photometric centroid source offsets ($y$-axis) 
from the target star catalog position.  Right:
The ratio PRF-fit offsets/photometric centroid source offsets ($y$-axis) vs. 
magnitude of the PRF-fit offsets ($x$-axis).  APO KOIs are marked by circles.  
The red line in both figures indicates equality between the
PRF-fit and photometric offsets.
We see that the photometric centroid estimate of the source distance agrees with the PRF estimate
for distances of a few arcsec from the target star.  As expected, the photometric centroid usually 
overestimates the offset for transit 
sources that are further from the target star (see \S\ref{section:fw_error}). }
\label{fig:prf_vs_fw}
\end{center}
\end{figure}

Figure~\ref{fig:prf_vs_fw} compares the PRF-fit and photometric centroid source offset estimates for all KOIs, and
shows that the photometric centroid estimate of the source offset is generally (but not always) larger than the PRF-fit estimate 
when the PRF-fit source location source is more than a few arcsec from the target.

\begin{figure}[htbp]
\begin{center}
\includegraphics[scale=0.9]{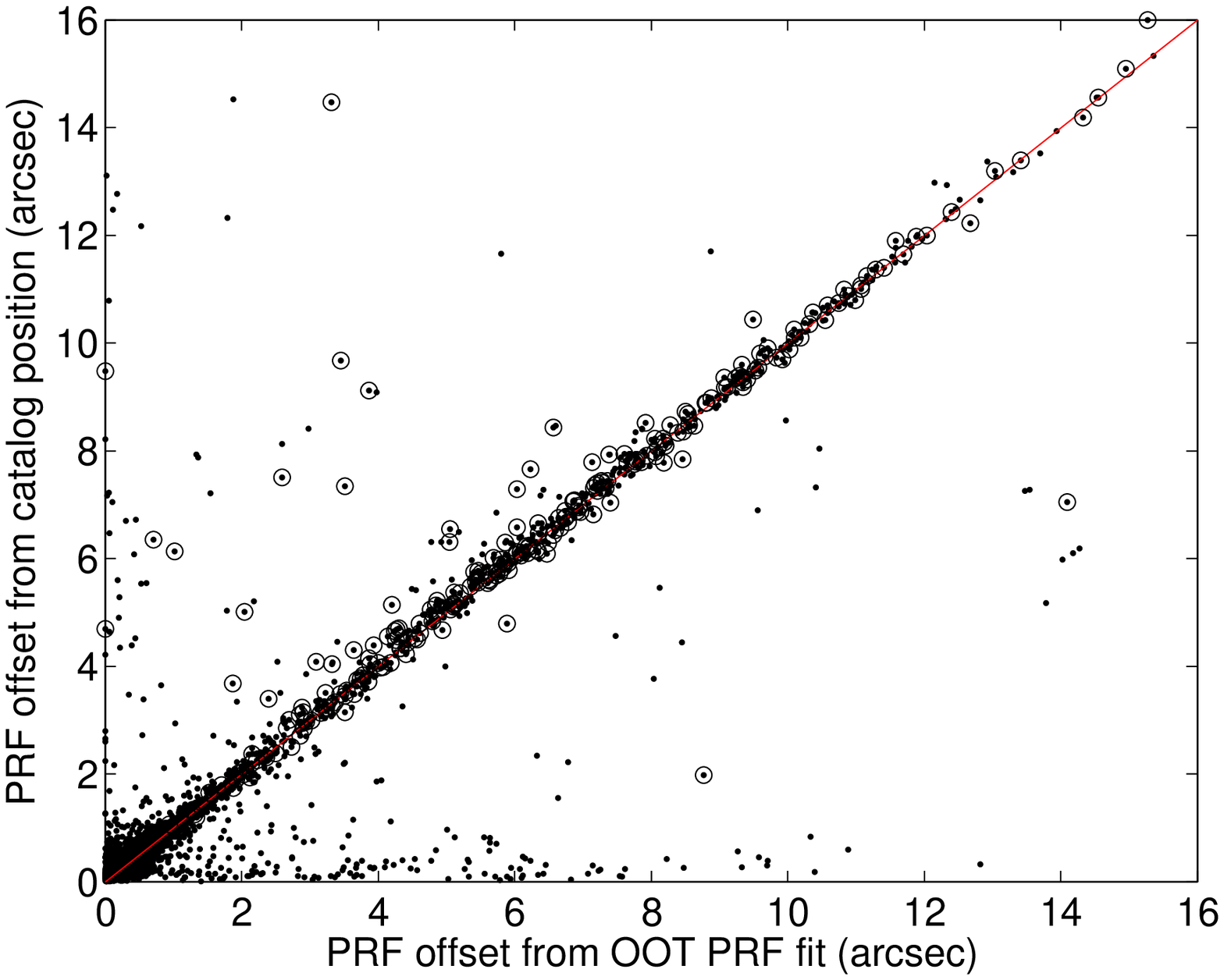} 
\caption{A comparison of the PRF-fit source offset relative to the PRF fit to the
out-of-transit pixel image ($x$-axis) and the PRF-fit source offset relative to the catalog position 
of the target star.  APO KOIs are marked by circles.  We see that most targets
with large offsets cluster along the diagonal indicating that the two offsets are generally in reasonable
agreement.  Outliers are likely due to crowding issues. }
\label{fig:prf_prf_vs_kic}
\end{center}
\end{figure}

Figure~\ref{fig:prf_prf_vs_kic}
compares the PRF-fit source offset relative to the target star catalog position with the PRF-fit source offset relative to the 
out-of-transit PRF-fit centroid.  These two offsets are similar for the majority of stars, with outliers that are likely due 
to bias due to crowding.  


\begin{figure}[htbp]
\begin{center}
\includegraphics[scale=0.45]{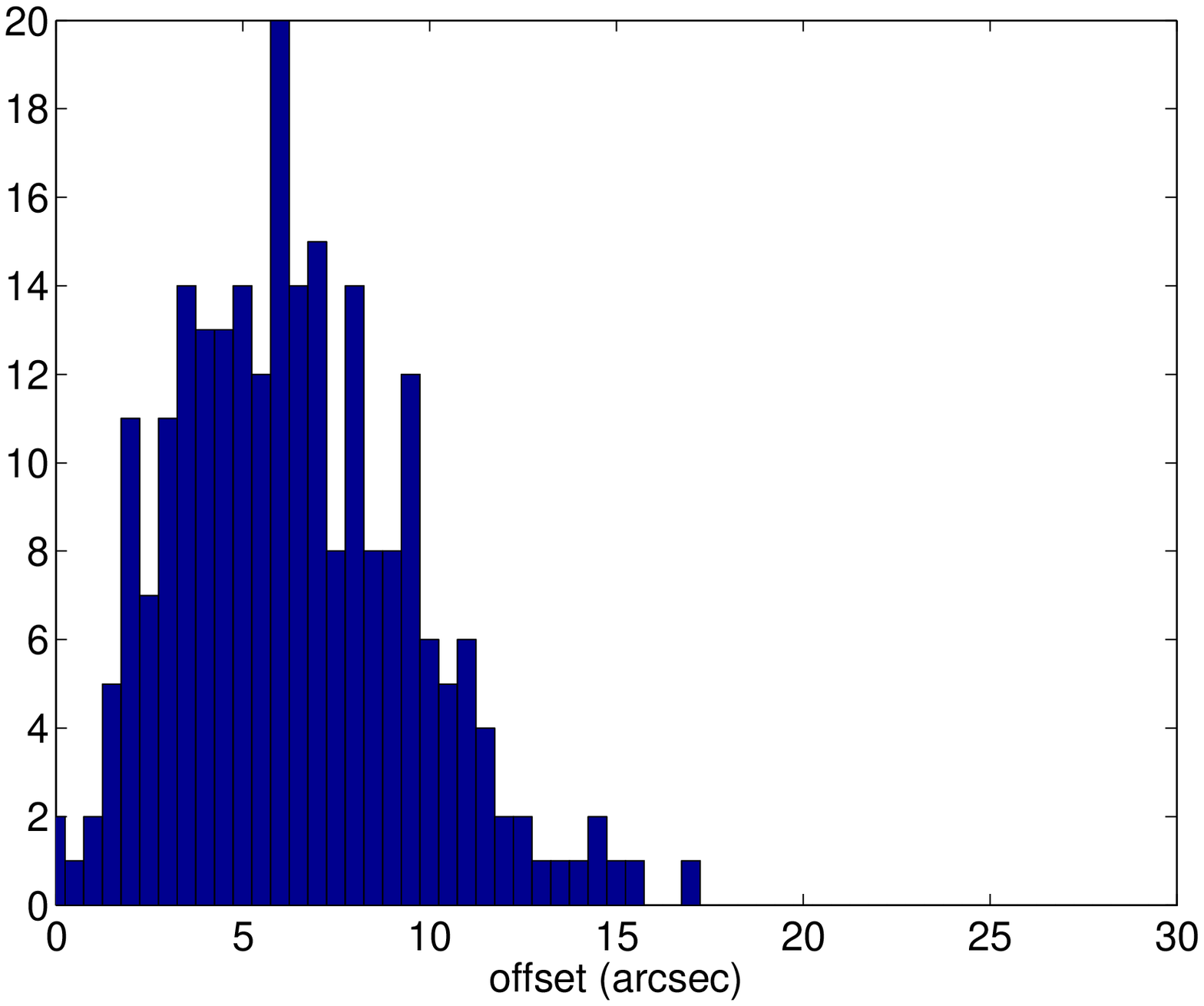} 
\includegraphics[scale=1.2]{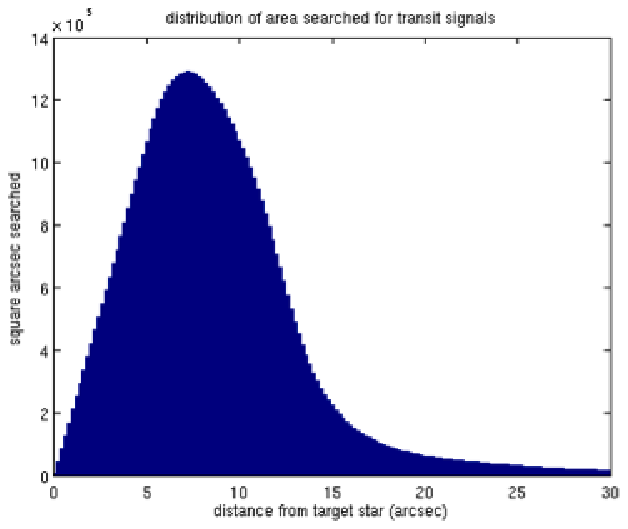} 
\caption{Left: the distribution of PRF-fit source offsets for targets identified as APOs.  
There is a strong peak at about 6-7 arcseconds.
This distribution is strongly dependent on the pixel aperture associated with each target star, which limits
the offset that can be detected.  Right: the distribution of pixel area as a function of distance from the target
star associated with each pixel, across the {\it Kepler} field of view.  This distribution also a peak at about
7 arcseconds.  The similarity between these two distributions is consistent with the identified APOs representing
a uniform distribution of background sources such as eclipsing binaries and large transiting planets.}
\label{fig:prf_prf_distribution}
\end{center}
\end{figure}

Figure~\ref{fig:prf_prf_distribution} compares the distribution of the APO KOIs and the distribution of observed pixel
area relative to target stars.  The fact that these two distributions have similar shapes with similar peaks is consistent
with the identified APOs representing a uniform background of eclipsing binaries and possibly large planetary transits.  
This consistency contributes to our confidence that the APOs are correctly identifying astrophysical false positives.


\subsection{Precision vs. SNR}

\begin{figure}[htbp]
\begin{center}
\includegraphics[scale=0.44]{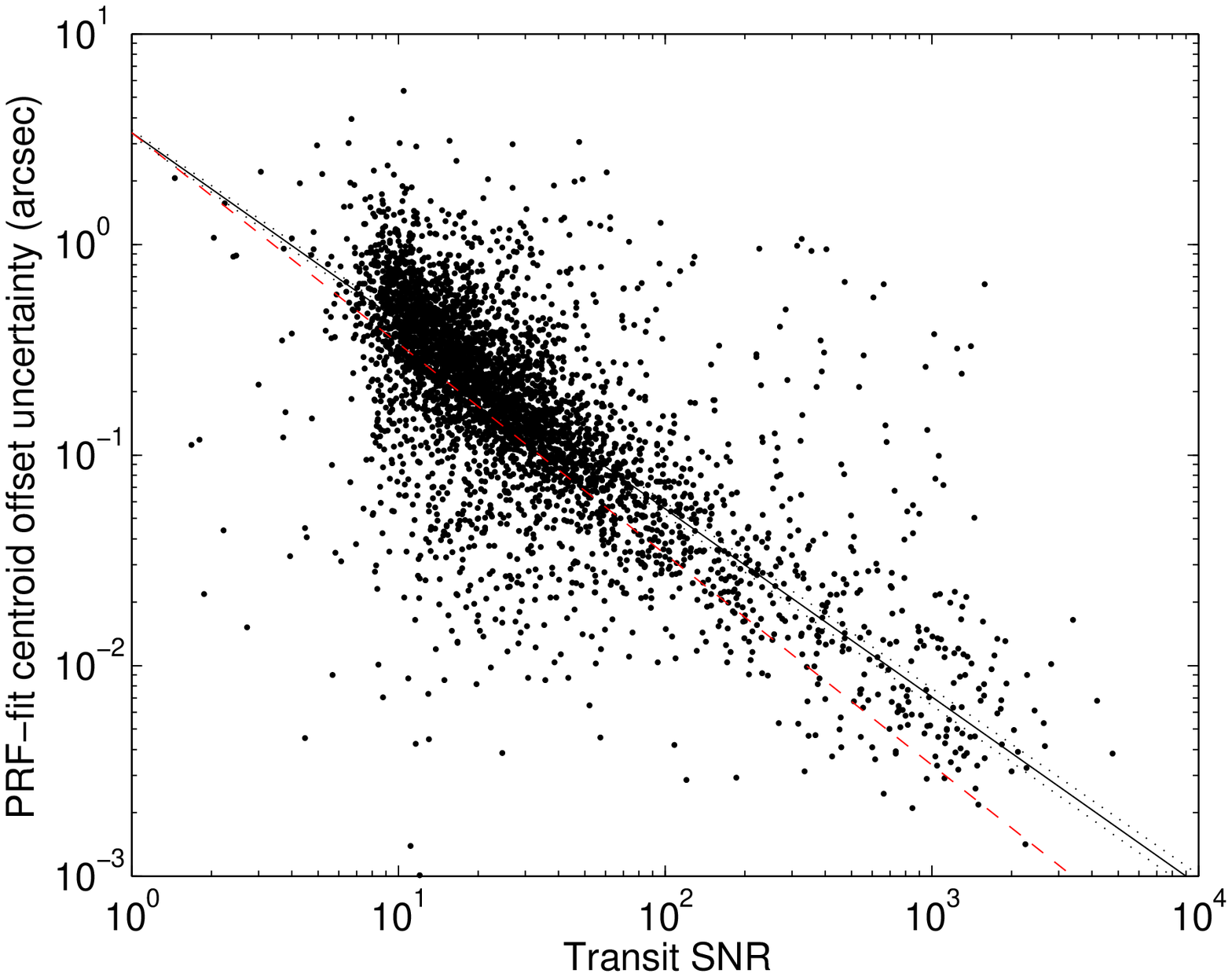} 
\includegraphics[scale=0.44]{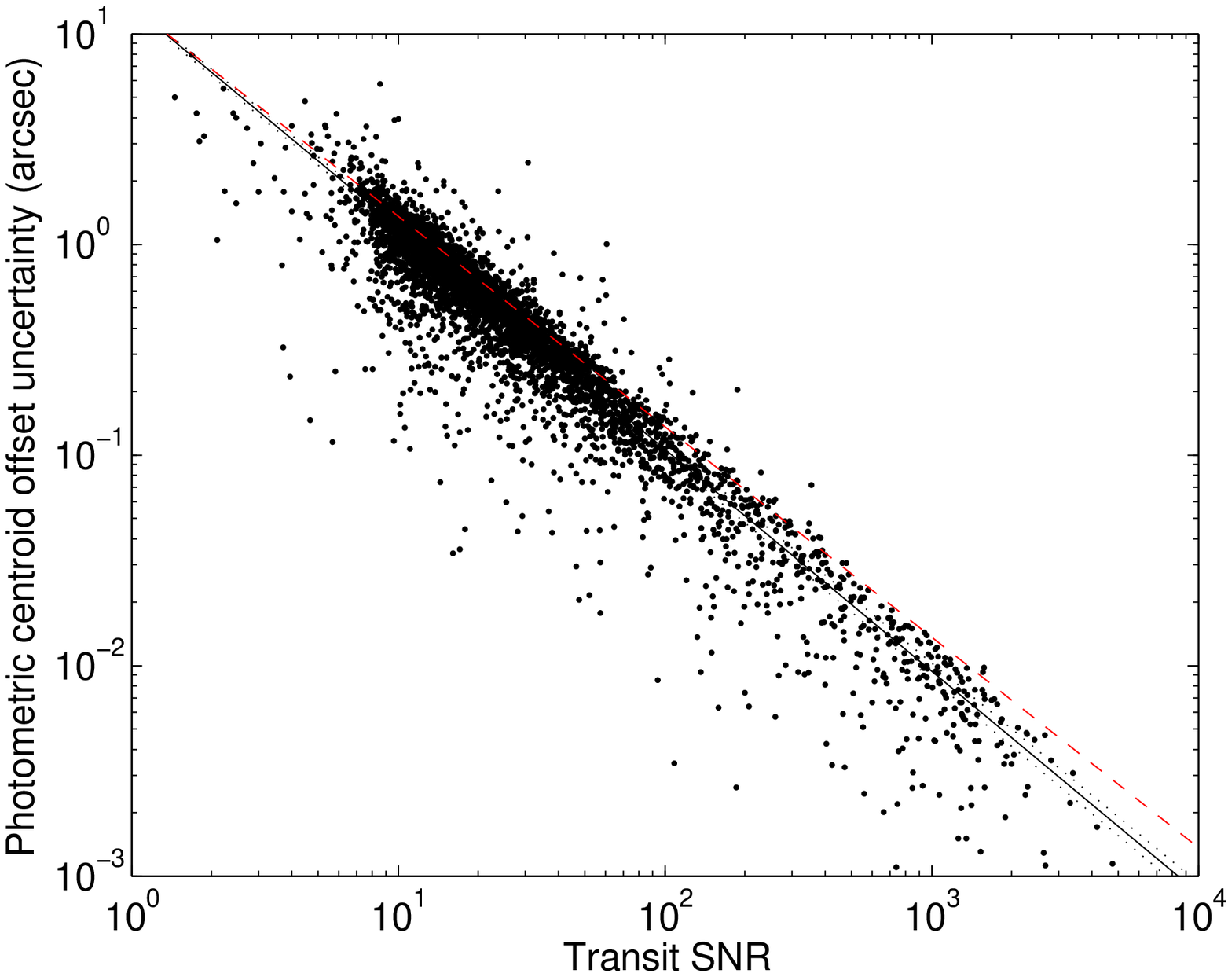} 
\caption{Formal offset uncertainty vs. transit SNR for PRF fit (left) and
photometric (right) centroids using 12 quarters of data.  The red dashed line
in both figures shows the 1/SNR dependency for comparison.  We see that the precision of the PRF-fit offsets
is somewhat better on average than the photometric centroid offsets.  This precision does not 
account for bias due to systematic error for either type of centroid.}
\label{fig:offset_vs_snr}
\end{center}
\end{figure}

\begin{figure}[htbp]
\begin{center}
\includegraphics[scale=0.9]{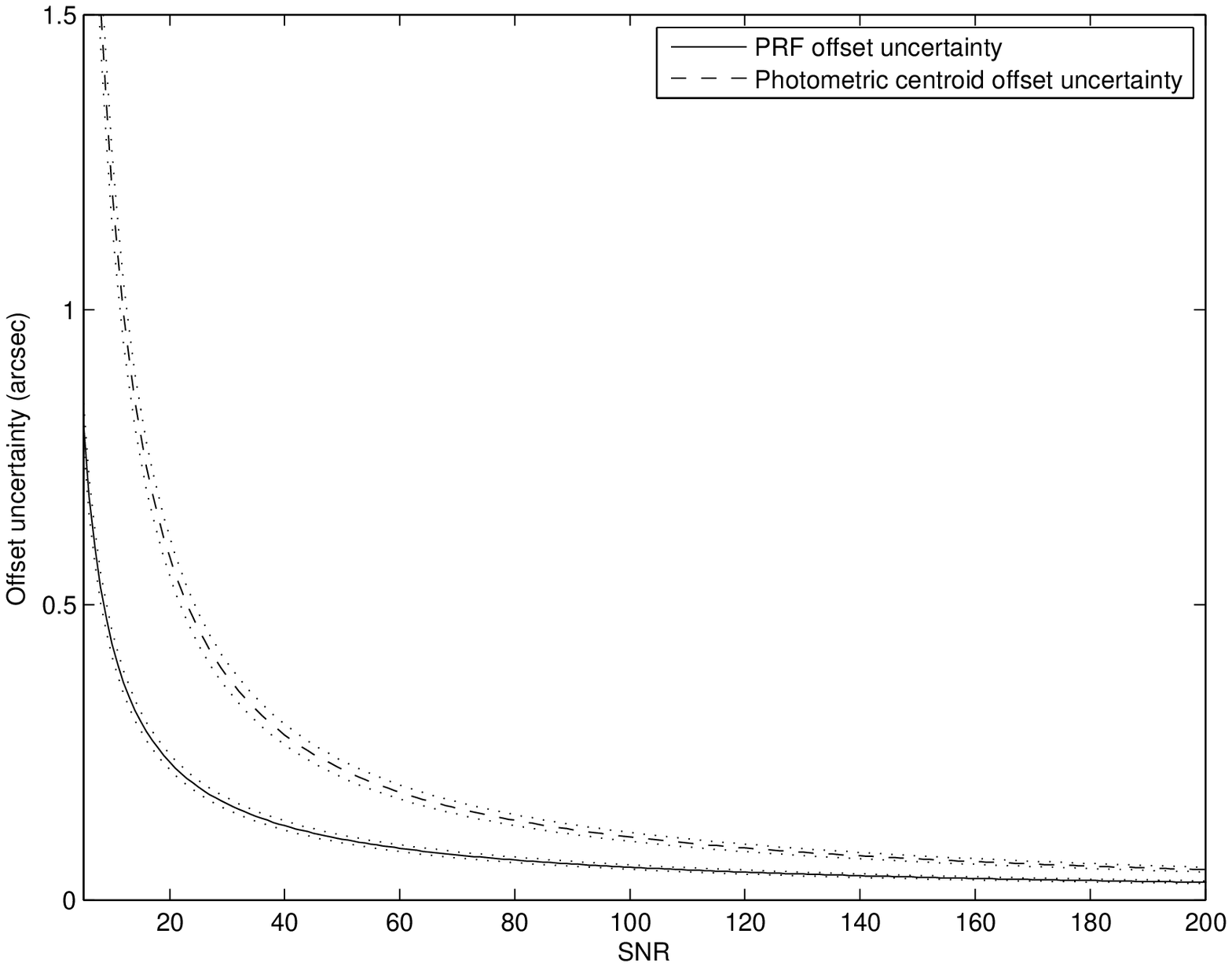} 
\caption{Uncertainty vs. SNR from the fits in Figure \ref{fig:offset_vs_snr} plotted on linear scales.
The dotted lines indicate the range of variation due to the 1-$\sigma$ uncertainties in the fit parameters.}
\label{fig:offset_vs_snr_fit}
\end{center}
\end{figure}

The precision of a centroid measurement is dependent on the strength of the transit signal in each pixel.  This strength
depends on the transit depth, host star brightness and number of transits among other factors.  All of these factors
contribute to the transit SNR, so we analyze precision as a function of transit SNR.
Figure~\ref{fig:offset_vs_snr} shows the dependence of formal centroid source offset uncertainty on transit SNR.  Both the 
PRF-fit and photometric centroid methods show similar dependencies, though the
uncertainties for the PRF-fit centroid method are somewhat smaller.  A linear fit to the log-log data gives the uncertainty of
the two methods as
\begin{equation}
\sigma_{\mathrm{photometric}} = \frac{13.6 \pm 0.16}{\mathrm{(SNR)}^{1.05 \pm 0.00}}, \qquad
\sigma_{\mathrm{PRF-fit}} = \frac{3.39 \pm 0.10}{\mathrm{(SNR)}^{0.89 \pm 0.01}}.
\end{equation}
These fits, along with the range of values implied by the 1-$\sigma$ uncertainties in the fit parameters, are shown in Figure~\ref{fig:offset_vs_snr_fit}.
The uncertainty of the photometric centroid method is inversely proportional to the SNR, as expected, while the PRF-fit method has 
a somewhat smaller dependence on inverse SNR.  The coefficient of these uncertainties (13.6 for photometric uncertainties and 
3.39 for the PRF fit) is larger than the full-width-half-max expected for centroid uncertainties because these uncertainties include 
contributions from the offset computation.
The uncertainties reported in this section are propagated formal
uncertainties, however, which are only valid if all noise sources are zero-mean Gaussian white noise.  As described in
this paper there are several sources of systematic error that impact transit source offset estimation.  These systematic
errors are not reflected in the formal uncertainty.

\begin{figure}[htbp]
\begin{center}
\includegraphics[scale=0.9]{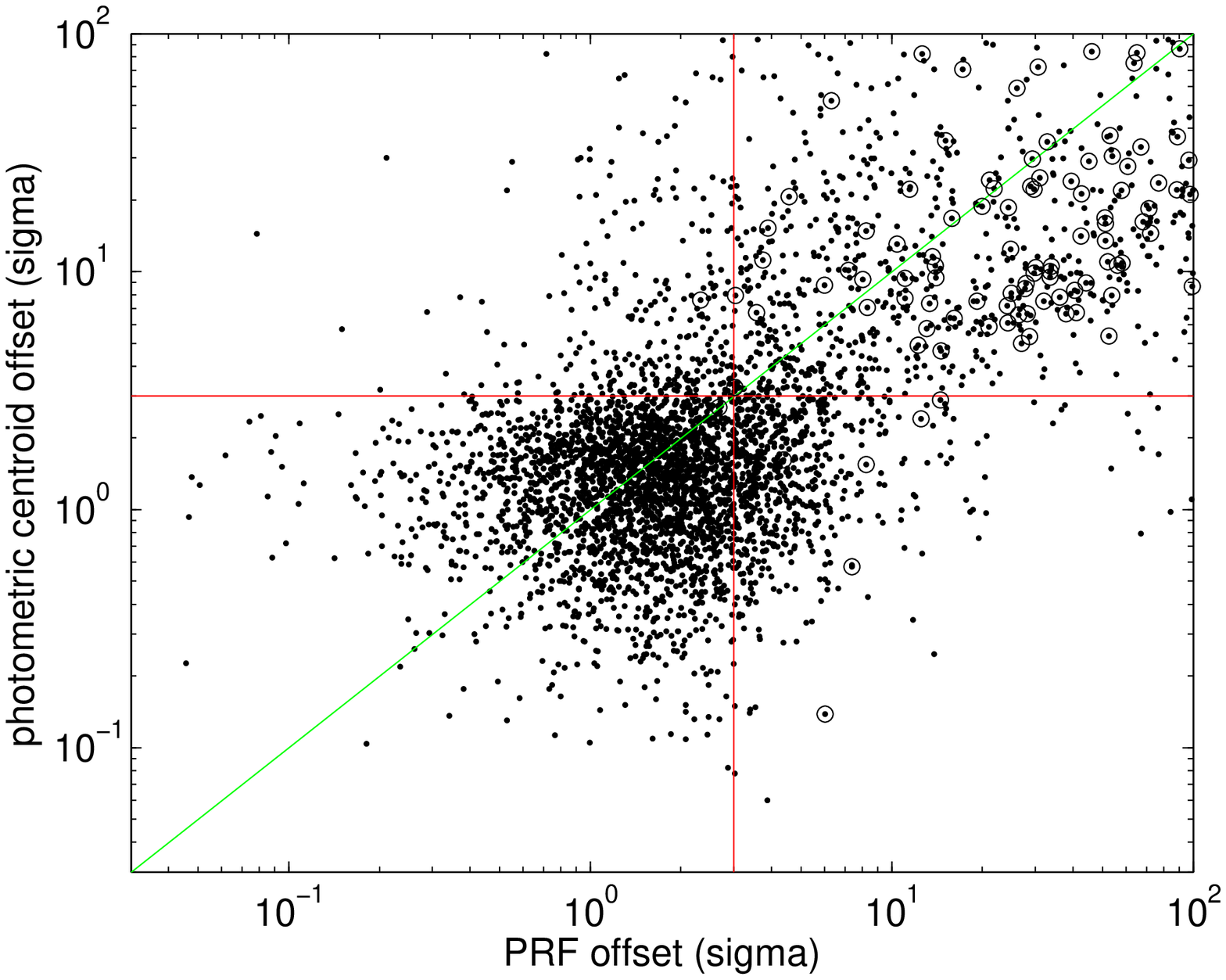} 
\caption{A comparison of the PRF-fit source offset relative to the catalog position of the target star
($x$-axis) and the photometric centriod source offset ($y$-axis), both in units of $\sigma$.
The vertical and horizontal lines mark where the offset $ = 3 \sigma$, above which the offset
is considered statistically significant. 
APO KOIs are marked by circles.  We see that most targets
have both offsets below $3 \sigma$, but there are a significant number of targets for which the photometric 
centroid source offset is less than $3 \sigma$ but the PRF-fit offset is $> 3 \sigma$ and vice versa.}
\label{fig:PRF_vs_fw_offset_sigma}
\end{center}
\end{figure}

Because the dependence of the PRF-fit and photometric centroid estimates of the source offset on SNR have similar 
log slopes we expect that if one technique indicates a significant offset then the other technique will as well.  This
is shown in Figure~\ref{fig:PRF_vs_fw_offset_sigma}, which indicates that for most targets the photometric
centroid and PRF-fit methods are in agreement as to whether there is a significant offset for a particular target.  But there are many
targets, including a few identified APOs, that have photometric centroid source offsets $< 3\sigma$ but PRF-fit source
offsets $> 3\sigma$ and vice versa.

Quantitatively, for 54.9\% of all KOIs the two techniques are in agreement that the source offset is $< 3\sigma$; 
24.7\% of all KOIs have agreement that the source offset is $>3\sigma$; 13.9\% of all KOIs have offsets $>3\sigma$
according to the PRF-fit technique but $<3\sigma$ according to photometric centroids; and 
6.45\% of all KOIs have offsets $< 3\sigma$ according to the PRF-fit technique but $>3\sigma$ according to photometric
centroids.  Therefore the two methods are in agreement on significance for about 80\% of the targets.  Most of the
targets for which the PRF-fit techniques indicate an offset $> 3\sigma$ but the photometric centroids 
have a shift $< 3\sigma$ have very small PRF-fit offsets, so they are at distances where residual bias
dominates as discussed in \S\ref{section:high_snr_bias}.  

The results described in the previous paragraph should only be taken as a comparison of the
photometric centroid and difference image techniques, rather than a statistical measurement of the APO population in 
the {\it Kepler} data.  When both the difference image and photometric centroid method agree that there is a significant
offset, while this offset is likely to indicate an APO due to a background false positive, each individual case must be examined to
assure that the offset is not actually due to the systemic errors described in this paper.  When one of the methods
indicates a significant offset but the other does not, it is less likely that the offset is due to a background false positive 
rather than systematic error.  However, an approximately 25\% significant APO rate is consistent with the
observed APO rate described in \S\ref{intro}, averaged over the {\it Kepler} field of view.  

\subsection{Residual Bias and High SNR Transits} \label{section:high_snr_bias}

\begin{figure}[htbp]
\begin{center}
\includegraphics[scale=0.45]{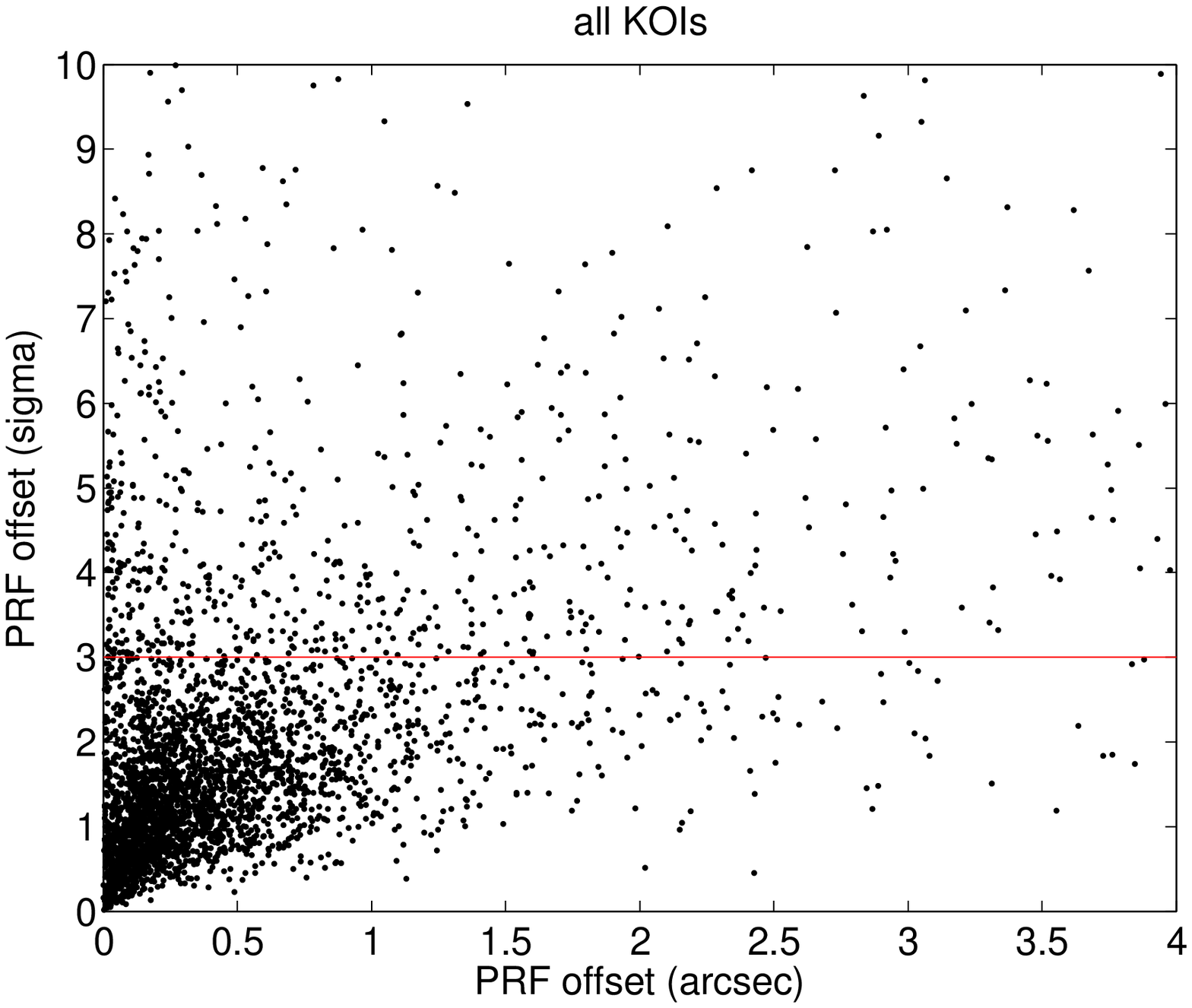} 
\includegraphics[scale=0.45]{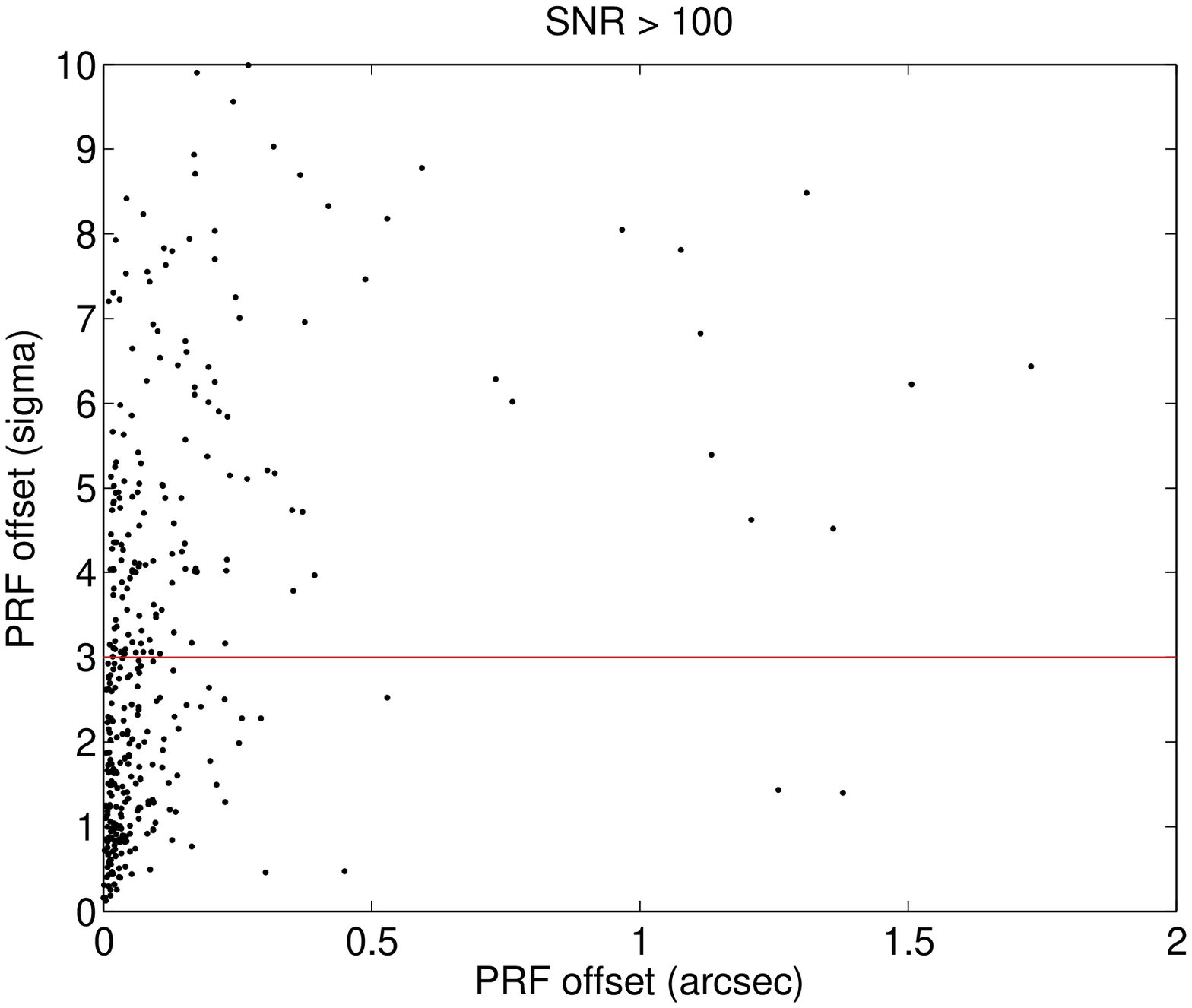} 
\caption{The relationship between the PRF-fit source offset ($x$-axis) and source offset in units of sigma 
($y$-axis).  Left: all KOIs.  Right: KOIs with transit SNR $> 100$.  On the left we see that for offsets 
$< 3$ arcseconds there seem to be an excess of targets with offset $> 3\sigma$ (red line).  On the
right we see that for high SNR targets the offset is small, but there is an excess of targets
with offset $> 3\sigma$.  This is likely due to residual bias from the errors discussed in
\S\ref{section:prf_error}.
}
\label{fig:prf_offset_vs_offset_in_sigma}
\end{center}
\end{figure}

As described in \S\ref{section:prf_error}, the computation of the PRF-fit source offset is subject
to various kinds of bias due to PRF error and crowding.  When the transit SNR is high, both centroid methods will
have very high formal precision with very small uncertainties.  The PRF-fit source offset estimate essentially hits
a noise floor, where the offsets are dominated by residual biases.  Figure~\ref{fig:prf_offset_vs_offset_in_sigma} shows that 
this noise floor begins to be apparent at source offsets of about 2 arcseconds, where there is a noticable
increase in objects with offsets between 3 and  $4\sigma$.  Below about 0.2 arcseconds there is a large
excess of objects with large offsets in units of $\sigma$.  The right panel of Figure~\ref{fig:prf_offset_vs_offset_in_sigma}
shows targets with high SNR.  In this population offsets are mostly very small, and we find most of the large excess 
of high-$\sigma$ offsets.  We interpret this to mean that residual biases in the PRF-fit source offset are dominant under 0.2 arcseconds.


\begin{figure}[htbp]
\begin{center}
\includegraphics[scale=0.45]{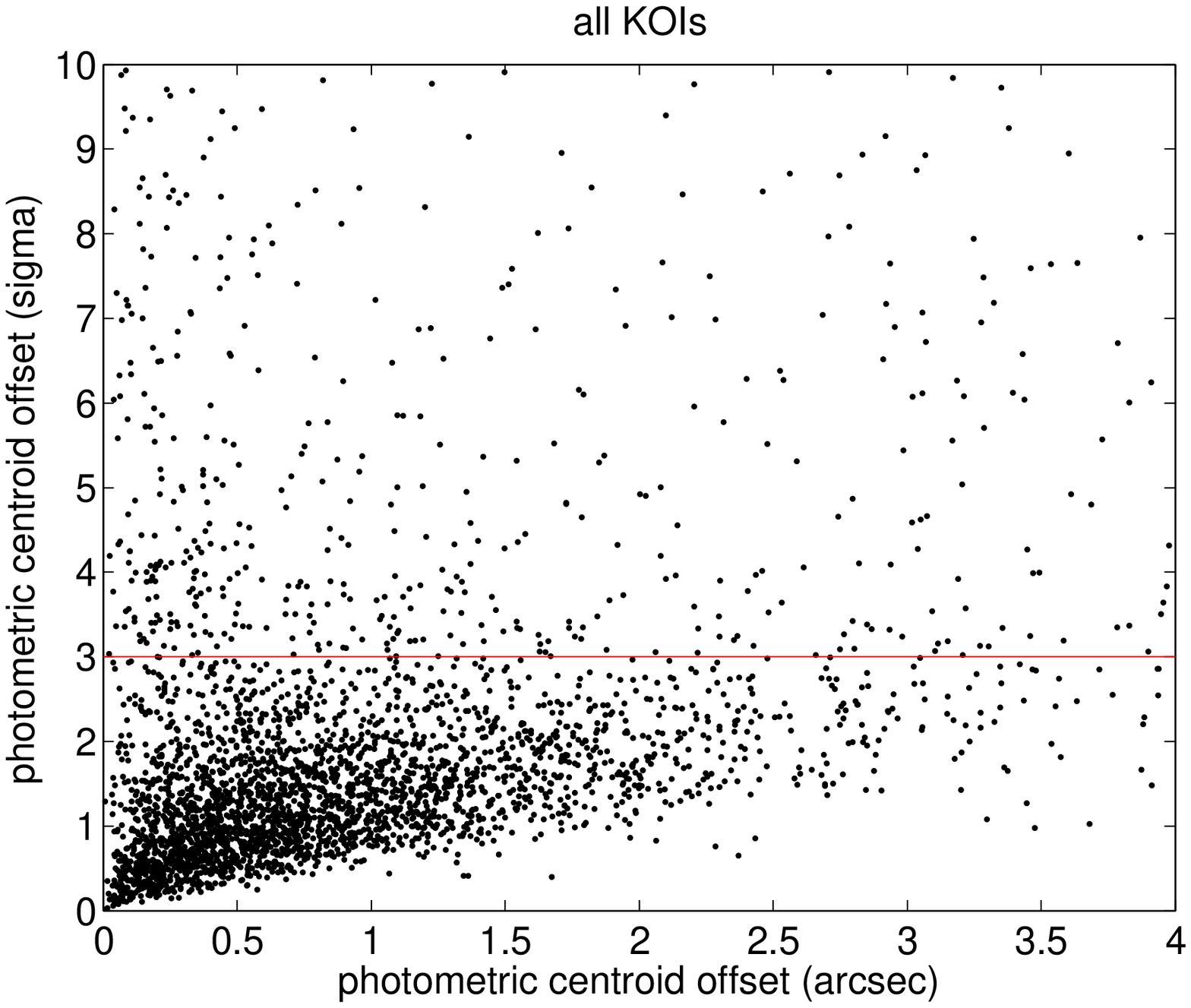} 
\includegraphics[scale=0.45]{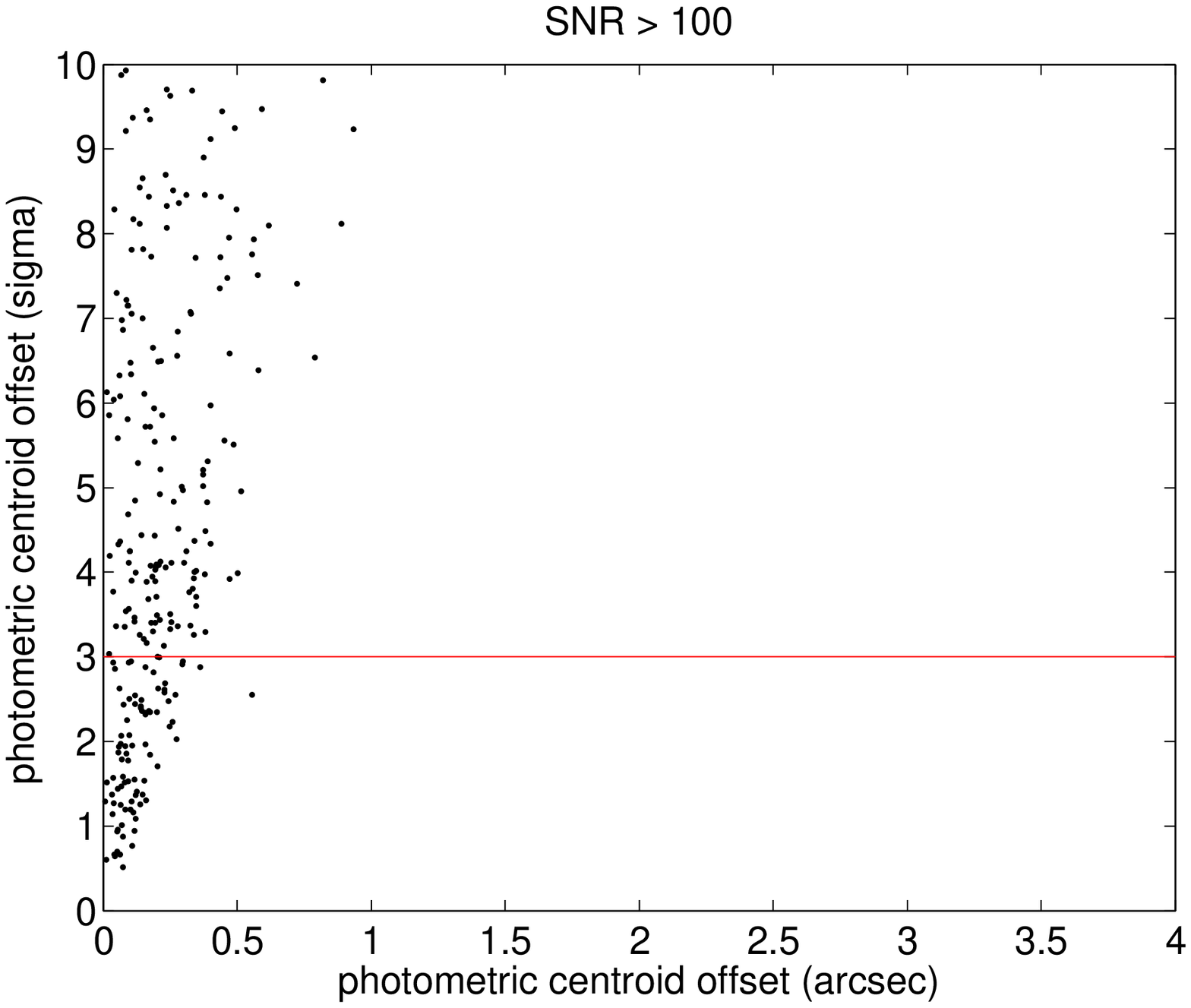} 
\caption{The relationship between the photometric centroid source offset ($x$-axis) and source offset in units of sigma 
($y$-axis).  Left: all KOIs.  Right: KOIs with transit SNR $> 100$.  Many KOIs fall outside 
the plot, but our interest is in small offset behavior.  On the left we see that for offset 
$< 0.2$ arcseconds there seem to be an excess of targets with offset $> 3\sigma$  (red line).  On the
right we see that for high SNR targets the offset is small, but there is an excess of targets
with offset $> 3\sigma$.  
}
\label{fig:fw_offset_vs_offset_in_sigma}
\end{center}
\end{figure}

Figure~\ref{fig:fw_offset_vs_offset_in_sigma} shows a similar analysis for photometric-centroid-based source offsets.  
The excess of significantly offset targets is apparent but less severe in this case.  

\begin{figure}[htbp]
\begin{center}
\includegraphics[scale=0.45]{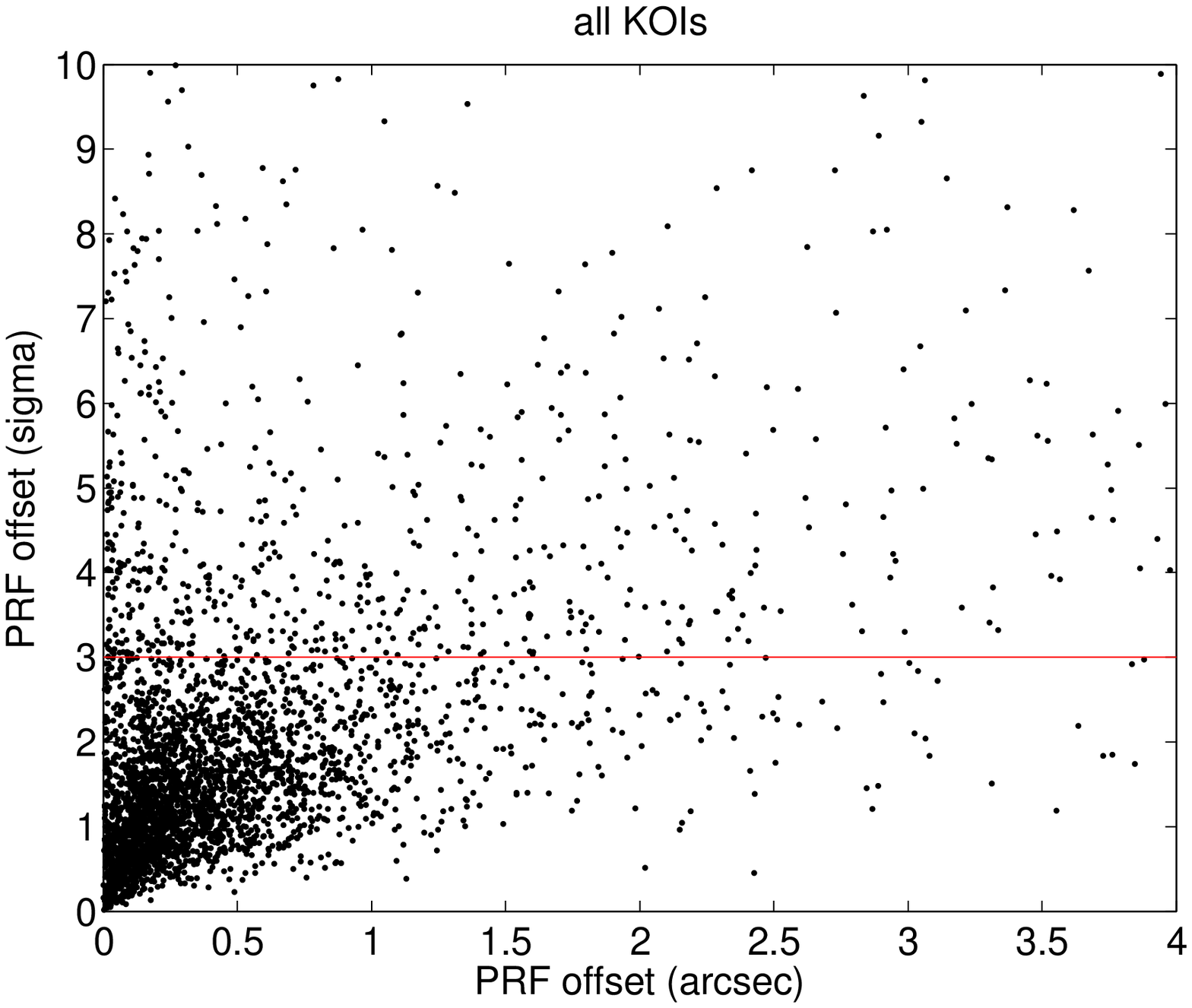} 
\includegraphics[scale=0.45]{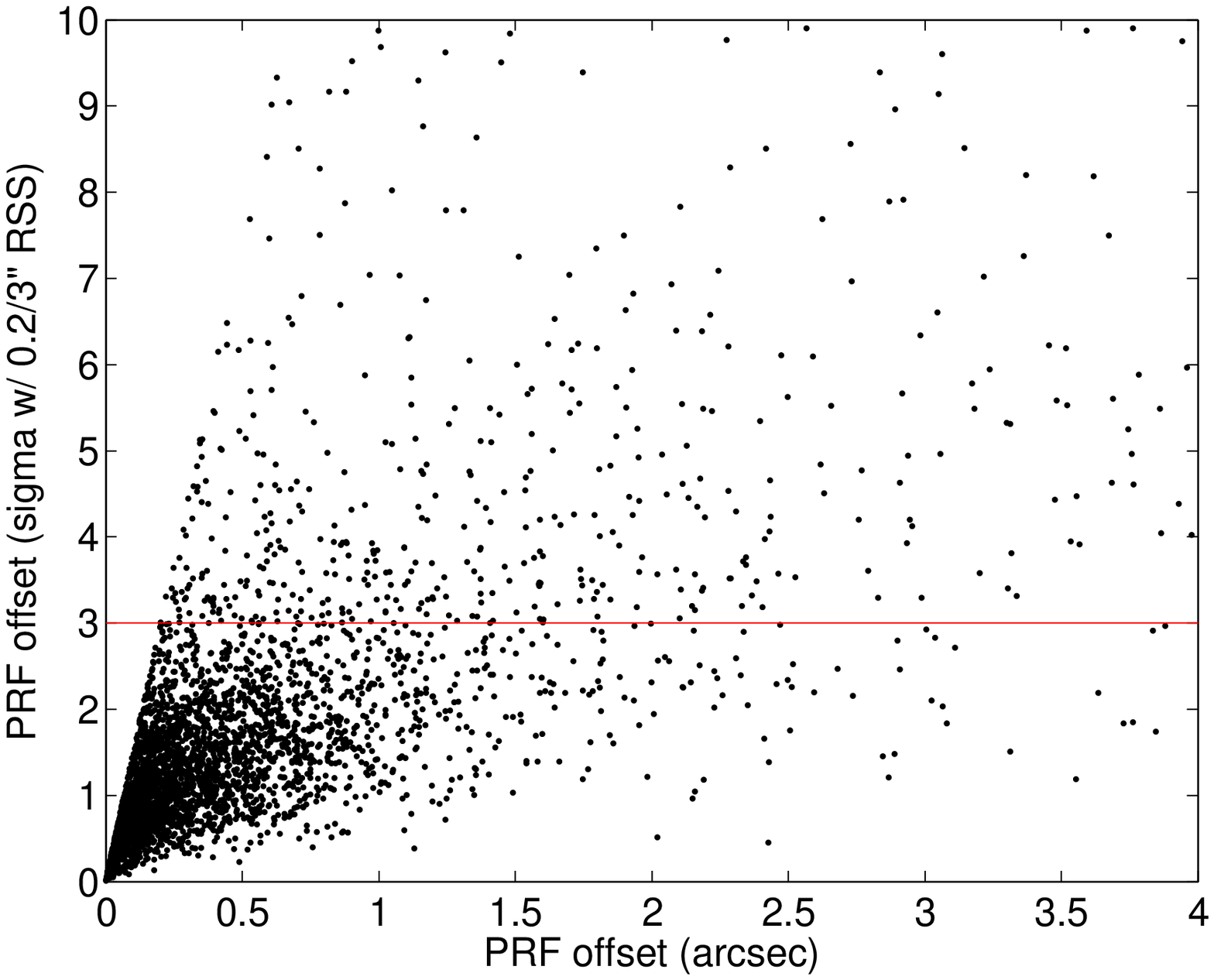} 
\caption{The effect of adding a small constant to the PRF-fit source offset uncertainty on 
the relationship between the PRF-fit source offset ($x$-axis) and source offset in units of sigma 
($y$-axis). Left: all KOIs from Figure~\ref{fig:prf_offset_vs_offset_in_sigma}.  Right: the same targets
with a constant 0.2/3 arcseconds added to the formal uncertainty in quadrature.  The excess of targets
exceeding $3\sigma$ at offset $< 0.2$ sigma has been removed.
}
\label{fig:prf_offset_vs_offset_in_sigma_augment}
\end{center}
\end{figure}

We mitigate the impact of residual bias on small offset / high SNR targets in PRF-fit estimates of the source offset in two ways:
\begin{itemize}
\item Adding a small constant "noise floor" to reflect the residual bias.  Because bias seems to dominate at less than 0.2 arcseconds, 
we want to avoid classifying any target with a source offset less than 0.2 arcseconds as an APO false positive.  Because this classification
is based on a $3\sigma$ threshold we add $\sigma_0 = 0.2/3$ arcseconds in quadrature to the formal uncertainty in each component: 
$\sigma_{\Delta \alpha} \rightarrow \sqrt{\sigma_{\Delta \alpha}^2 + \sigma_0^2}$, 
$\sigma_{\Delta \delta} \rightarrow \sqrt{\sigma_{\Delta \delta}^2 + \sigma_0^2}$.  (This has the same effect on
the offset distance uncertainty $\sigma_{D}$ as adding 
$\sigma_0$ to $\sigma_{D}$ in quadrature).  
The impact of adding this noise floor is shown in Figure~\ref{fig:prf_offset_vs_offset_in_sigma_augment}.
\item Special treatment is given to vetting targets with small source offsets.  An example simple set of rules for manual vetting for false positives 
is the following: 
	\begin{itemize}
	\item pass all targets with offsets $< 0.2$ arcseconds (this happens automatically when using the above noise floor)
	\item for targets with offsets $< 1$ arcsecond, manually investigate those targets with offsets $> 3\sigma$
	\item for targets with offsets between 1 and 2 arcseconds, manually investigate those targets with offsets between 3 and $4\sigma$
	\item for targets with offsets between 1 and 2 arcseconds, declare as APO targets with offsets above $4\sigma$
	\item for targets with offsets $> 2$ arcseconds, declare as APO targets with offsets above $3\sigma$
	\end{itemize}
\end{itemize}

\section{Conclusions}

Many background astrophysical false positives can be identified through centroid analysis of {\it Kepler} pixel data.
The high photometric precision of the {\it Kepler} data provides opportunities to identify such objects 
close to the target star, but great care must be taken to account for various systematic biases.
We have presented three different techniques, two of which were analyzed in detail.  This ensemble 
provides a power arsenal of tools for dispositioning nearly all KOIs. 

The PRF fit technique provides the best accuracy in the localization of transit sources that are not on the target star.
The photometric centroid technique behaves best when the target star is isolated and the transit source is close
to (or is) the target star.  The photometric centroid technique is therefore useful for confirming that the transit is on
the target star when this is also indicated by the PRF fit technique.  The photometric centroid technique can indicate when the 
transit source is separated from the target star, but when the separation is more than a few arcseconds the source location determined
by photometric centroids is unreliable.

When the SNR is low or there is significant crowding, the PRF technique can break down.  In this case the photometric
technique may provide the best evidence that the centroid is on the target star.  The pixel correlation images can also
be useful in this circumstance, though the pixel correlation technique is fragile.

We find that we often use all three techniques when
investigating a difficult target.  This toolbox of techniques is a critical component of the {\it Kepler} planet candidate 
vetting process and makes a significant contribution to the reliability of the {\it Kepler} planet candidate list.

\acknowledgements

\section{Acknowledgements}
We gratefully acknowledge the outstanding work of the entire {\it Kepler} team that performs the data acquisition and analysis,
and delivers the precision that makes the techniques described in this paper possible.  We particularly thank the 
{\it Kepler} Science Operations Center and Science Office for their support and creativity while these techniques 
were being developed.  We thank Martin Still, Susan Thompson and Jeff Coughlin for valuable comments
on early drafts of this paper.  Finally we thank Bill Borucki, Ted Dunham, Dave Latham, Nick Gautier and the wider
{\it Kepler} science community for constant support and encouragement.

     {\it Kepler} was competitively selected as the tenth Discovery mission. Funding for 
this mission is provided by NASA's Science Mission Directorate. 

\noindent{\it Facilities:} \facility{The Kepler Mission}

\appendix

\section{Derivation of the formula relating centroid shifts to transit source location} \label{appendix_a}

Assume that we are observing a target star with flux $b_0$ at $\left( \alpha_0, \delta_0 \right)$, 
with $N$ nearby stars at RA and Dec $\left( \alpha_j, \delta_j \right)$, $j = 1, \ldots
N,$ and flux $b_j$.  Assume the star $k$, with $k \neq 0$, is a background eclipsing binary 
with fractional eclipse depth
$d_{\mathrm{back}}$ (so the flux of star $k$ in mid eclipse is $\left( 1 - d_{\mathrm{back}} \right) b_k$). 
We model the PSF of the star with a function $f \left(\alpha, \delta \right)$ that has the following
properties, where the integral is taken over the domain where $f >0$:
\begin{itemize}
\item $f \left(\alpha, \delta \right)$ has finite support ($f = 0$ outside of a finite area).
\item $\int f \left(\alpha, \delta \right) d \alpha \, d \delta = 1$.  In other words $f$ has unit flux so
$b_j f$ has the total flux 
\begin{equation*}
\int b_j f \left(\alpha, \delta \right) d \alpha \, d \delta = b_j.
\end{equation*}
\item $\int \alpha f\left(\alpha - \alpha_j, \delta - \delta_j \right) d \alpha \, d \delta = \alpha_j$ 
and $\int \delta f\left(\alpha - \alpha_j, \delta - \delta_j \right) d \alpha \, d \delta  = \delta_j$
so, for example,
\begin{equation*}
\frac{\int \alpha b_j f\left(\alpha - \alpha_j, \delta - \delta_j \right) d \alpha \, d \delta}
{\int b_j f\left(\alpha - \alpha_j, \delta - \delta_j \right) d \alpha \, d \delta}
= \frac{\alpha_j b_j}{b_j} = \alpha_j,
\end{equation*}
so the centroid of an isolated star is the same as that star's position.
\end{itemize}
We now consider an aperture on the sky that may not completely capture all flux 
from stars in the aperture, and may contain flux from stars outside the aperture.  
Therefore
$\int_{\mathrm{ap}} b_j f \left(\alpha, \delta \right) d \alpha \, d \delta \neq b_j$,  
$\int_{\mathrm{ap}} \alpha f \left(\alpha - \alpha_k, \delta - \delta_k \right) d \alpha \, d \delta \neq \alpha_k$ 
and $\int_{\mathrm{ap}} \delta f \left(\alpha - \alpha_k, \delta - \delta_k \right) d \alpha \, d \delta  \neq \delta_k$,
where $\int_{\mathrm{ap}}$ denotes an integral over the aperture.
We model the background flux as an arbitrary function $B\left(\alpha, \delta \right)$.
We denote the total flux in the aperture by 
\begin{equation*}
F^{\mathrm{ap}} = \int_{\mathrm{ap}} \left( \sum_{j = 1}^N b_j f \left(\alpha - \alpha_j, \delta - \delta_j \right) 
	+  B\left(\alpha, \delta \right) \right) d \alpha \, d \delta.
\end{equation*}
To simplify the following discussion, we define the notation
\begin{eqnarray*}
I^{\mathrm{ap}}_j &:=& \int_{\mathrm{ap}} f \left(\alpha - \alpha_j, \delta - \delta_j \right) d \alpha \, d \delta, \qquad
B^{\mathrm{ap}} := \int_{\mathrm{ap}}  B\left(\alpha, \delta \right) d \alpha \, d \delta, \\
I^{\mathrm{ap}, \alpha}_j &:=& \int_{\mathrm{ap}}  \alpha f \left(\alpha - \alpha_j, \delta - \delta_j \right) d \alpha \, d \delta, \qquad
I^{\mathrm{ap}, \delta}_j := \int_{\mathrm{ap}}  \delta f \left(\alpha - \alpha_j, \delta - \delta_j \right) d \alpha \, d \delta, \\
B^{\mathrm{ap}, \alpha} &:=& \int_{\mathrm{ap}}  \alpha B\left(\alpha, \delta \right) d \alpha \, d \delta, \qquad 
B^{\mathrm{ap}, \delta} := \int_{\mathrm{ap}}  \delta B\left(\alpha, \delta \right) d \alpha \, d \delta\\
\end{eqnarray*}
So $b_j I^{\mathrm{ap}}_j$ is the flux from star $j$ in the aperture, $B^{\mathrm{ap}}$ is the background flux in the aperture, and
the superscript $\alpha$ or $\delta$ indicates the first moment in RA or Dec of these quantities. 
Then $F^{\mathrm{ap}} = \sum_{j = 1}^N b_j I^{\mathrm{ap}}_j + B^{\mathrm{ap}}$.

The out-of-transit centroid (including all flux in the aperture) is given by
\begin{equation*}
  C^{\mathrm{out}}_\alpha = \frac{\sum_{j = 1}^N b_j I^{\mathrm{ap}, \alpha}_j + B^{\mathrm{ap}, \alpha}}{F^{\mathrm{ap}} }, \qquad
  C^{\mathrm{out}}_\delta = \frac{\sum_{j = 1}^N b_j I^{\mathrm{ap}, \delta}_j + B^{\mathrm{ap}, \delta}}{F ^{\mathrm{ap}}}.
  \end{equation*}
The in-transit centroid is given by
\begin{eqnarray*}
   C^{\mathrm{in}}_\alpha &=& \frac{\sum_{j = 1, j \neq k}^N b_j I^{\mathrm{ap}, \alpha}_j + B^{\mathrm{ap}, \alpha} 
   	+ \left( 1 - d_{\mathrm{back}} \right) b_k I^{\mathrm{ap}, \alpha}_k }
	{\sum_{j = 1, j \neq k}^N b_j I^{\mathrm{ap}}_j + B^{\mathrm{ap}} + \left( 1 - d_{\mathrm{back}} \right) b_k I^{\mathrm{ap}}_k} \\
	&=& \frac{C^{\mathrm{out}}_\alpha F^{\mathrm{ap}} 
	- d_{\mathrm{back}} b_k I^{\mathrm{ap}, \alpha}_k }{F^{\mathrm{ap}} - d_{\mathrm{back}} b_k I^{\mathrm{ap}}_k}, \\
   C^{\mathrm{in}}_\delta &=&  \frac{C^{\mathrm{out}}_\delta F^{\mathrm{ap}} - d_{\mathrm{back}} b_k \delta_k }{F^{\mathrm{ap}} - d_{\mathrm{back}} b_k}.
\end{eqnarray*}

The observed depth is
defined so that the observed flux in mid eclipse is $\left( 1 - d_{\mathrm{obs}} \right) F^{\mathrm{ap}}$. 
Assuming that the eclipse is the only cause of a change in flux, the observed flux in
mid eclipse is also given by $F^{\mathrm{ap}} - d_{\mathrm{back}} b_k I^{\mathrm{ap}}_k$.  Therefore 
$\left( 1 - d_{\mathrm{obs}} \right) F^{\mathrm{ap}} = F^{\mathrm{ap}} - d_{\mathrm{back}} b_k I^{\mathrm{ap}}_k$, so
$d_{\mathrm{obs}} = \frac{d_{\mathrm{back}} b_k I^{\mathrm{ap}}_k}{F^{\mathrm{ap}}}$. 

The centroid shift is given by
\begin{eqnarray*}
  \frac{\Delta C_\alpha}{\cos \delta} & = & C^{\mathrm{in}}_\alpha - C^{\mathrm{out}}_\alpha\\
  & = & \frac{C^{\mathrm{out}}_\alpha F - d_{\mathrm{back}} b_k I^{\mathrm{ap}, \alpha}_k -
  	C^{\mathrm{out}}_\alpha F^{\mathrm{ap}} + C^{\mathrm{out}}_\alpha d_{\mathrm{back}} b_k I^{\mathrm{ap}}_k }
  	{F^{\mathrm{ap}} - d_{\mathrm{back}} b_k I^{\mathrm{ap}}_k}\\
  & = & - \frac{d_{\mathrm{back}} b_k }{F^{\mathrm{ap}}} \frac{I^{\mathrm{ap}, \alpha}_k - C^{\mathrm{out}}_\alpha I^{\mathrm{ap}}_k}{1 - d_{\mathrm{obs}}} \\
  & = & - \frac{d_{\mathrm{obs}} }{1 - d_{\mathrm{obs}}}  \frac{I^{\mathrm{ap}, \alpha}_k - C^{\mathrm{out}}_\alpha I^{\mathrm{ap}}_k}{I^{\mathrm{ap}}_k} \\
  & = & - \frac{d_{\mathrm{obs}} }{1 - d_{\mathrm{obs}}}  \left( \frac{I^{\mathrm{ap}, \alpha}_k }{I^{\mathrm{ap}}_k} - C^{\mathrm{out}}_\alpha \right), \\
  \Delta C_\delta & = & - \frac{d_{\mathrm{obs}} }{1 - d_{\mathrm{obs}}}  \left( \frac{I^{\mathrm{ap}, \delta}_k }{I^{\mathrm{ap}}_k} - C^{\mathrm{out}}_\delta \right).
\end{eqnarray*}
We define 
\begin{equation*}
C^{\mathrm{ap},\alpha}_k =  \frac{I^{\mathrm{ap}, \alpha}_k }{I^{\mathrm{ap}}_k}, \qquad
C^{\mathrm{ap},\delta}_k =  \frac{I^{\mathrm{ap}, \delta}_k }{I^{\mathrm{ap}}_k}, 
\end{equation*}
which are the RA and Dec of the centroid of the flux of the transit source $k$ in the aperture when all other flux is absent
(alternatively this is the centroid of the difference image formed by subtracting in-transit pixels from out-of-transit
pixels when all other flux is constant).
Therefore this centroid is given by
\begin{equation}
  C^{\mathrm{ap},\alpha}_k := C^{\mathrm{out}}_\alpha - \left( \frac{1}{d_{\mathrm{obs}}} - 1 \right) \frac{\Delta C_\alpha}{\cos \delta}, \qquad 
  C^{\mathrm{ap},\delta}_k := C^{\mathrm{out}}_\delta - \left( \frac{1}{d_{\mathrm{obs}}} - 1 \right) \Delta C_\delta.
  \label{eqn:transit_source_centroid_location}
\end{equation}
$\left( C^{\mathrm{ap},\alpha}_k, C^{\mathrm{ap},\delta}_k \right)$ approximate the transit source location $\left( \alpha_k, \delta_k \right)$, 
with the error in this approximation decreasing as more flux from the transit source is captured in the aperture.  
When all flux from the transit source is captured
in the aperture, $\left( C^{\mathrm{ap},\alpha}_k, C^{\mathrm{ap},\delta}_k \right) = \left( \alpha_k, \delta_k \right)$.  

In the {\it Kepler} pipeline implementation, the transit depth is estimated using the optimal aperture \citep{bryson_TAD} while the 
centroids are measured using the optimal aperture plus a one-pixel ring around the optimal aperture.  This is because 
some optimal apertures consist of only a single pixel, which cannot be usefully centroided.  This use of one aperture for centroid 
computation and a smaller aperture to estimate observed transit depth invalidates the conclusion of the above analysis
because $d_{\mathrm{obs}}$ in Equation~(\ref{eqn:transit_source_centroid_location}) is different from the depth
$d_{\mathrm{obs}}^{\mathrm{optAp}}$ determined using the optimal aperture.  

We can estimate the difference in these observed depths and predict the impact on the estimated transit source position.  For the
aperture used for centroiding, we have the relation $d_{\mathrm{obs}} F^{\mathrm{ap}} = d_{\mathrm{back}} b_k I^{\mathrm{ap}}_k$,
while for the optimal aperture we have the same relation: 
$ d_{\mathrm{obs}}^{\mathrm{optAp}} F^{\mathrm{optAp}} = d_{\mathrm{back}} b_k I^{\mathrm{optAp}}_k$.  
Solving both relations for $d_{\mathrm{back}} b_k$
and equating, we find 
\begin{equation}
\frac{d_{\mathrm{obs}} F^{\mathrm{ap}}}{I^{\mathrm{ap}}_k} 
	= \frac{d_{\mathrm{obs}}^{\mathrm{optAp}} F^{\mathrm{optAp}}}{I^{\mathrm{optAp}}_k}
	\Rightarrow
	d_{\mathrm{obs}}^{\mathrm{optAp}} = d_{\mathrm{obs}} \frac{F^{\mathrm{ap}}}{F^{\mathrm{optAp}}}
	\frac{I^{\mathrm{optAp}}_k}{I^{\mathrm{ap}}_k}.
\end{equation}
Because the optimal aperture is contained within the aperture used for centroiding, $F^{\mathrm{ap}}/F^{\mathrm{optAp}} > 1$ 
while $I^{\mathrm{optAp}}_k/I^{\mathrm{ap}}_k < 1$.  In the typical case where the background star is much dimmer than the 
target star, $F^{\mathrm{ap}}/F^{\mathrm{optAp}}$ will be not much greater than 1, while 
$I^{\mathrm{optAp}}_k/I^{\mathrm{ap}}_k$ can be very close to zero, for example when the core of star $k$ is in the 
pixel ring and only its wings are in the optimal aperture.  Therefore $d_{\mathrm{obs}}^{\mathrm{optAp}}$ can be much 
smaller than $d_{\mathrm{obs}}$, resulting in a significant overshoot of star $k$'s position in Equation~(\ref{eqn:transit_source_centroid_location}).  
This overshoot is particularly likely to happen when star $k$ is
outside the optimal aperture, in other words for background stars further from the target star.  
When star $k$ is brighter than stars in the optimal aperture, including the target star, the overshoot is reduced because the 
the flux in the aperture is dominated by the flux from star $k$.
When star $k$ is in the optimal aperture, the impact on Equation~(\ref{eqn:transit_source_centroid_location})
is much less dramatic and it can provide a very good estimate
of the transiting star's position.

\clearpage

\includeFigureFile

\end{document}